\documentclass{aa}  

\usepackage{graphics,graphicx}
\usepackage{txfonts}
\usepackage{natbib}
\usepackage{lscape}
\usepackage{subfig}
\usepackage{longtable}

\newcommand{\phn}     {\phantom{0}}
\newcommand{\phnn}    {\phantom{0}\phantom{0}}
\newcommand{\phnnn}   {\phantom{0}\phantom{0}\phantom{0}}

\newcommand{\et}      {et al.}
\newcommand{\ie}      {i.\,e.,}
\newcommand{\eg}      {e.\,g.,}

\newcommand{\kms}     {km~s$^{-1}$}

\newcommand{\hii}     {H{\small II}}
\newcommand{\uchii}   {UC~H{\small II}}

\newcommand{\mo}      {$M_{\sun}$} 
\newcommand{\MonR}    {Mon~R2}

\begin{document}
\title{Dynamics of cluster-forming hub-filament systems}
\subtitle{The case of the high-mass star-forming complex Monoceros\,R2}

\author{S.~P.~Trevi\~no-Morales\inst{1}
        \and
        A.~Fuente\inst{2}
        \and
        \'A.~S\'anchez-Monge\inst{3}
        \and
        J.~Kainulainen\inst{1,5}
        \and
        P.~Didelon\inst{4}
        \and
        S.~Suri\inst{3,5} 
        \and
        N.~Schneider\inst{3}
        \and
        J.~Ballesteros-Paredes\inst{6}
        \and
        Y.-N.~Lee\inst{7}
        \and
        P.~Hennebelle\inst{4}
        \and
        P.~Pilleri\inst{8}
        \and
        M.~Gonz\'alez-Garc\'ia\inst{9}
        \and  
        C.~Kramer\inst{10}
        \and
        S.~Garc\'ia-Burillo\inst{2}
        \and
        A.~Luna\inst{11}
        \and
        J.~R. Goicoechea\inst{12}
        \and 
        P.~Tremblin\inst{4}
        \and
        S.~Geen\inst{13}
        }
\authorrunning{S.~P.~Trevi\~no-Morales, A.~Fuente, \'A.~S\'anchez-Monge, et al.}
\institute{
   Chalmers University of Technology, Department of Space, Earth and Environment, SE-412 93 Gothenburg, Sweden
   \email{sandra.trevino@chalmers.se}
   \and 
   Observatorio Astron\'omico Nacional, Apdo.\ 112, 28803 Alcal\'a de Henares Madrid, Spain
   \and 
   I.\ Physikalisches Institut, Universit\"at zu K\"oln, Z\"ulpicher Str.\ 77, 50937 K\"oln, Germany
   \and
   Laboratoire AIM, Paris-Saclay, CEA/IRFU/SAp – CNRS – Universit\'e Paris Diderot, 91191 Gif-sur-Yvette Cedex, France
   \and
   Max-Planck-Institute for Astronomy, K\"onigstuhl 17, 69117 Heidelberg, Germany
   \and
   Instituto de Radioastronom\'ia y Astrof\'isica, Universidad Nacional Aut\'onoma de M\'exico, P.O.\ Box 3-72, 58090 Morelia, Mexico
   \and
   Institut de Physique du Globe de Paris, Sorbonne Paris Cit\'e, Universit\'e Paris Diderot, UMR 7154 CNRS, 75005 Paris, France
   \and
   IRAP, Universit\'e de Toulouse, CNRS, UPS, CNES, 9 Av.\ colonel Roche, BP 44346, 31028 Toulouse Cedex 4, France
   \and
   Instituto de Astrof\'isica de Andaluc\'ia, IAA-CSIC, Glorieta de la Astronom\'ia s/n, 18008 Granada, Spain
   \and
   Institut de Radioastronomie Millim\'etrique (IRAM), 300 rue de la Piscine, 38406 Saint Martin d'H\`eres, France
   \and
   Instituto Nacional de Astrof\'isica, \'Optica y Electr\'onica, Luis Enrique Erro \#1, 72840 Tonantzintla, Puebla, Mexico
   \and
   Instituto de F\'isica Fundamental (CSIC). Calle Serrano 121, E-28006, Madrid, Spain.
   \and
   Zentrum f\"ur Astronomie, Institut f\"ur Theoretische Astrophysik, Universit\"at Heidelberg, Albert-Ueberle-Str.\ 2, 69120 Heidelberg, Germany
   }

   \date{Received ????; accepted ????}

\abstract
{High-mass stars and star clusters commonly form within hub-filament systems.
Monoceros~R2 (hereafter \MonR), at a distance of 830~pc, harbors one of the closest such systems, making it an excellent target for case studies.}
{We investigate the morphology, stability and dynamical properties of the \MonR\ hub-filament system.}
{We employ observations of the $^{13}$CO and C$^{18}$O 1$\rightarrow$0 and 2$\rightarrow$1 lines obtained with the IRAM-30m telescope. We also use H$_2$ column density maps derived from \textit{Herschel} dust emission observations.}
{We identified the filamentary network in \MonR\ with the DisPerSE algorithm and characterized the individual filaments as either main (converging into the hub) or secondary (converging to a main filament) filaments. The main filaments have line masses of 30--100~\mo~pc$^{-1}$ and show signs of fragmentation, while the secondary filaments have line masses of 12--60~\mo~pc$^{-1}$ and show fragmentation only sporadically. In the context of Ostriker's hydrostatic filament model, the main filaments are thermally super-critical. If non-thermal motions are included, most of them are trans-critical. Most of the secondary filaments are roughly trans-critical regardless of whether non-thermal motions are included or not. From the morphology and kinematics of the main filaments, we estimate a mass accretion rate of $10^{-4}$--$10^{-3}$~\mo~yr$^{-1}$ into the central hub. The secondary filaments accrete into the main filaments with a rate of 0.1--0.4$\times10^{-4}$~\mo~yr$^{-1}$. The main filaments extend into the central hub. Their velocity gradients increase towards the hub, suggesting acceleration of the gas. We estimate that with the observed infall velocity, the mass-doubling time of the hub is $\sim2.5$~Myr, ten times larger than the free-fall time, suggesting a dynamically old region. These timescales are comparable with the chemical age of the {H{\small II}} region. Inside the hub, the main filaments show a ring- or a spiral-like morphology that exhibits rotation and infall motions. One possible explanation for the morphology is that gas is falling into the central cluster following a spiral-like pattern.}
{}

    \maketitle
 
%
\section{Introduction\label{s:intro}}

\begin{figure*}[h!]
\centering
\includegraphics[width=0.4\textwidth]{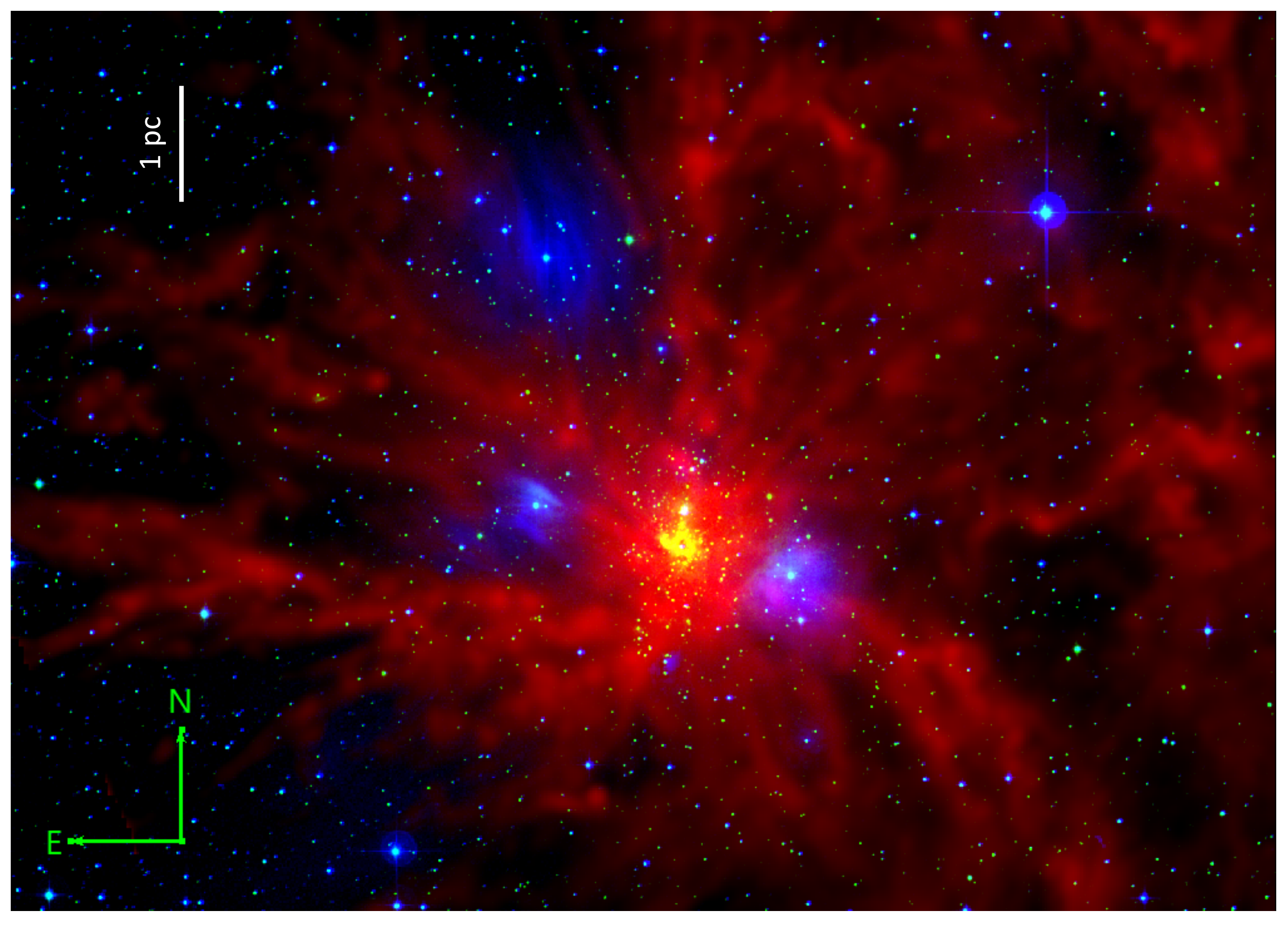}
\includegraphics[width=0.5\textwidth]{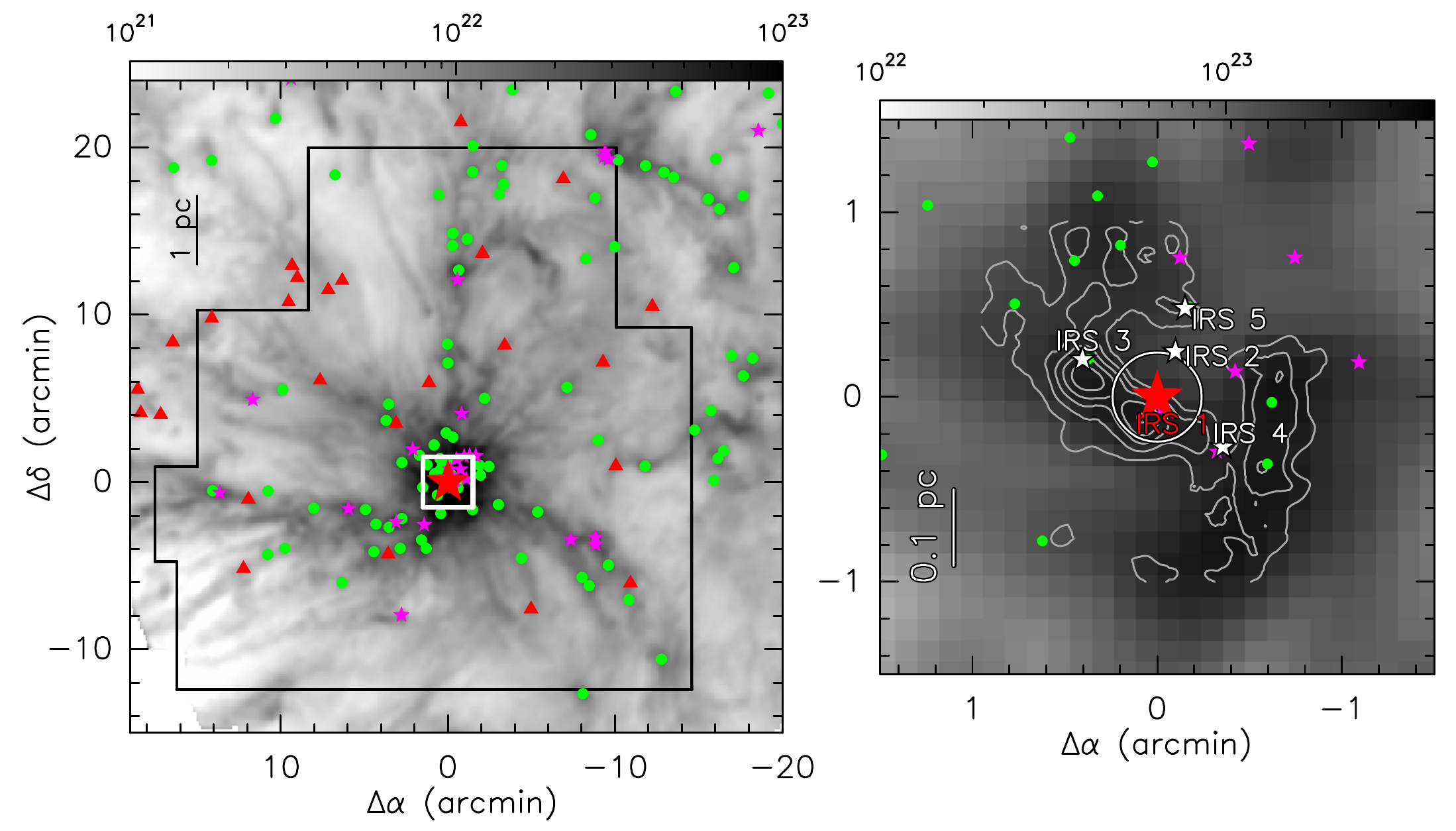}
\caption {\textit{Left}: Three-color image of the \MonR\ cluster-forming hub-filaments system. Red: H$_2$ column density map derived from \textit{Herschel} SPIRE and PACS observations \citep{Didelon+2015}, green: 1.65~$\mu$m band of
2MASS (Two micron all sky survey; \citealt{Skrutskie+2006}), and blue: 560~nm band of DSS (Digitalized Sky Survey; \citealt{Lasker+1990}). \textit{Center}: \textit{Herschel} H$_2$ column density (in cm$^{-2}$, \citealt{Didelon+2015}). The black polygon shows the area surveyed with the IRAM-30m telescope, while the white box corresponds to the inner $0.7$~pc$\times0.7$~pc around the central hub and zoomed in the right panel. \textit{Right}: \textit{Herschel} H$_2$ column density (in cm$^{-2}$) of the central hub of \MonR. Grey contours show the H$^{13}$CO$^{+}$\,(3$\rightarrow$2) emission tracing the high density molecular gas \citep{Trevino-morales+2014}. The red star marks the position of IRS~1 (with coordinates $\alpha$(J2000) $=06^{\mathrm{h}}07^{\mathrm{m}}46.2^{\mathrm{s}}$, $\delta$(J2000) $= -06^{\circ}23'08.3''$). White stars indicate the positions of infrared sources. The white circle indicates the beam size of the IRAM-30m telescope at 100~GHz (see Section~\ref{sec:MonR2Observations}). The colored symbols the sources identified by \cite{Rayner+2017}: Pink stars are protostars, green circles are bound clumps, and red triangles are unbound clumps.}
\label{fig:MonR2-intro}
\end{figure*} 

In the last decades, our view of star-forming regions has been going under a revolution thanks to the new observational facilities. Space telescopes such as \emph{Spitzer} and \textit{Herschel} had provided
observations of a large number of molecular clouds that reveal an ubiquity of filamentary structures containing stars in different evolutionary stages \citep[\eg][]{Schneider-Elmegreen1979, Loren1989a, Loren1989b, Nagai+1998, Myers2009, Andre+2010, Molinari+2010, Schneider+2010, Busquet+2013, Stutz+2013, Kirk+2013, Peretto+2014, Feher+2016, Abreu-Vicente+2016}. Filamentary structures pervading clouds are unstable against both radial collapse and fragmentation \citep[\eg][]{Larson1985, Miyama+1987a, Miyama+1987b, Inutsuka-Miyama1997}, and although their origin or formation process is still unclear, turbulence and gravity \citep[\eg][]{Klessen+2000, Andre+2010} can produce, together with the presence of magnetic fields \citep[\eg][]{Molina+2012, Kirk+2015}, the observed structures. It is thought that star formation occurs preferentially along the filaments, with high-mass stars forming in the highest density regions where several filaments converge, called \emph{ridges} or \emph{hubs} ($N_\mathrm{H}\sim10^{23}$~cm$^{-2}$ and $n_\mathrm{H_2}\sim 10^{6}$~cm$^{-3}$, \eg\ \citealt{Schneider+2010, Schneider+2012, Liu+2012, Peretto+2013, Peretto+2014, Louvet+2014}). This suggests that filaments precede the onset of star formation, funneling interstellar gas and dust into increasingly denser concentrations that will contract and fragment leading to gravitationally bound prestellar cores that will eventually form both low and high-mass stars. Following this process, high-mass stars can inject large amounts of radiation and turbulence in the surrounding medium, that may affect the structural properties of filaments leading to a different level of fragmentation \citep[\eg][]{Csengeri+2011, Seifried-Walch2015, Seifried-Walch2016}.

In the last years, an increasing number of works have focused on the study of the dynamics and fragmentation of filamentary structures from both, observational and theoretical points of view \citep[see \eg][]{Andre+2010, Schneider+2010, Schneider+2012, Hennemann+2012, Busquet+2013, Galvan-Madrid+2013, Hacar+2013, Hacar+2018, Peretto+2013, Louvet+2014, Tafalla+2015, Smith+2014, Henshaw+2014, Tackenbergt+2014, Seifried-Walch2016, Kainulainen+2017, Seifried+2017, Arzoumanian+2019, Williams+2018, Clarke+2019}. However, few of these works are focus on massive star forming regions within hub-filament system, and little is still known about the dynamics of filamentary networks (\eg\ cluster-forming hub filament systems) and their role in the accretion processes that regulate the formation of high-mass star-forming clusters. In addition, most of the research on high-mass star-forming regions focus on the study of one particular cloud: the Orion A molecular cloud \citep[\eg][]{Hacar+2018, Suri+2019}. Thus, and with the goal of having a better understanding of the filament properties in high-mass star-forming regions, it is necessary to study other massive clouds. For this, the Monoceros star-forming complex appears as an ideal target.

Located at a distance of only 830~pc (Racine 1968), Monoceros~R2 (hereafter \MonR) is an active massive star forming cloud that hosts one of the closest ultracompact (UC)~\hii\ regions. Recently, \textit{Herschel} observations have revealed an intriguing look of the cloud with several filaments converging into the central area ($\sim2.25$~pc$^2$, see left panel in Fig.~\ref{fig:MonR2-intro}; \citealt{Didelon+2015, Pokhrel+2016, Rayner+2017}). A number of hot bubbles and already-developed \hii\ regions are identified throughout the region (visible in blue in the image shown in Fig.~\ref{fig:MonR2-intro}-left) mainly in the outskirts of the central and densest region, where a cluster of young high-mass stars is found to be forming at the junction (or hub) of the filamentary structures. The most massive star of this infrared cluster is IRS~1, at $\alpha$(J2000) $= 06^{\mathrm{h}}07^{\mathrm{m}}46.2^{\mathrm{s}}$, $\delta$(J2000) $= -06^{\circ}23'08.3''$, with a mass of $\sim$12~M$_\odot$ \citep[\eg][]{Thronson+1980, Giannakopoulou+1996}. This source is driving an UC~\hii\ region that has created a cavity free of molecular gas extending for about 30$^{\prime\prime}$ (or 0.12~pc, \eg\ \citealt{Choi+2000, Dierickx+2015}) and surrounded by a number of photon-dominated regions (PDRs) with different physical and chemical conditions \citep[\eg][]{Ginard+2012, Pilleri+2012, Trevino-morales+2014, Trevino-morales+2016}. Based on \textit{Herschel} PACS and SPIRE maps, \citet{Didelon+2015} determined that the central region hosting the UC~\hii\ region shows a power-law density profile of $\rho(r)\propto r^{-2.5}$. This density profile was attributed to an external pressure certainly associated with global collapse. \citet{Rayner+2017} studied the distribution of dense cores and young stellar objects in the region and proposed that the hub may be sustaining its star formation by filamentary accretion of material from the large-scale mass reservoir \citep[see also][]{Trevino-morales2016}.

In summary, and thanks to its morphology, proximity and general characteristics, \MonR\ appears as one of the clearest examples of a hub-filament system, thus being an excellent target to study in detail the physical properties of these systems. In this paper, we report observations of the \MonR\ star-forming region conducted with the IRAM-30m telescope. We observed different molecular line transitions that allow us to study the molecular gas content in the region, and for the first time, study the large-scale gas dynamics of its filamentary structure. The observational data are introduced in Sect.~\ref{sec:MonR2Observations}. In Sect.~\ref{sec:MonR2Res}, we present the large-scale (at parsec scales) structure of the molecular gas, while in Sect.~\ref{sec:analysis} we analyze the filamentary structure in \MonR, giving special emphasis on the kinematic properties and zooming into the central hub. A general discussion and a summary of the main results are presented in Sect.~\ref{sec:discussion} and~\ref{sec:summary}, respectively.

%
\section{Observations and data reduction\label{sec:MonR2Observations}}

\begin{figure}[t]
\centering
\includegraphics[width=0.4\textwidth]{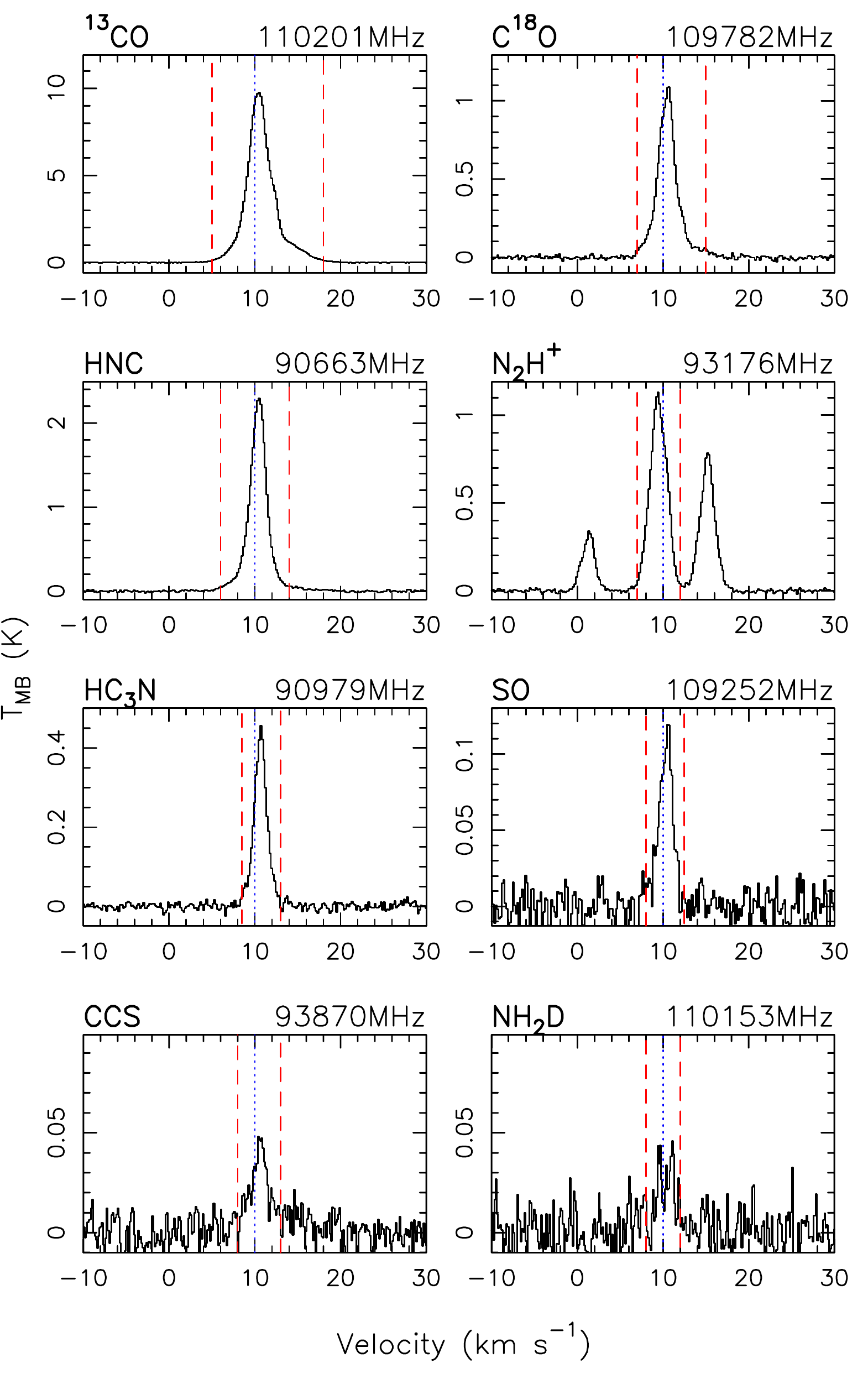}
\caption{Spectra averaged over an area of $0.7$~pc$\times0.7$~pc centered at the position of IRS~1 (corresponding to the area shown in Fig.~\ref{fig:MonR2-intro}-right). The blue, dotted vertical line indicates the source velocity ($v_{\mathrm{LSR}}=10$~\kms). The red, dashed vertical lines indicate the velocity range where the $S/N$ ratio is above 3$\sigma$ for the molecular emission. These ranges are used to generate the integrated intensity maps presented in Fig.~\ref{fig:large-scale-moments}: 5--18~\kms\ for $^{13}$CO, 7--15~\kms\ for C$^{18}$O, 6--14~\kms\ for HNC, and 7--12~\kms\ for N$_2$H$^{+}$.}
\label{fig:02714spectra}
\end{figure} 

\begin{table}[t]
\centering
\caption{Observational parameters of the main detected lines.}
\begin{tabular}{l c c c c c c}
\hline\hline\noalign{\smallskip}
&
&Freq.
&HPBW\tablefootmark{a}
&$B_{\rm{eff}}$\tablefootmark{a}
&rms\tablefootmark{b}
\\
Species
&Transition
&(GHz)
&(arcsec)
&(\%)
&(mK)
\\
\hline
\noalign{\smallskip}
HNC           &1$_{0,0}$--0$_{0,0}$ &\phn90.664  &28.6 &80   &0.15 \\
HC$_3$N       &10--9                &\phn90.979  &28.5 &80   &0.15 \\
N$_2$H$^{+}$  &1--0                 &\phn93.173  &27.8 &80   &0.14 \\
CCS           &7$_{8}$--6$_{7}$     &\phn93.870  &27.6 &80   &0.15 \\
HC$_3$N       &12--11               &109.174     &23.8 &80   &0.20 \\
SO            &3$_{2}$--2$_{1}$     &109.252     &23.7 &80   &0.20 \\
C$^{18}$O     &1--0                 &109.782     &23.6 &80   &0.22 \\
NH$_2$D       &1$_{1,1}$--1$_{0,1}$ &110.154     &23.5 &79   &0.23 \\
$^{13}$CO     &1--0                 &110.201     &23.5 &79   &0.24 \\
C$^{18}$O     &2--1                 &219.560     &10.5 &58   &0.15 \\
$^{13}$CO     &2--1                 &220.399     &10.5 &58   &0.15 \\
\hline
\end{tabular} 
\tablefoot{
\tablefoottext{a}{The values of HPBW (half-power beam width), $F_{\rm{eff}}$ (forward efficiency: 95\% between 90 and 110~GHz, and 92\% at 220~GHz) and $B_{\rm{eff}}$ (beam efficiency) are taken from \url{http://www.iram.es/IRAMES/mainWiki/Iram30mEfficiencies}.}
\tablefoottext{b}{Rms noise level over the whole surveyed area. The rms is given at the nominal resolutions of the spectrometers used, as described in Sect.~\ref{sec:MonR2Observations}.}
}
\label{tab:observations}
\end{table}

We observed the \MonR\ star-forming region with the IRAM-30m telescope (Pico Veleta, Spain). The observations were conducted between July 2014 and December 2016\footnote{Under the project numbers 027-14, 035-15 and D03-16; PI: A. Fuente and S. P. Treviño-Morales} under good weather conditions, with precipitable water vapor (pwv) between 1 and 3~mm and $\tau\sim$ 0.06--0.18\footnote{The atmospheric opacity $\tau$ at 225~GHz is calculated from the expression $\tau(225) = 0.058\times$pwv$+0.004$}. We used the on-the-fly (OTF) mapping technique to cover a field of view of 855~arcmin$^{2}$ at 3~mm in dual polarization mode using the EMIR receivers \citep{Carter+2012}, with the Fast Fourier Transform spectrometer (FTS) at 50~kHz of resolution \citep{Klein+2012}. The observed area is indicated with a black polygon in the middle panel of Fig.~\ref{fig:MonR2-intro}, where the offset [0\arcsec,0\arcsec] corresponds to the position of the IRS~1 star. The molecular spectral lines covered and detected within our spectral setup are listed in Table~\ref{tab:observations}. During the observations, the pointing was corrected by observing the strong nearby quasar 0605$-$058 every 1--2~h, and the focus by observing a planet every 3--4~h. Pointing and focus corrections were stable throughout all the runs.

The data were reduced with a standard procedure using the CLASS/GILDAS package\footnote{See http://www.iram.fr/IRAMFR/GILDAS for information on the GILDAS software.} \citep{Pety+2005}. For each molecular transition listed in Table~\ref{tab:observations}, we created individual data cubes centered at the source velocity ($v_{\mathrm{LSR}}=10$~\kms), and spanning a velocity range of $\pm60$~\kms. The native spectral resolution across the whole observed frequency band varies between 0.13 and 0.16~\kms. In order to perform a proper comparison of the line profiles of every molecule, we smoothed it to a common value of 0.17~\kms. A two-order polynomial baseline was applied for baseline subtraction. The final data do not show platforming effects and/or spikes (bad channels) in the observed sub-bands. The emission from the sky was subtracted using different reference positions, which were observed every 2~min for a duration of 20~s. Single-pointing observations of the reference positions revealed the presence of weak $^{13}$CO\,(1$\rightarrow$0) emission ($T_\mathrm{MB}<300$~mK), but not from the other transitions included in the setup. We corrected the $^{13}$CO\,(1$\rightarrow$0) emission data-cube of \MonR\ by adding synthetic spectra derived from Gaussian fits to the emission found in the reference positions. Throughout this paper, we use the main beam brightness temperature ($T_\mathrm{MB}$) as intensity scale, while the output of the telescope is usually calibrated in antenna temperature ($T_{\mathrm{A}}^{*}$). The conversion between $T_{\mathrm{A}}^{*}$ and $T_\mathrm{MB}$ is done by applying the factor $F_{\rm{eff}}/B_{\rm{eff}}$, where $F_{\rm{eff}}$ is the forward efficiency which equals 95\%, and $B_{\rm{eff}}$ is the beam efficiency (see Table~\ref{tab:observations}).

In addition to the IRAM-30m data at 3~mm, we also make use of complementary C$^{18}$O and $^{13}$CO\,(2$\rightarrow$1) maps. These maps were obtained with the IRAM-30m telescope during 2013 (PI: P.\ Pilleri). The observations were performed using the same technique described above, but combining the EMIR receivers with the FTS backed at 200~kHz of resolution. The $J$=2$\rightarrow$1 maps cover an area of about 10~arcmin$^{2}$ around the IRS~1 star. The data were processed following the strategy described above.

%
\section{Parsec-scale molecular emission\label{sec:MonR2Res}}

Figure~\ref{fig:02714spectra} shows the spectra for the detected species averaged over an area of $3\arcmin\times3\arcmin$ (or 0.7~pc$\times$0.7~pc at the distance of \MonR), corresponding to the inner part of the hub (see Fig.~\ref{fig:MonR2-intro} right). Among all the detected species, $^{13}$CO, C$^{18}$O, HNC and N$_2$H$^+$ are the brightest with $T_\mathrm{MB}\ge1$~K. For these species, the emission spans a velocity range of $\sim$13~\kms\ for $^{13}$CO, $\sim$8--10~\kms\ for C$^{18}$O and HNC, and $\sim$5~\kms\ for N$_2$H$^+$. The emission from the other species (\ie\ HC$_3$N, SO, CCS and NH$_2$D) spans a velocity range of 4--6~\kms\ and presents weaker intensities with $T_{\mathrm{MB}}<1$~K. In Fig.~\ref{fig:large-scale-moments}, we show the integrated intensity (left column), velocity centroid (middle column) and linewidth (right column) maps for the $^{13}$CO\,(1$\rightarrow$0), C$^{18}$O\,(1$\rightarrow$0), HNC\,(1$\rightarrow$0) and N$_2$H$^+$\,(1$\rightarrow$0) molecular lines, from top to bottom rows. The velocity range considered includes emission above 3$\sigma$ (see red, dashed vertical lines in Fig.~\ref{fig:02714spectra}). 

As seen in the top panels of Fig.~\ref{fig:large-scale-moments}, the CO isotopologues show extended emission distributed across all the surveyed area revealing a set of filaments coming from all directions to flow into the central hub. For clarity, we refer to various relevant structures seen in the maps as \textit{N} for the north-south elongated structure, \textit{NE} for the structure to the north-east of the central hub, \textit{E} for the structure extending to the east, and \textit{SW} for the emission towards the south-west of the central area. For the HNC and N$_2$H$^{+}$ species (see bottom panels), the emission is mainly found in the central region. However, these species also show faint extended emission coincident with the elongated structures identified in the $^{13}$CO and C$^{18}$O maps. The lack of N$_2$H$^+$ emission within the elongated structures might mean that CO could be frozen-out outside the central hub. These structures are also traced by HNC and N$_2$H$^{+}$, but their lower abundances result in a lower S/N ratio which challenges their detection. In the following, we use the $^{13}$CO and C$^{18}$O\,(1$\rightarrow$0) lines to study the physical properties and kinematics of the extended structures in \MonR.

The central area around IRS~1 is bright in all the observed species, but some different features can be distinguished. The emission of most of the detected species appears mainly in an arc/shell structure surrounding the central cluster of infrared stars (see red star in Fig.~\ref{fig:large-scale-moments}, see also right panel of Fig.~\ref{fig:MonR2-intro}) that pinpoint the location of newly-formed stars in \MonR. The arc structure points toward the south of the infrared cluster, in agreement with the cometary shape of the \hii\ region as revealed in previous works \citep[\eg][]{Ginard+2012, Pilleri+2012, Marti+2013}. The observed species present their strongest emission to the north-east and south-west of the infrared cluster. HNC and N$_2$H$^+$ maps show a third bright peak to the south of the cluster, where the CO intensity decreases. This spatial differentiation may be due to different physical conditions causing $^{13}$CO and C$^{18}$O to be depleted onto dust grains and/or a high opacity that results in self-absorption of the CO lines. However, the spectra at these positions show Gaussian profiles with no signatures of self-absorption. A more detailed study of the chemical properties in this region is the subject of a forthcoming paper.

The middle-column panels in Fig.~\ref{fig:large-scale-moments} show the velocity field as determined from the first-order moment analysis. The region presents complex kinematics with different velocity components and velocity gradients. At large scales, there is a global velocity gradient ($\sim$1.5~\kms/pc) from east to west. At smaller scales, we do not find a clear velocity gradient along the N structure, with most of the emission at systemic velocities ($\sim$10~\kms). The NE structure is mainly blue-shifted, with a velocity $\sim$8.5~\kms. The E structure shows a velocity gradient of $\sim$3~\kms from east (at 7.5~\kms) to the center of the region (at 10.5~\kms). Finally, the southern part of SW is red-shifted (11~\kms), but shows a velocity gradient towards the central part, reaching a velocity of 9.5~\kms. In addition to the longitudinal gradients, these four structures also show signatures of smaller velocity gradients ($\sim$1~\kms) across them. The velocity features of these structures are studied in more detail in Sect.~\ref{sec:kinematics}. The velocity structure around the hub is similar in all the species with a prominent Northeast-Southwestern velocity gradient. Interestingly, the blue-shifted gas is reminiscent of an elongated curved structure that starts to the west of IRS~1 and approaches the center through the north. The red-shifted emission, although not as clear as for the blue-shifted component, also seems to converge towards the IRS~1 position from the east and then south, constituting a complementary curved structure to the blue-shifted one (see Sect.~\ref{sec:hub} for a detailed discussion).

The right-column panels of Fig.~\ref{fig:large-scale-moments} show the velocity dispersion as determined from the second-order moment analysis. The extended emission has a constant, relatively narrow linewidth of $\sim$1--1.5~\kms, which increases towards the central part, reaching a maximum value of $\sim$6~\kms\ for $^{13}$CO, $\sim$4~\kms\ for C$^{18}$O, $\sim$4~\kms\ for HNC, and $\sim$2.0~\kms\ for N$_2$H$^+$. These large linewidths are more likely the consequence of the complex kinematics in the inner region which is not resolved by the IRAM-30m beam.

%
\section{The filamentary network of \MonR\label{sec:analysis}}

In the following section we analyze the structure of the dense gas in \MonR, concentrating on the characterization of the filamentary structure previously seen in dust continuum emission maps with \textit{Herschel} and now, for the first time, resolved in velocity in different molecular species. In Sects.~\ref{sec:columndensity} and \ref{sec:identification}, we derive column density maps from molecular line emission and identify filamentary structures from the position-position-velocity datacubes. The stability of the filaments is explored in Sect.~\ref{sec:stability}, and their kinematic properties are discussed in Sect.~\ref{sec:kinematics}. We study the convergence of the filaments into the central hub in Sect.~\ref{sec:hub}.

%
\subsection{Column density structure \label{sec:columndensity}}

\begin{figure*}[ht!]
\centering
\begin{tabular}{c c}
  \vspace{-0.85cm}
  \hspace{-0.55cm}
  \includegraphics[angle=0, width=0.34\textwidth]{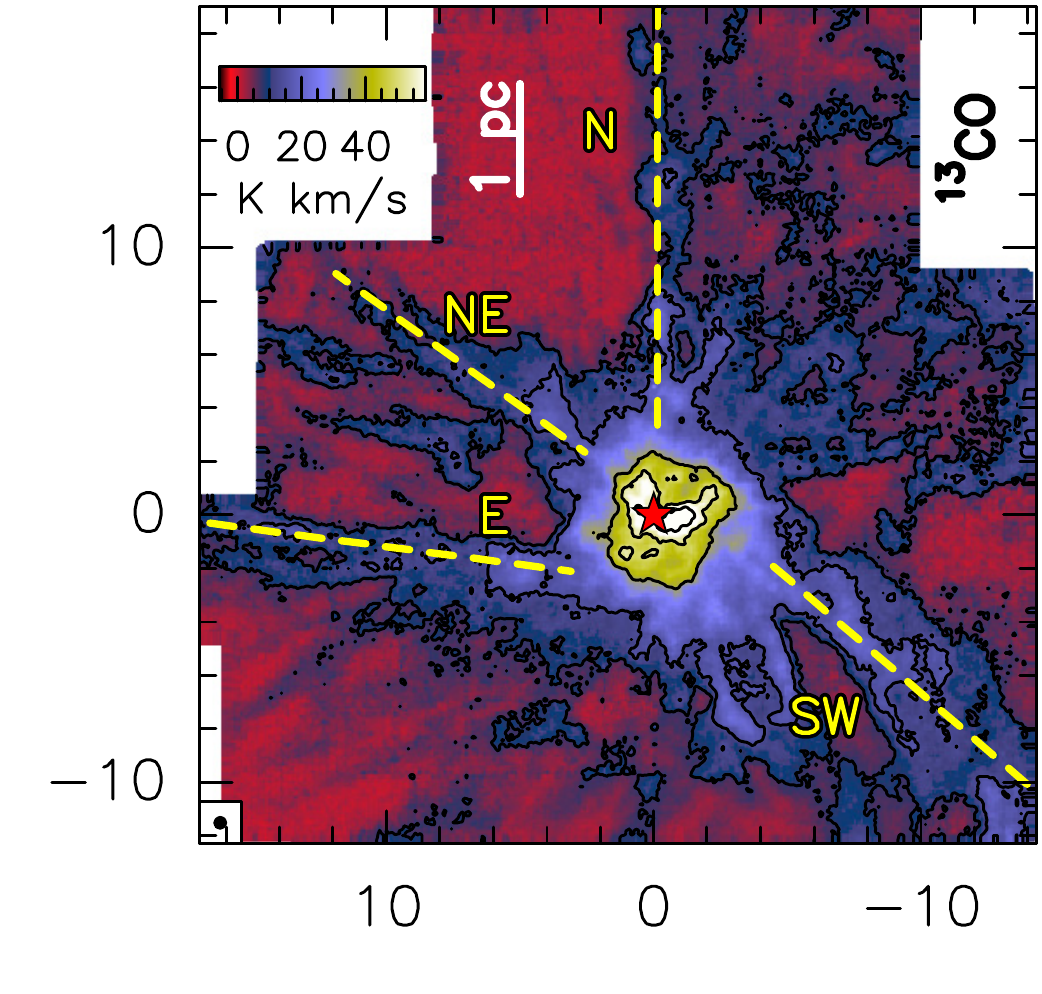} &
  \hspace{-0.5cm}
  \includegraphics[angle=0, width=0.583\textwidth]{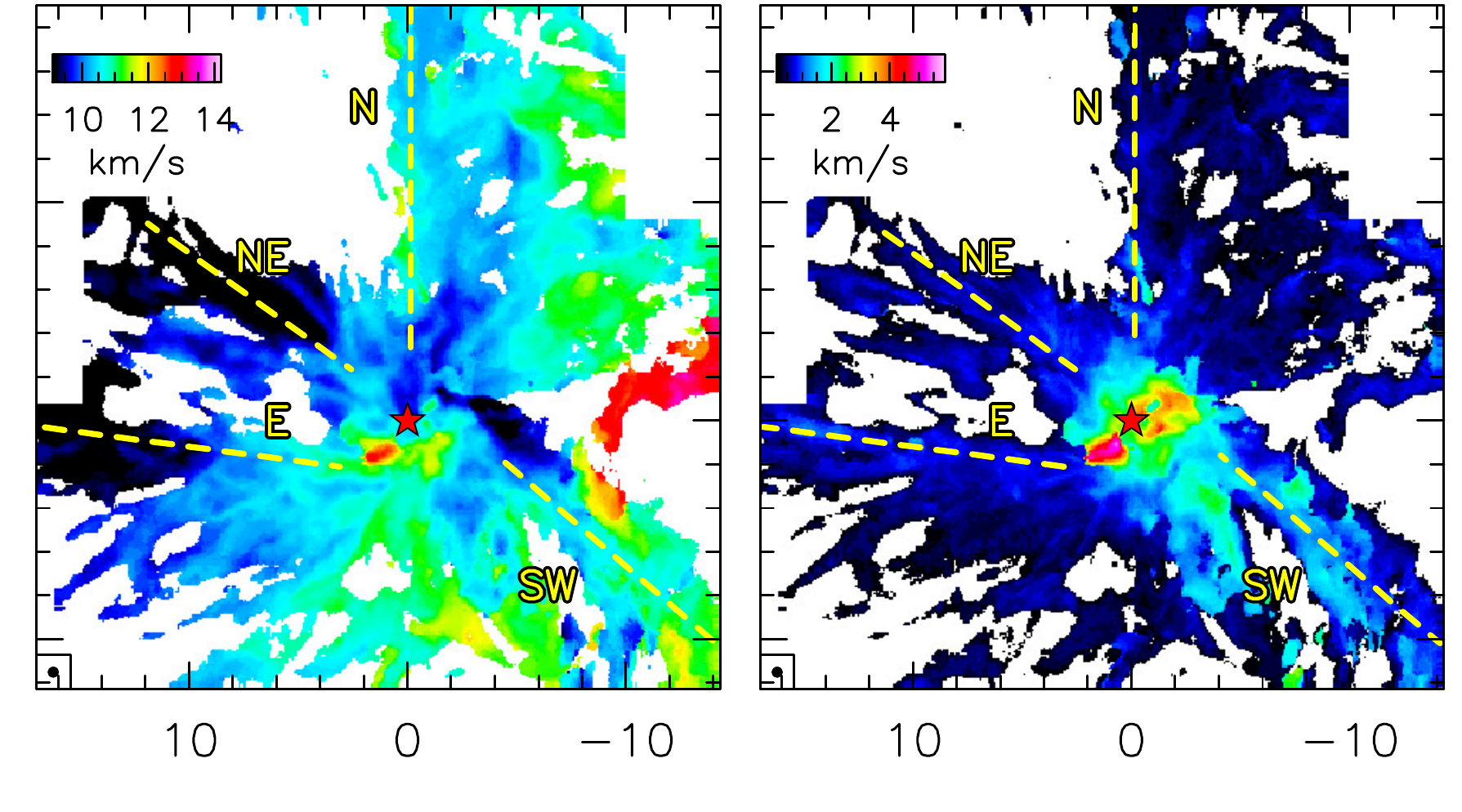} \\
  \vspace{-0.85cm}
  \hspace{-0.55cm}
  \includegraphics[angle=0, width=0.34\textwidth]{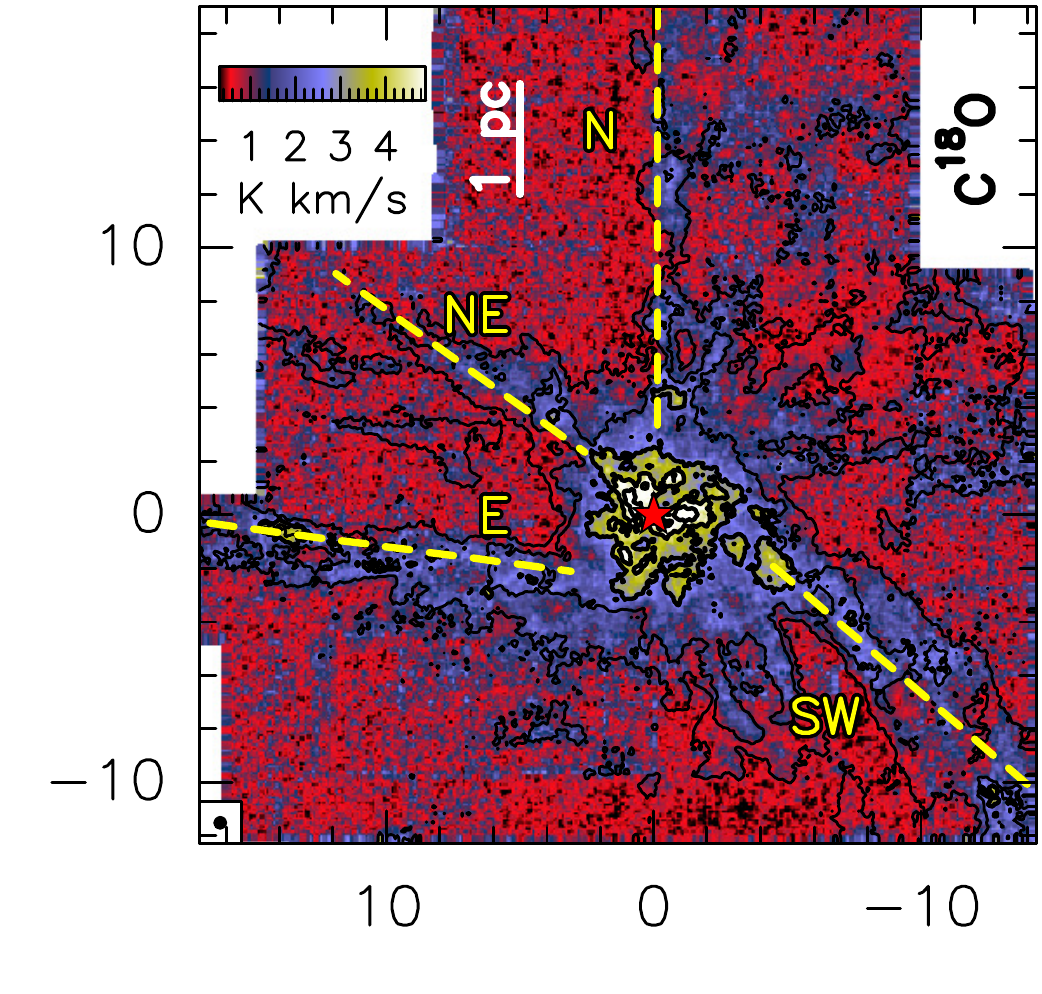} &
  \hspace{-0.5cm}
  \includegraphics[angle=0, width=0.583\textwidth]{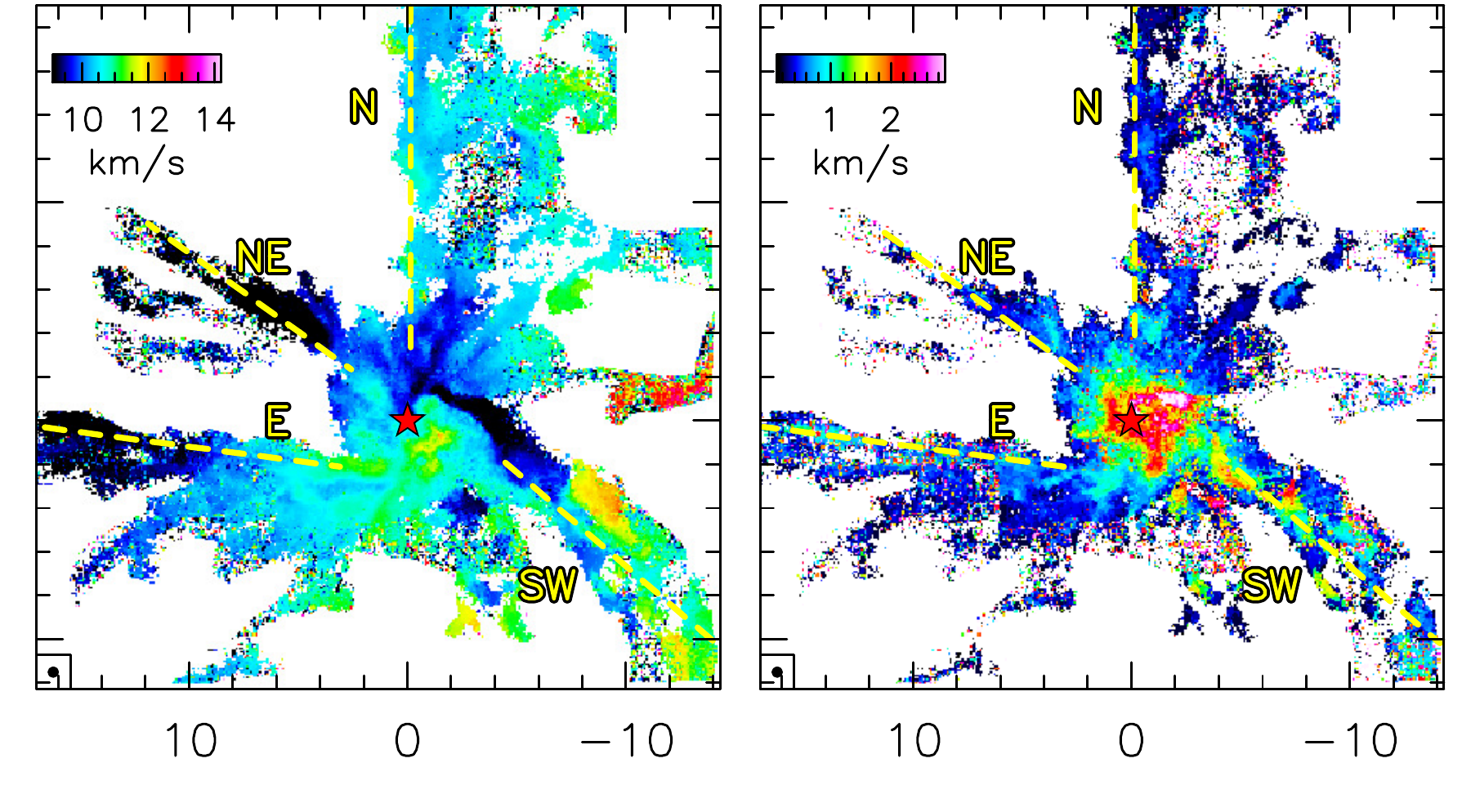} \\
  \vspace{-0.85cm}
  \hspace{-0.55cm}
  \includegraphics[angle=0, width=0.34\textwidth]{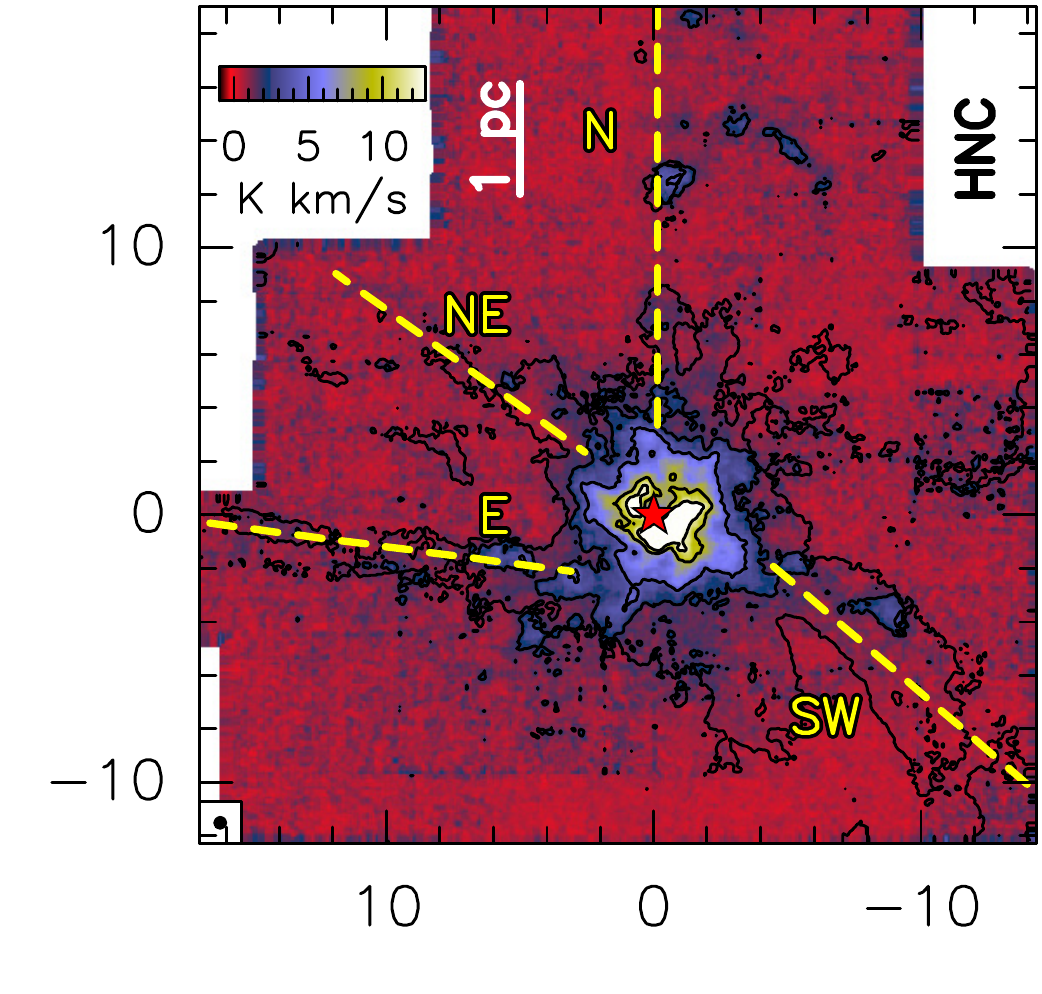} &
  \hspace{-0.5cm}
  \includegraphics[angle=0, width=0.583\textwidth]{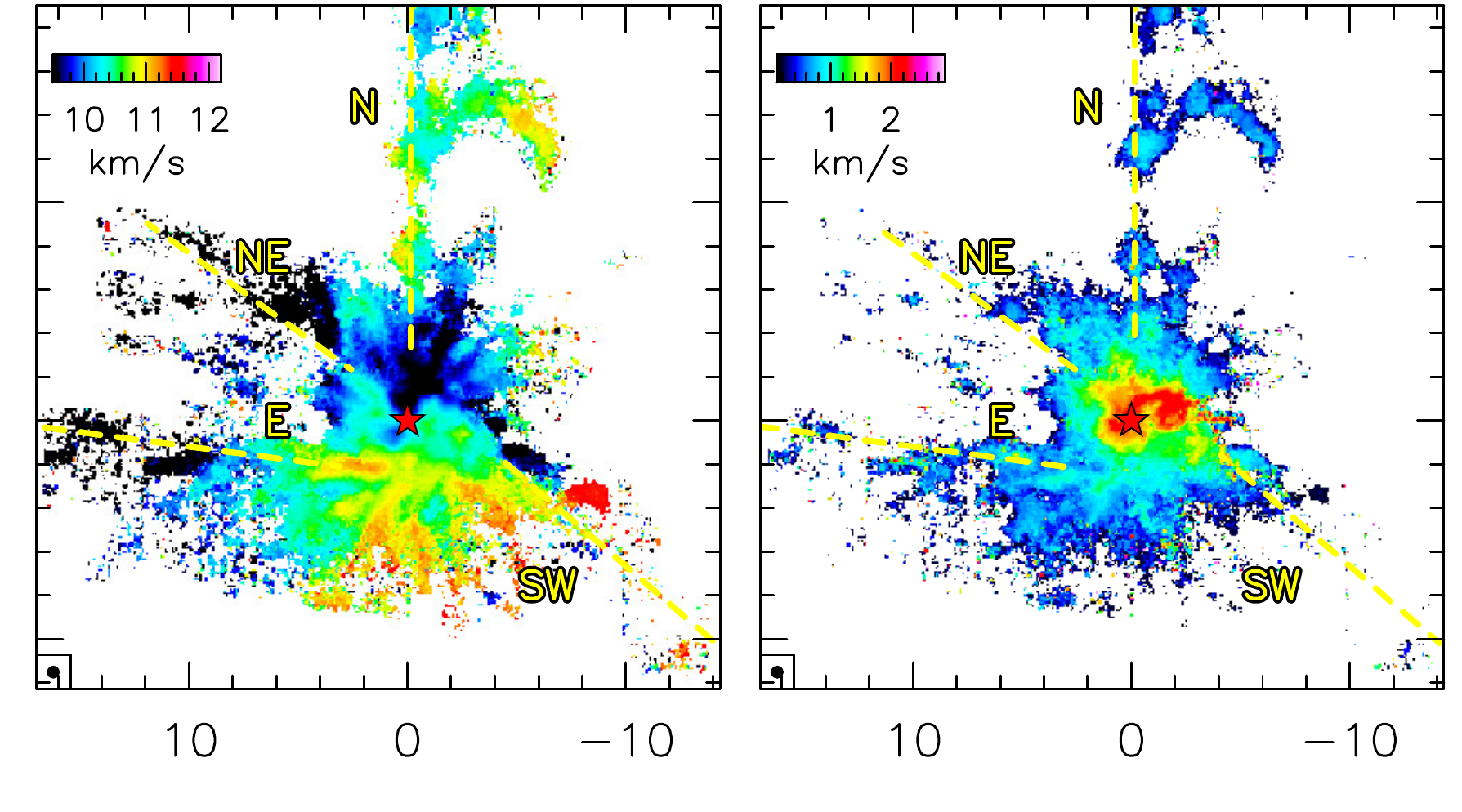} \\
  \vspace{-0.85cm}
  \hspace{-0.55cm}
  \includegraphics[angle=0, width=0.34\textwidth]{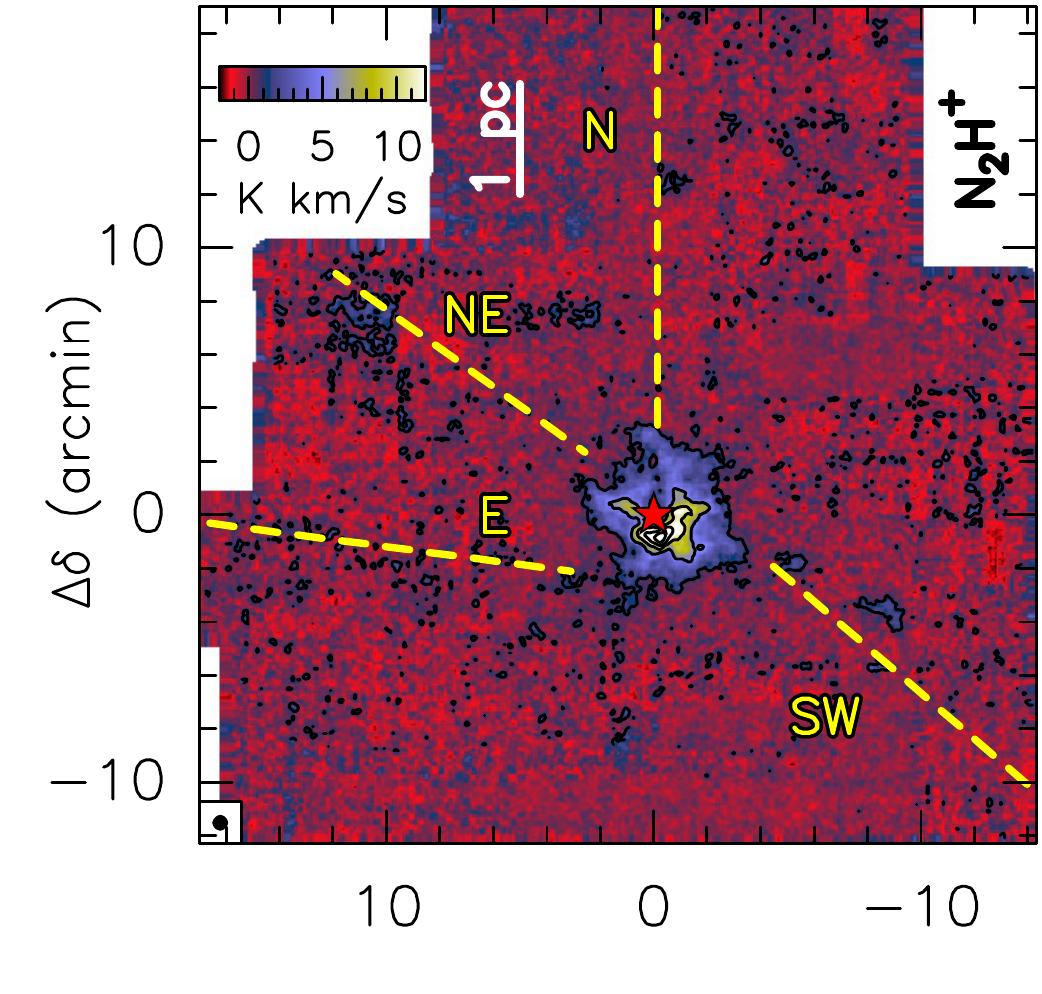} &
  \hspace{-0.5cm}
  \includegraphics[angle=0, width=0.583\textwidth]{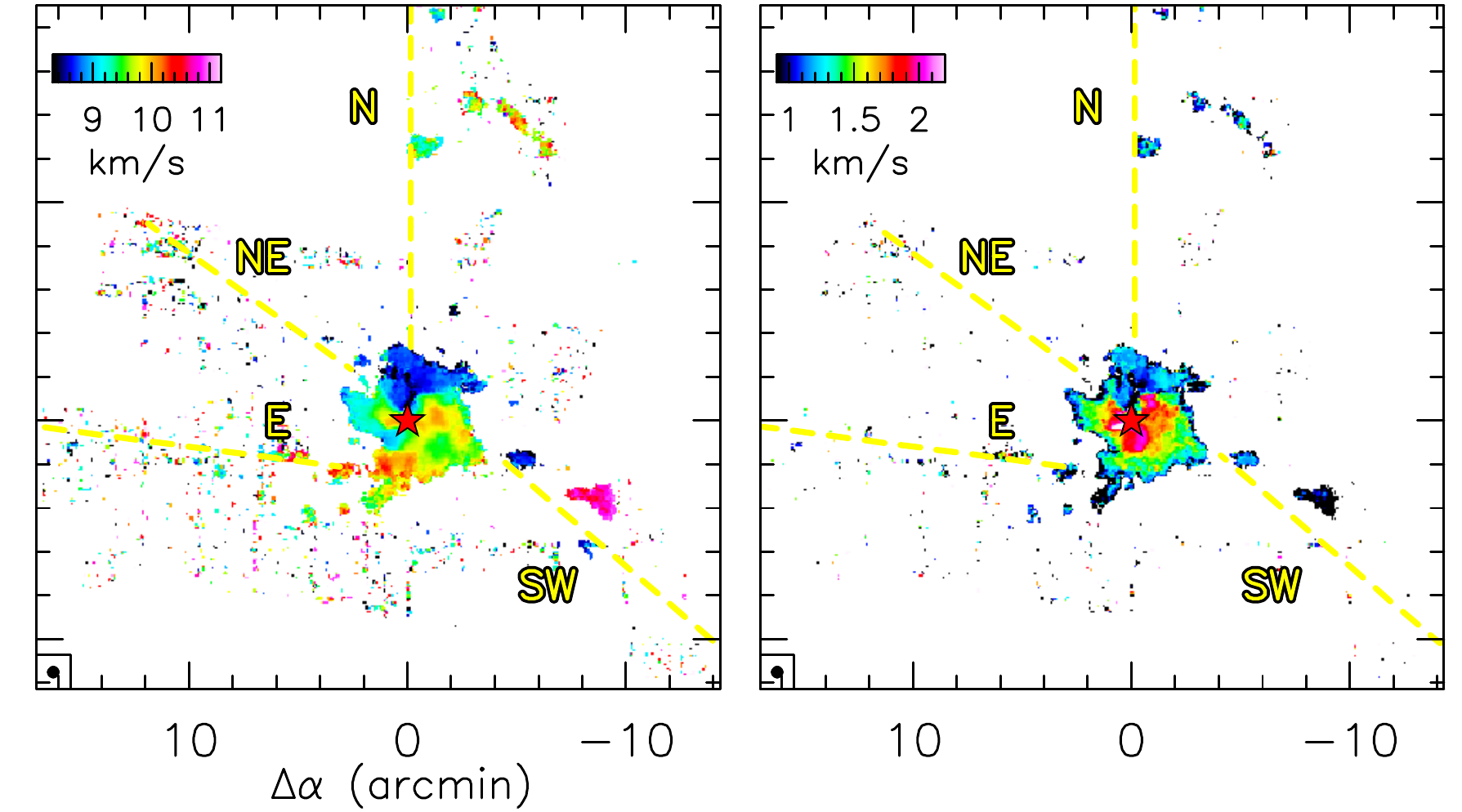} \\
  \vspace{0.35cm}
\end{tabular}
\caption{Left column panels show the integrated intensity maps over the whole surveyed area for the (1$\rightarrow$0) transition lines of the $^{13}$CO, C$^{18}$O, HNC and N$_2$H$^{+}$ molecules. Middle column panels present the velocity centroid. Right column panels show the linewidth. The maps have been produced by computing the zero (left panels), first (middle panels) and second (right panels) order moments in the velocity range defined in Fig.~\ref{fig:02714spectra}. The yellow labels, and the dotted lines, indicate the main features identified in the region. The red star at (0\arcsec,0\arcsec) offset marks the position of IRS~1.}
\label{fig:large-scale-moments}
\end{figure*}

The integrated intensity maps of the $^{13}$CO\ and C$^{18}$O\,(1$\rightarrow$0) lines reveal the existence of several filamentary structures converging into the central hub (see Fig.~\ref{fig:large-scale-moments}). These filamentary structures are also detected in the H$_2$ column density map derived from the \emph{Herschel} continuum emission maps \citep[see][]{Didelon+2015}. Complementary to the H$_2$ column density maps, we derive column density maps for the $^{13}$CO and C$^{18}$O species. Assuming local thermodynamic equilibrium (LTE) and optically
%
%
\begin{figure*}[ht!]
\centering
\begin{tabular}{c c}
  \vspace{-0.7cm}
  \hspace{-0.9cm}
  \includegraphics[angle=0, width=0.4\textwidth]{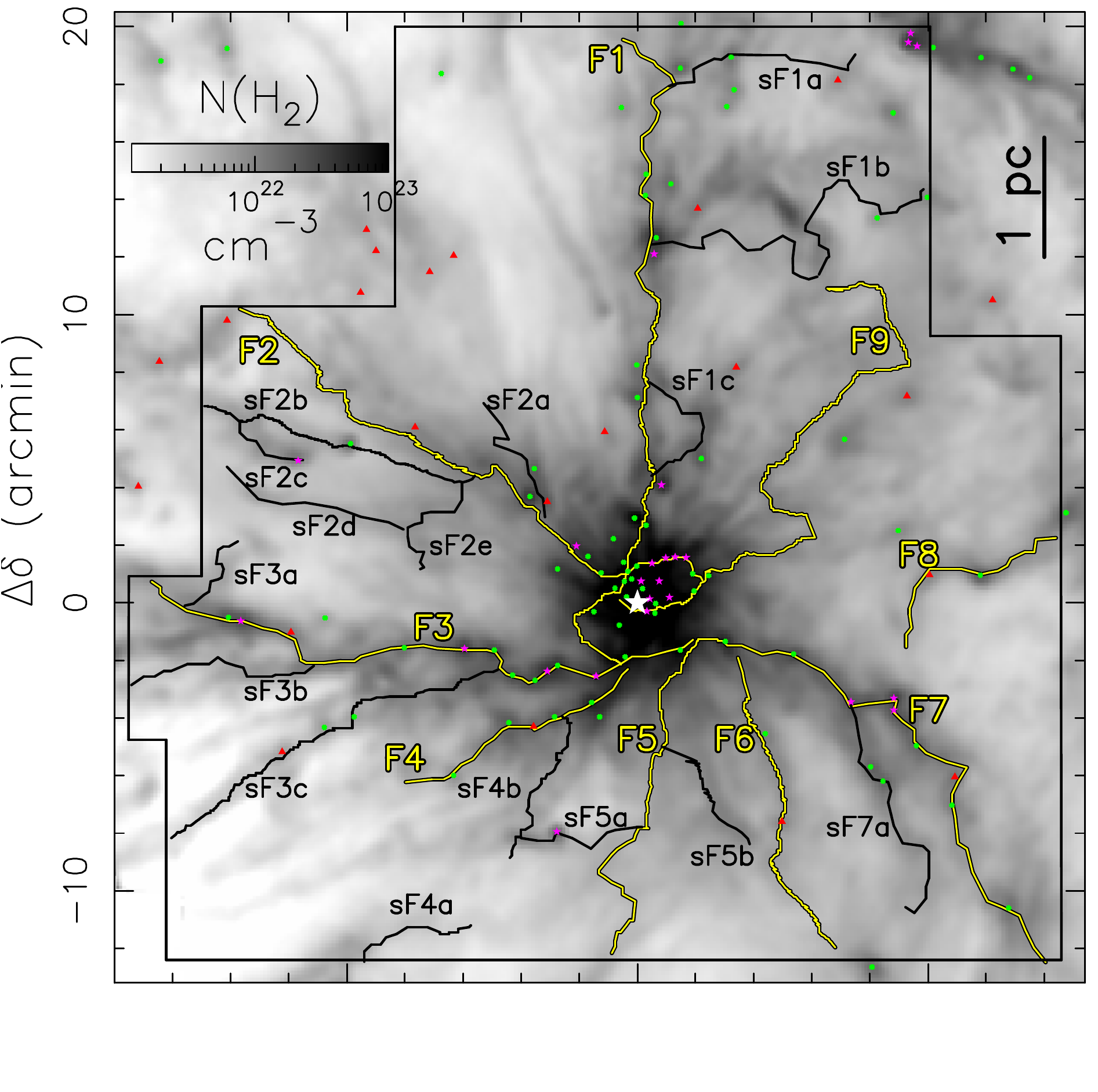} &
  \hspace{-0.75cm}
  \includegraphics[angle=0, width=0.4\textwidth]{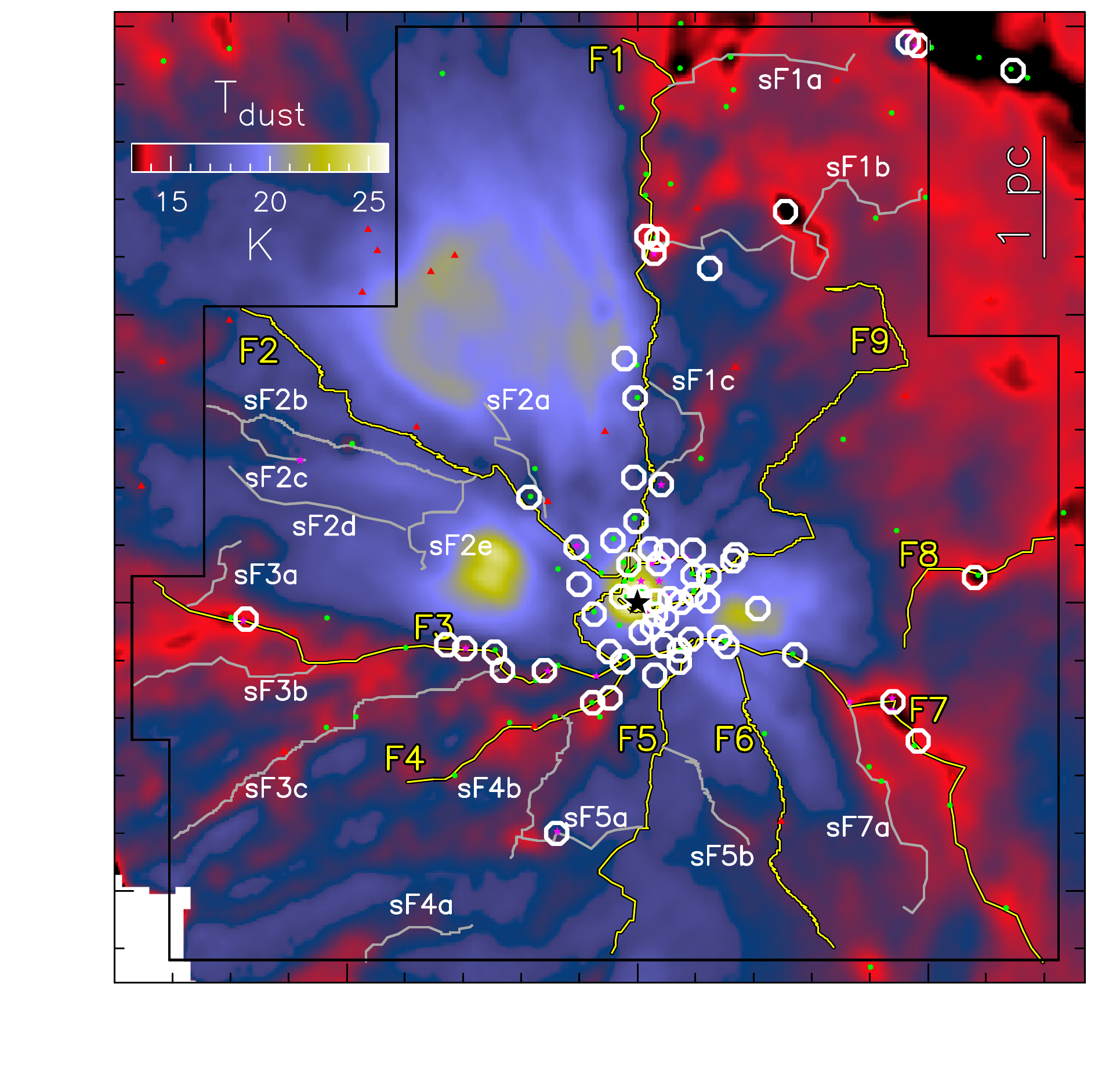} \\
  \vspace{-0.7cm}
  \hspace{-0.9cm}
  \includegraphics[angle=0, width=0.4\textwidth]{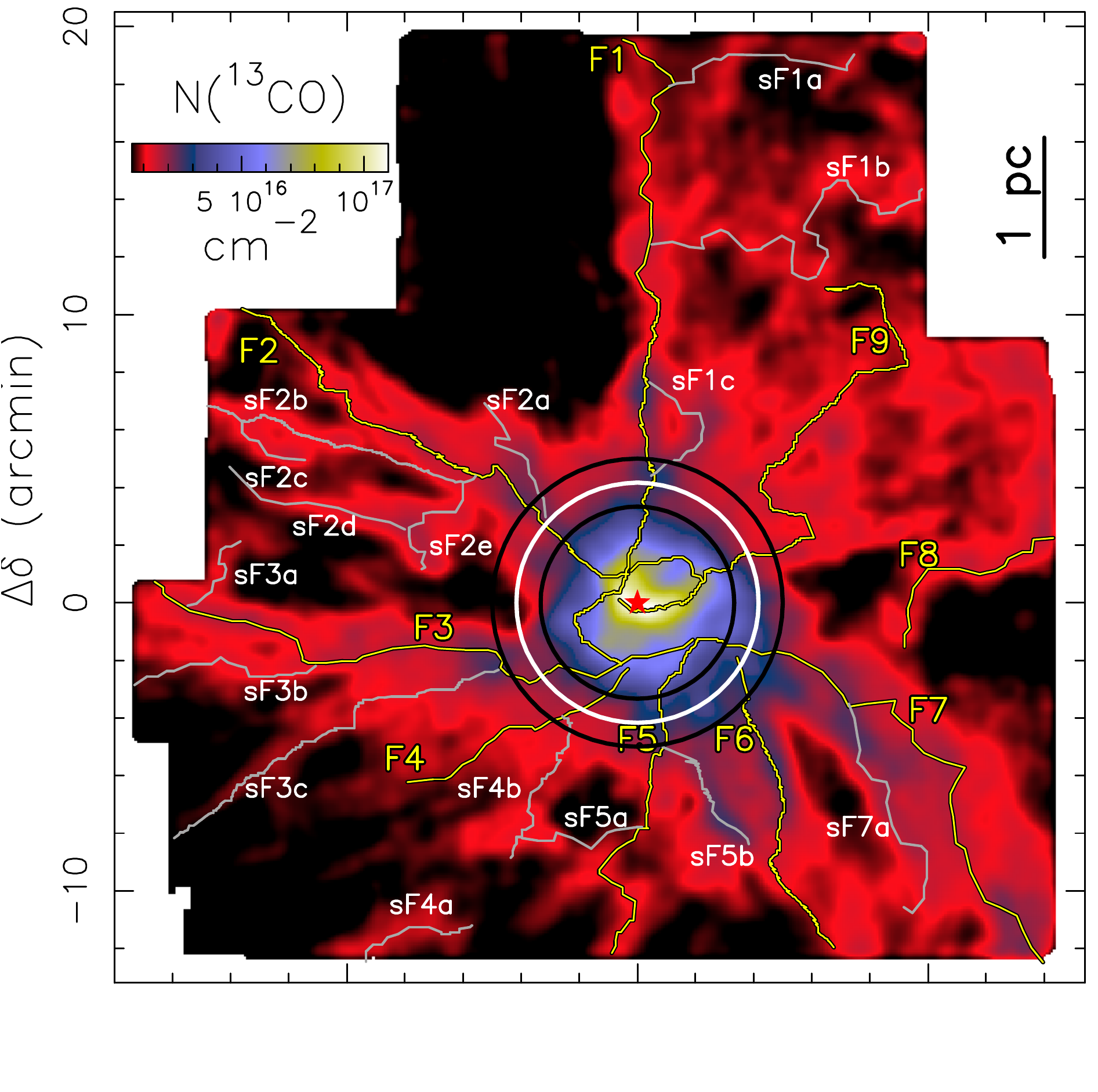} &
  \hspace{-0.75cm}
  \includegraphics[angle=0, width=0.4\textwidth]{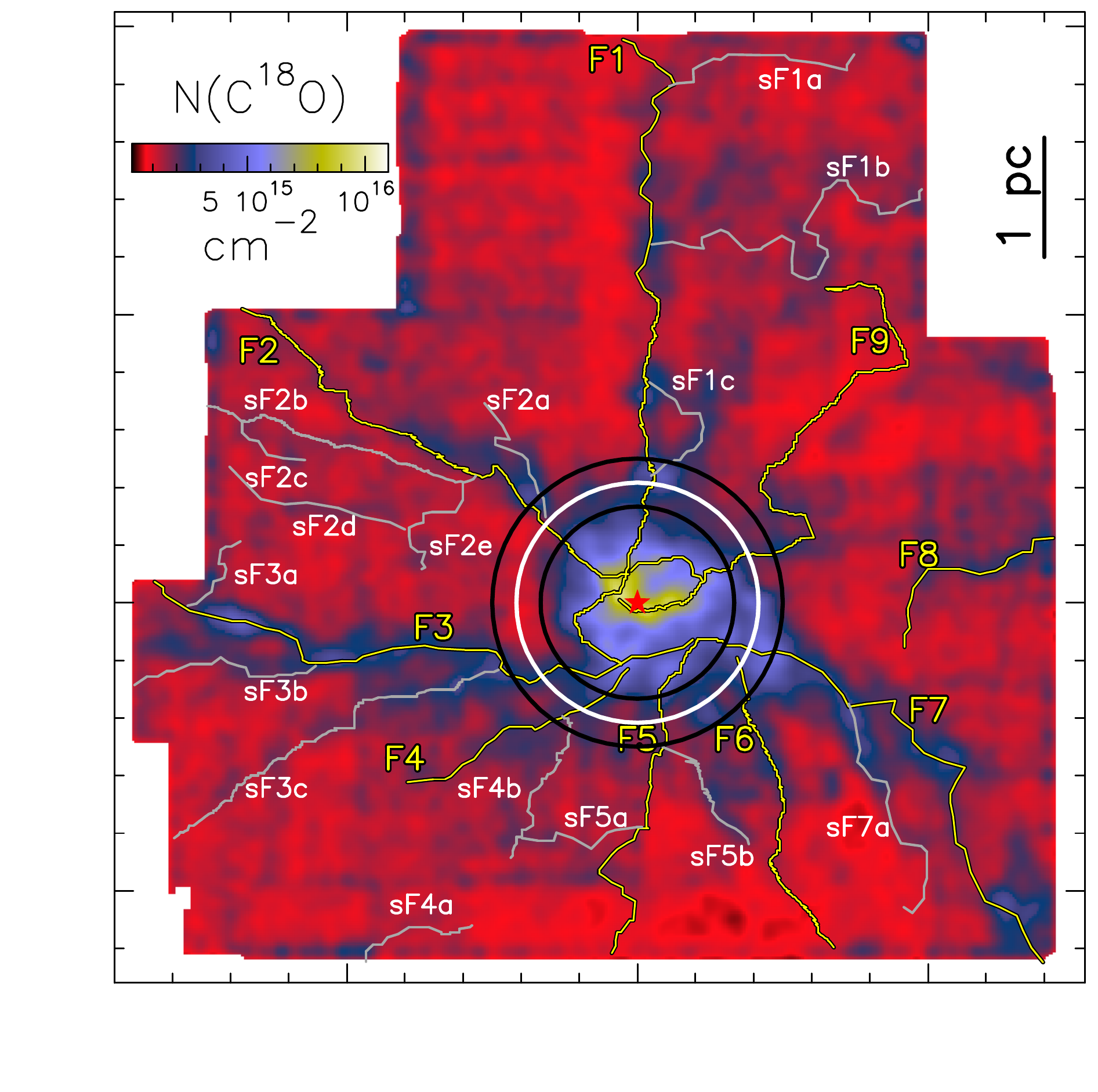} \\
  \vspace{-0.7cm}
  \hspace{-0.9cm}
  \includegraphics[angle=0, width=0.4\textwidth]{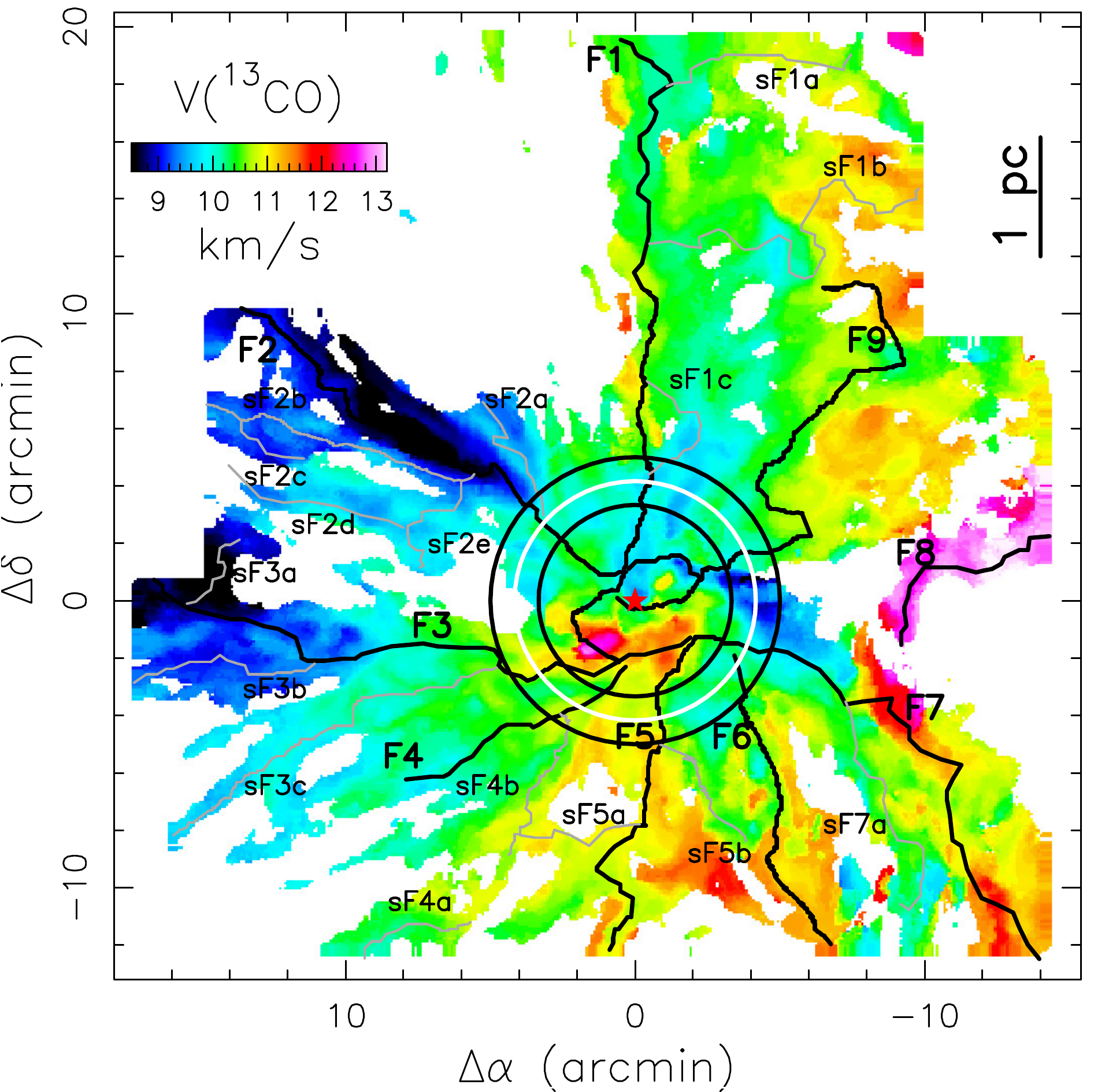} &
  \hspace{-0.75cm}
  \includegraphics[angle=0, width=0.4\textwidth]{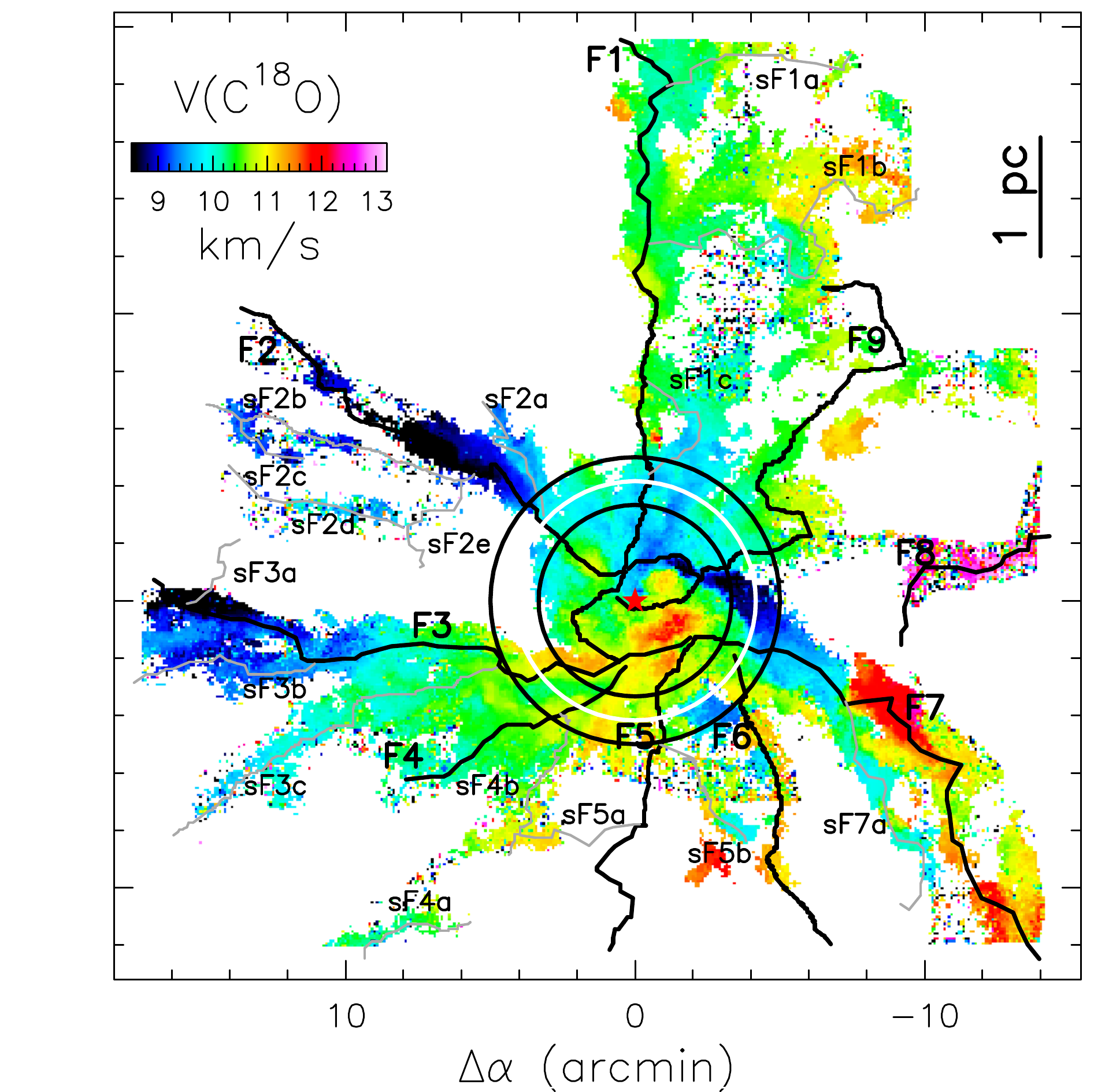} \\
   \vspace{0.1cm}
\end{tabular}
\caption{Top panels show the H$_2$ column density (left) and the dust temperature (right) maps from \textit{Herschel} \citep{Didelon+2015}. Middle panels show the $^{13}$CO (left) and C$^{18}$O (right) column density maps. Bottom panels show the velocity centroid for $^{13}$CO (left) and C$^{18}$O (right). The `skeleton' of identified filaments are marked with solid white, black or yellow lines. The black/white circles corresponding to the radii at 200\arcsec, 250\arcsec\ and 300\arcsec\ (transition between the hub and the filaments, see Fig.~\ref{fig:azimuthal}). The white circles in the top-right panels show sources identified by \cite{Sokol+2019}, the colored symbols show the sources identified by \cite{Rayner+2017}.}
\label{fig:large-skeleton}
\end{figure*} 
%
thin molecular emission, the column densities are calculated (see Appendix~\ref{A:ColDens}) as
\begin{equation}\label{eq:N13CO}
\left[\frac{N(^{13}\mathrm{CO})}{\mathrm{cm}^{-2}}\right] =
4.69\times10^{13}\phn
T_\mathrm{ex}\phn
e^{\frac{5.30}{T_\mathrm{ex}}}\phn
\left[\frac{\int T(v)~dv}{\mathrm{K~km~s}^{-1}}\right],
\end{equation}
and
\begin{equation}\label{eq:NC18O}
\left[\frac{N(\mathrm{C^{18}O})}{\mathrm{cm}^{-2}}\right] = 
4.73\times10^{13}\phn
T_\mathrm{ex}\phn
e^{\frac{5.28}{T_\mathrm{ex}}}\phn
\left[\frac{\int T(v)~dv}{\mathrm{K~km~s}^{-1}}\right],
\end{equation}
%
\begin{figure*}[ht!]
\centering
\includegraphics[width=0.95\textwidth]{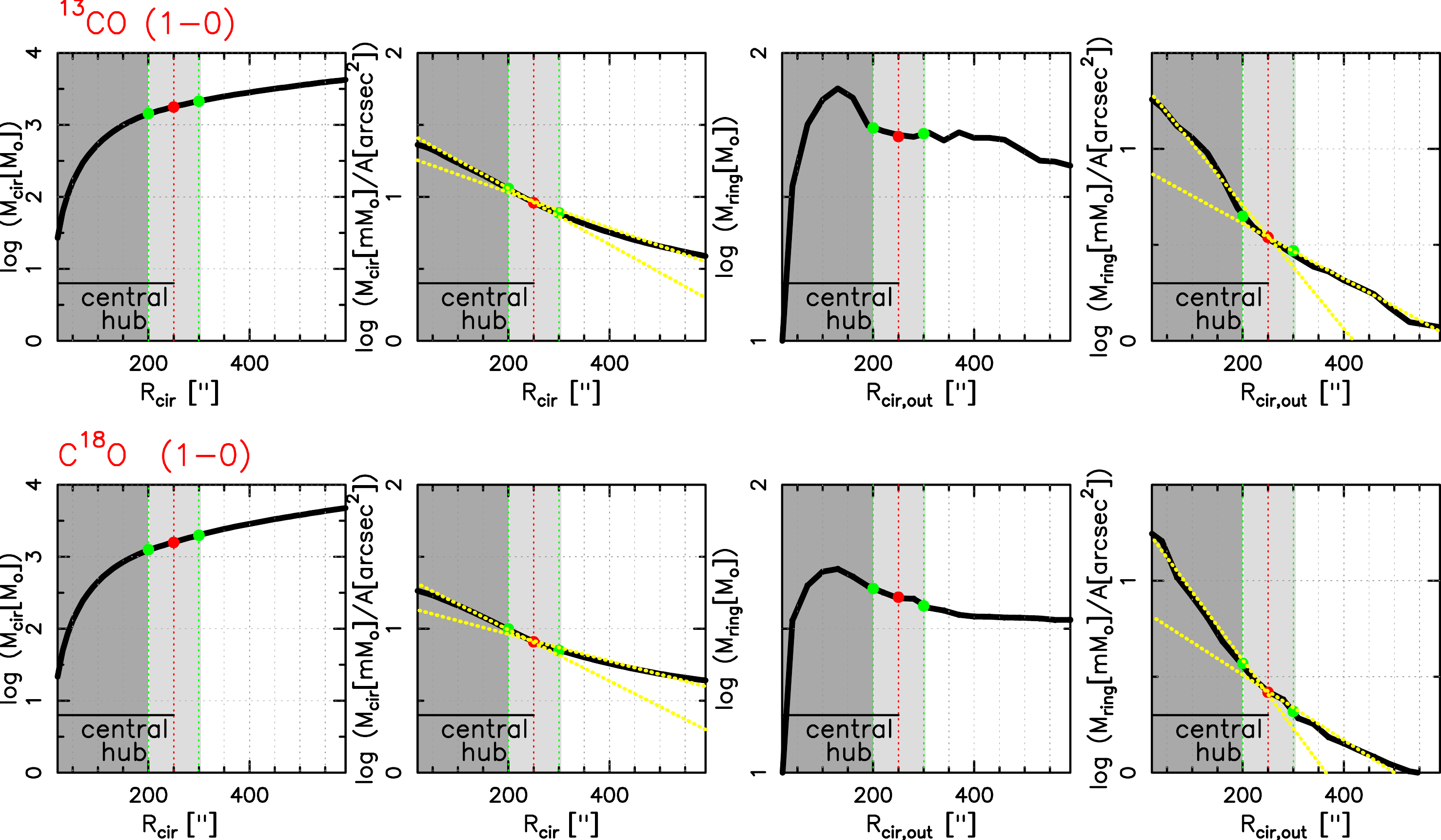}
\caption{Azimuthally mass and surface density derived from the $^{13}$CO (top) and C$^{18}$O (bottom) column density maps. From left to right each column shows: \textit{i)} radially integrated mass, \textit{ii)} radially integrated mass divided by the area of the circle with radius $R_{\mathrm{cir}}$ (\ie\ radially integrated surface density), \textit{iii)} concentric annular mass, \textit{iv)} concentric annular surface density. The radially integrated mass and surface density were calculated within circles of radius $R_{\mathrm{cir}}$ centered on IRS~1 from $R=25"$ ($\sim$0.1~pc) to $R=600"$ ($\sim$2.4~pc). The concentric annular mass and surface density were calculated within concentric rings of radius $R_{\mathrm{ring}} = R_{\mathrm{cir,out}}-R_{\mathrm{cir,in}}$ = $36''$ (corresponding to the \textit{Herschel} beam size). In order to do a direct comparison of the profiles, the x-axis in the concentric annular mass and surface density profiles correspond to $R_{\mathrm{cir,out}}$. The yellow dotted lines mark the different slopes in the surface density profiles. The light gray zone indicates the transition between the hub and the filaments, from 200\arcsec\ to 300\arcsec. The dark gray area marks the central hub with $R_{\mathrm{hub}}=250"=1$~pc.}
\label{fig:azimuthal}
\end{figure*}
%
where $T_\mathrm{ex}$ is the excitation temperature in K, and the term $\int T(v)~dv$ is the integrated flux of the (1$\rightarrow$0) line in K~\kms. We assume that the lines are thermalized with the excitation temperature being equal to the gas kinetic temperature, \ie\ $T_\mathrm{ex}=T_\mathrm{k}$, and that this equals the dust temperature, $T_\mathrm{dust}$, as derived in \citet[see top-right panel in Fig.~\ref{fig:large-skeleton}]{Didelon+2015}. This assumption is only accurate in dense regions ($n>10^4$~cm$^{-3}$) shielded from the UV radiation. Hence, in the surroundings of the central UC \hii\ region and the PDRs, the UV radiation will increase the gas temperature ($T_\mathrm{gas}$), and $T_\mathrm{gas}=T_\mathrm{dust}$ should be considered as a lower limit to the real one. We have smoothed the IRAM-30m molecular maps to the angular resolution of the \textit{Herschel}-derived $T_\mathrm{dust}$ map (\ie\ 36\arcsec) and used Eqs.~\ref{eq:N13CO} and \ref{eq:NC18O} to derive the molecular column density maps shown in Fig.~\ref{fig:large-skeleton}. The largest column densities are found towards the central hub with $N$($^{13}$CO)$>$$5\times10^{16}$~cm$^{-2}$. Outside the hub, we find a constant column density of $N$($^{13}$CO)$\approx$$1\times10^{16}$~cm$^{-2}$ with local enhancements associated with the filamentary structures. For C$^{18}$O, we derive column densities $\simeq10$ times smaller than for $^{13}$CO.

We next study the internal structure of the cluster-forming region, specifically aiming at determining if a well-defined hub can be identified, and if so, measuring its size and average radial parameters. For this, we study the azimuthally-averaged mass and surface density of the cloud within concentric circles and rings centered at IRS~1. The circles radius $R_{\mathrm{cir}}$ ranges from 0.1~pc to 2.4~pc (or 25\arcsec-- 600\arcsec, the radius of the UC~\hii\ region is 12.5\arcsec). While, the ring radius $R_{\mathrm{ring}}$ is the difference of an external circle $R_{cir,out}$ and an inner circle $R_{cir,in}$. In Fig.~\ref{fig:azimuthal}, we plot the azimuthally-averaged radial profiles for $^{13}$CO (top panels) and C$^{18}$O (bottom panels). We first consider the radially integrated gas mass $M_{\mathrm{cir}}$ (first column panels) calculated in circles of radius $R_{\mathrm{cir}}$, and then, we calculate the gas mass $M_{\mathrm{ring}}$ (third column panels) over concentric rings with radius $R_{\mathrm{ring}} = R_{\mathrm{cir,out}}-R_{\mathrm{cir,in}}$. The gas mass $M$ within each circle/ring is given by
\begin{equation}\label{eq:mass}
\centering
M=\frac{N}{X}A(2.8\,m_\mathrm{H}),
\end{equation}
where $N$ is the total column density of the molecule (as derived in Eqs.~\ref{eq:N13CO} and \ref{eq:NC18O}), $X$ is the relative abundance of the molecule with respect to H$_2$, $A$ is the surface area of the circle, and $m_\mathrm{H}$ is the hydrogen atom mass. We use the typical \MonR\ abundances $X$($^{13}$CO)$=1.7\times10^{-6}$ and $X$(C$^{18}$O)$=1.7\times10^{-7}$ \citep[\eg][]{Ginard+2012}. These values are consistent with the average abundances that can be derived by comparing the H$_2$ (from \textit{Herschel}) and the $^{13}$CO and C$^{18}$O column density maps (see Fig.~\ref{A:abundances}). Figure~\ref{fig:azimuthal} also shows the radially integrated gas mass divided by the circles surface area (second column panels) and the concentric rings mass divided by the rings surface area (fourth column panels), \ie\ the surface densities profiles. The radial profiles of the surface density in Fig.~\ref{fig:azimuthal} show two different slopes (yellow dotted lines) with the turnover point occurring at a radius between 200\arcsec\ and 300\arcsec\ (or 0.8 to 1.2~pc). This change of slope may result from a transition between a denser region in the center and a more diffuse component in the outside. We therefore consider that there is a well-defined hub-structure with a radius of about 250\arcsec, or 1~pc. Hereafter, we refer to this as the hub radius, $R_\mathrm{hub}$. We notice that the radial mass and surface density profiles do not correspond to the initial mass distribution of the cloud. They are just a tool to investigate the morphology of the current evolutionary stage of the cloud.

\begin{table}[t!]
\caption{Gas and dust mass derived from different tracers for the different structures in \MonR\ (see Sect.~\ref{sec:columndensity} for details).}
\begin{tabular}{l c c c c}
\hline\hline\noalign{\smallskip}
&\multicolumn{3}{c}{Mass derived from}
&
\\
\cline{2-4}\noalign{\smallskip}
&$^{13}$CO
&C$^{18}$O
&dust
&Average 
\\
&(\mo)
&(\mo)
&(\mo)
&mass
\\
\hline
\noalign{\smallskip}
total cloud 	          & 6200 & 8300 & 8400 & 100\%    \\  
hub                       & 1800 & 1600 & 3600 & \phn32\% \\  
main/secondary filaments  & 2400 & 3200 & 2500 & \phn35\% \\
diffuse medium            & 2000 & 3500 & 2300 & \phn33\% \\
\hline
\end{tabular} 
\label{tab:hub-masses}
\end{table}

From the $^{13}$CO and C$^{18}$O column density maps, we estimate a mass of $\sim$1700~\mo\ within the $R_\mathrm{hub}=1$~pc, which corresponds to about 24\% of the total mass ($\sim$7200~\mo) of the surveyed area. From the H$_2$ column density maps obtained with \textit{Herschel} observations \citep{Didelon+2015}, we derive the mass of $\sim$3600~\mo\ for the hub and $\sim$8300~\mo\ for the surveyed area. These are in a reasonable agreement with the values derived from the molecular species (see Table~\ref{tab:hub-masses}). In summary, and considering the different tracers, we find that about 32\% of the mass in the surveyed area is contained in the central hub.

%
\subsection{Filament identification\label{sec:identification}}

As shown in Fig.~\ref{fig:large-skeleton}, \MonR\ has a filamentary structure outside the central hub. Making use of our three-dimensional data cubes (position-position-velocity) we have used the structure identification algorithm DisPerSE \citep[Discrete Persistent Structures Extractor,][]{Sousbie2011} to define filaments. DisPerSE was originally developed to search for filamentary structures in large scale cosmological simulations, but it has been successfully applied to identify filaments from molecular clouds and from numerical simulations of star forming regions \citep[\eg][]{Arzoumanian+2011, Schneider+2012, Palmeirim+2013, Smith+2014, Panopoulou+2017, Zamora-Aviles+2017, Chira+2018, Suri+2019}. DisPerSE identifies critical points in a dataset where the gradient of the intensity goes to zero and connects them with arcs; the arcs are then called filaments. The critical point pairs that form an arc with low significance can be eliminated with two thresholds; the persistence threshold and the detection threshold. The persistence is expressed as the difference between the intensities of critical points in a pair. The higher the persistence, the more contrast the structure has. The detection threshold eliminates the critical points that are below the noise. We used the $^{13}$CO emission map for filament identification with DisPerSE, and set both the persistence and the detection thresholds to be 5 times the noise level per channel. These thresholds assure that we select filaments with high significance.

Complementing the identification of filaments with DisPerSE, we have visually inspected the correspondence between the DisPerSE-identified filaments and elongated structures visible in the $^{13}$CO\,(1$\rightarrow$0) and C$^{18}$O\,(1$\rightarrow$0) data sets. Most of the structures identified with DisPerSE are clearly visible in at least one velocity interval and appear contiguous in successive velocity channels, which further supports the picture that they are coherent entities in the position-position-velocity space. Only few structures are not clearly identified in the molecular channel maps and have been discarded. Thus, our final set of filaments consists of those DisPerSE-identified structures that are confirmed via visual inspection in both $^{13}$CO and C$^{18}$O emission through different velocity intervals.

The skeletons of the identified filaments are shown in Fig.~\ref{fig:large-skeleton}. A comparison of the filaments with the \textit{Herschel} maps confirms that most of them trace H$_2$ column density structures (see top-left panel). Some of the filaments extend beyond the area surveyed with the IRAM-30m telescope. In total, we have identified nine filaments, which are named F1 to F9, counter-clock-wise from the North. Filaments F1 to F7 and F9 converge to the central hub, while F8 seems to be spatially and kinematically isolated from the other filaments (see Sect.~\ref{sec:kinematics}). In addition to these nine `main' filaments, DisPerSE identified other filaments that do not converge into the central hub, but merge into one of the `main' filaments. These structures are more prominent in $^{13}$CO than in C$^{18}$O. We call these structures \textit{secondary filaments}, and use labels like sF1a to indicate to which main filament they are connected with. The last letter in the label is an increasing index for the secondary filaments associated with one main filament. A total of 16 secondary filaments are identified.

On the basis of C$^{18}$O\,(2$\rightarrow$1) line observations, \citet{Rayner+2017} performed an identification and analysis of the filamentary structure in the inner area of \MonR\ (about 7~pc$^{2}$). They found eight filaments with about 1~pc of length converging into the \MonR\ hub. Six of them\footnote{The nomenclature in this work has been chosen to be consistent with the previous analysis presented in \citet{Trevino-morales2016}. The correspondence between our nomenclature and the one adopted by \citet{Rayner+2017} can be found in Fig.~\ref{fig:mass_length}.} seem to correspond to filaments identified in this work, extending into the hub. However, there are some differences between the filament skeletons presented by us and \citet{Rayner+2017}. We attribute these differences to the identification techniques and the difference in the resolution of the data-cube used by \citet{Rayner+2017} and the ones used in this paper.

%
\subsection{Physical properties of the filaments\label{sec:stability}}

\begin{figure}[t!]
\hspace{-0.5cm}
\centering
\includegraphics[width=0.9\columnwidth]{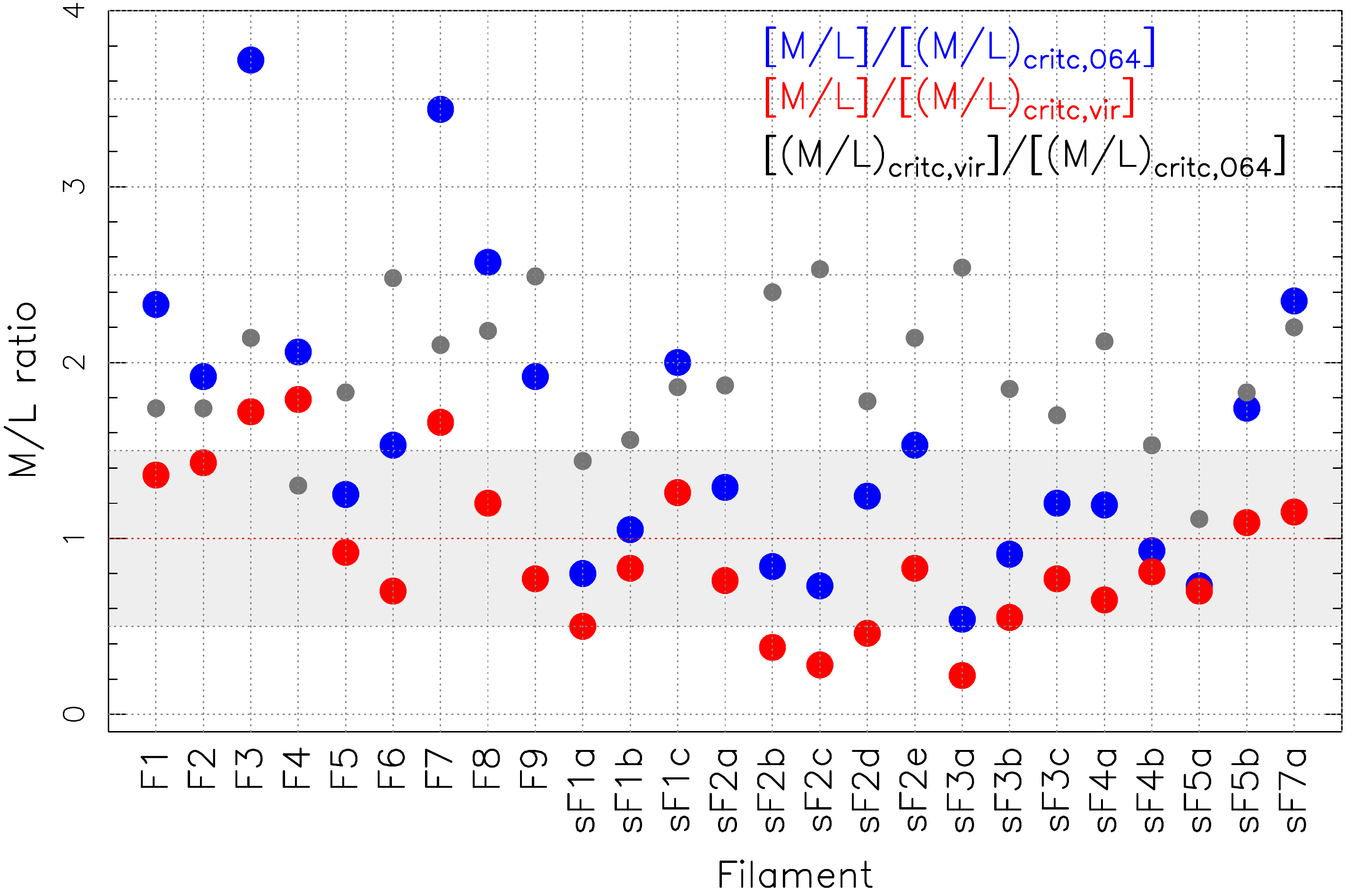}
\caption{Comparison of the observed $M/L$ values with the critical ones. The gray circles correspond to the $[(M/L)_{\mathrm{crit,vir}}]/[(M/L)_{\mathrm{crit,O64}}]$ ratio, the blue ones show the $[(M/L)]/[(M/L)_{\mathrm{crit,O64}}]$ values and the black ones correspond to the $[(M/L)]/[(M/L)_{\mathrm{crit,vir}}]$ ratio. The gray band indicates the trans-critical range, between 0.5 and 1.5.}
\label{fig:Mlin_ratio}
\end{figure} 

\begin{figure*}[t!]
\centering
\includegraphics[width=0.9\textwidth]{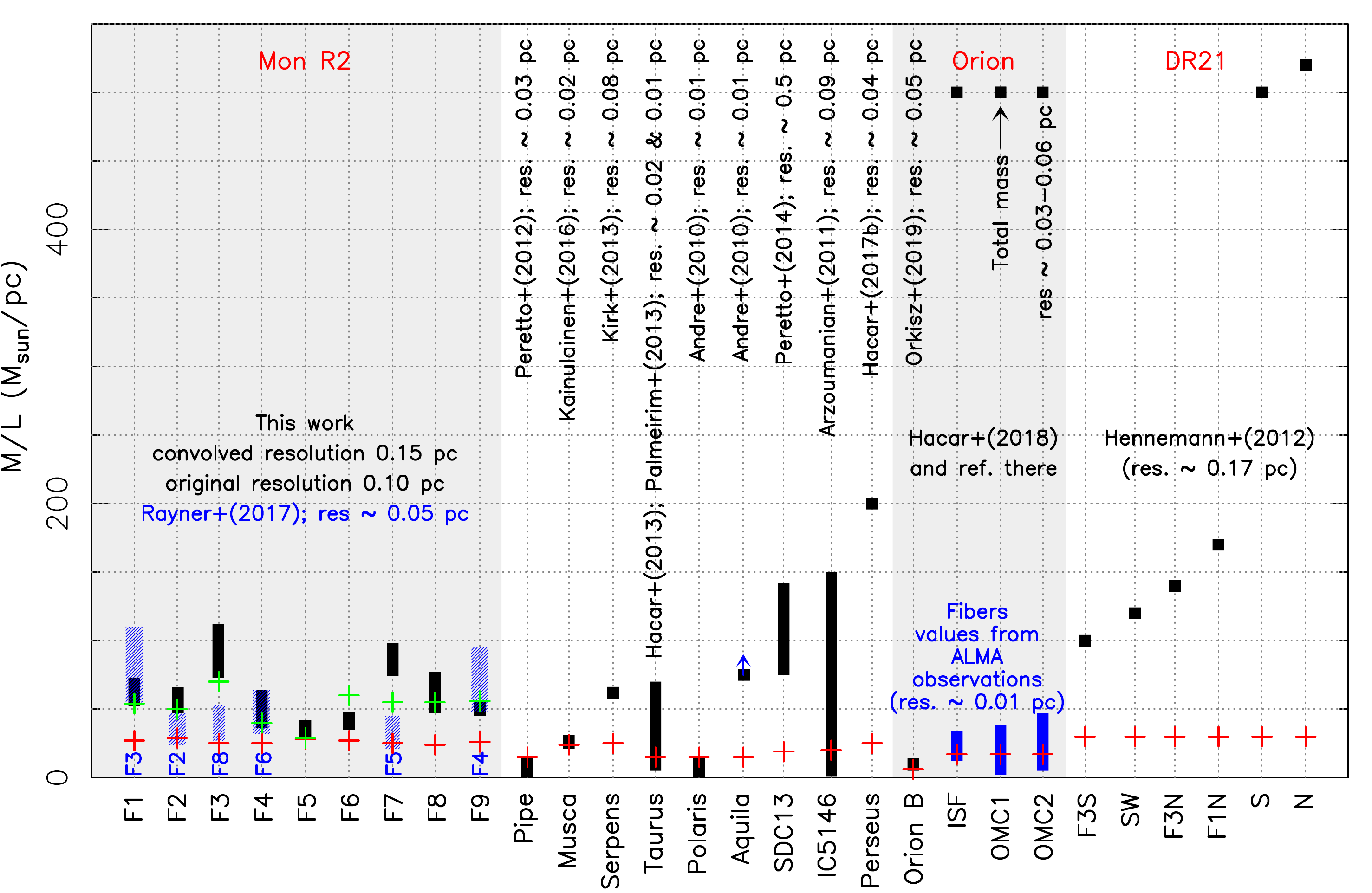}
\caption{Comparison of the observed $M/L$ for the filaments F1 to F9 with different low-mass and high-mass star forming regions. The gray band located at the left of the panel present the \MonR\ range values obtained in this work, in black, and by \citet{Rayner+2017}, in blue. The blue labels corresponds to \citet{Rayner+2017} nomenclature. The gray band located at the right of the plot separate the Orion values. The blue solid bars at the Orion band indicate the $M/L$ range found in the fibers within each filament. The black ones present the total filament $M/L$ reported by \cite{Bally+1987} and \cite{Johnstone-Bally1999}. The red crosses indicate the value of the $(M/L)_{\mathrm{crit,O64}}$ for each region, while the green ones indicate the value of the $(M/L)_{\mathrm{crit,vir}}$ for each filament in \MonR. The plot indicates the resolution used in each work to obtain the observational $M/L$.}
\label{fig:mass_length}
\end{figure*} 

\begin{figure*}[t!]
\centering
\includegraphics[width=0.9\textwidth]{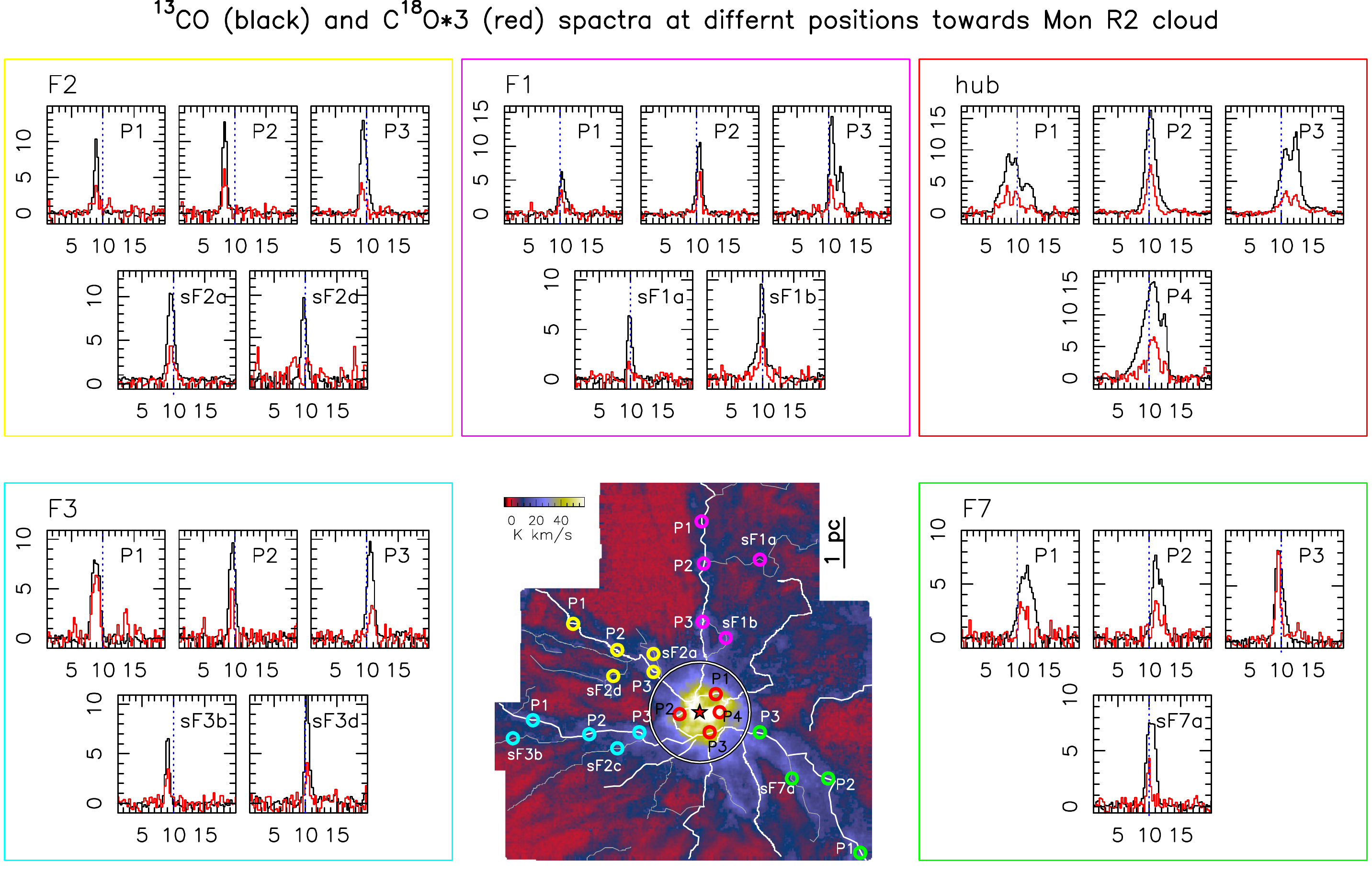}
\caption{Averaged spectra of the $^{13}$CO (black) and C$^{18}$O (red) molecules at different positions towards the hub (red box), and the filaments/secondary filaments F1 (pink box), F2 (yellow box), F3 (light-blue box) and, F7 (green box). The positions corresponding to each spectra are indicated in the central bottom panel.}
\label{fig:kinematics}
\end{figure*} 

One possible way to gain insight into the stability of filaments is to study their line mass, $M/L$ (mass per unit length). In the case of an isolated, infinitely long filament in which gravity and thermal pressure are the only forces, an equilibrium solution exists at the line mass \citep{ostriker1964}
\begin{equation}\label{eq:M_L_ther}
(M/L)_{\mathrm{crit,O64}}= \frac{2c_{\mathrm{s}}^{2}}{G} = 16.7\left(\frac{T}{10 ~\mathrm{K}}\right)~\mathrm{M}_{\sun}~\mathrm{pc}^{-1},
\end{equation}
where $c_{\mathrm{s}}=(kT/\mu m_{\mathrm{H}})^{1/2}$ is the sound speed, which is linked to the thermal velocity dispersion, and $G$ is the gravitational constant. Equation~\ref{eq:M_L_ther} only depends on the gas temperature. Linear perturbation analyses have shown that this equilibrium solution is prone to fragmentation due to gravitational fragmentation \citep[see, e.g.,][hereafter IM97]{Inutsuka-Miyama1997}. The fragmentation leads to clumps that are separated by a distance
\begin{equation*}
\lambda_{\mathrm{cl,IM97}}=
c_\mathrm{s}\left(\frac{\pi}{G\rho}\right)^{1/2}=
0.066~\mathrm{pc}
\left[\frac{T}{10~\mathrm{K}}\right]^{1/2}
\left[\frac{n_\mathrm{c}}{10^{5}~\mathrm{cm}^{-3}}\right]^{-1/2},
\end{equation*}
and have masses given by
\begin{equation*}
M_{\mathrm{cl,IM97}}=
(M/L)_\mathrm{crit}\times\lambda_\mathrm{cl}=
0.877~M_{\odot}
\left[\frac{T}{10~\mathrm{K}}\right]^{3/2}
\left[\frac{n_\mathrm{c}}{10^{5}~\mathrm{cm}^{-3}}\right]^{-1/2},
\end{equation*}
where $n_\mathrm{c}$ is the number density of gas at the filament center.

The above models only consider the thermal gas pressure as the force opposing gravity. It is possible, and commonly assumed in literature, that turbulence within gas can also provide a supporting pressure. (\citealt{Chandrasekhar1951}, hereafter C51, see also \citealt{Wang+2014}). This pressure can be simplistically taken into account by replacing the sound speed in Eq.~\ref{eq:M_L_ther} by an effective sound speed that results from the combination of thermal and non-thermal motions (or velocity dispersion). In this case, the critical line mass is given by \citep{Wang+2014}
\begin{equation}
\label{eq:M_L_non_ther}
(M/L)_{\mathrm{crit,vir}}= \frac{2\sigma_{\mathrm{tot}}^{2}}{G} = 465\left(\frac{\sigma_{\mathrm{tot}}}{1~\mathrm{km~s^{-1}}}\right)^{2}~\mathrm{M}_{\sun}~\mathrm{pc}^{-1},
\end{equation}
where $\sigma_{\mathrm{tot}} = \Delta v/\sqrt{8ln2}$ is the total velocity dispersion, which in our case is obtained from the $^{13}$CO and the C$^{18}$O linewidths (see Fig.~\ref{fig:large-scale-moments}). The separations and masses of the clumps are given by
\begin{equation*}
\lambda_{\mathrm{cl,vir}}=
1.24~\mathrm{pc}
\left[\frac{\sigma_{\mathrm{tot}}}{1~\mathrm{km~s}^{-1}}\right]
\left[\frac{n_\mathrm{c}}{10^{5}~\mathrm{cm}^{-3}}\right]^{-1/2},
\end{equation*}
\begin{equation*}
M_{\mathrm{cl,vir}}=
575.3~M_{\odot}
\left[\frac{\sigma_{\mathrm{tot}}}{1~\mathrm{km~s}^{-1}}\right]^{3}
\left[\frac{n_\mathrm{c}}{10^{5}~\mathrm{cm}^{-3}}\right]^{-1/2}.
\end{equation*}

One should note that the above models represent a simplistic case of an isolated and highly idealized gas cylinder. Effects of various additional physical processes on the filament stability and fragmentation have been studied by several works \citep[\eg][]{Fiege-Pudritz2000a, Fiege-Pudritz2000b, Fischera-Martin2012, Heitsch2013a, Heitsch2013b, Recchi+2014, Zamora-Aviles+2017}. Also, simulations have analysed the evolution of filaments in various setups \citep[\eg][]{Clarke+2016, Clarke+2017, Chira+2018, Kuznetsova+2018}. Regardless, we employ here the simplistic framework to gain the first insight into the stability of the filaments and to compare the filaments in \MonR\ with other works that have analyzed filaments using the same framework.

Making use of Eqs.~\ref{eq:M_L_ther} and~\ref{eq:M_L_non_ther}, we calculated the $(M/L)_{\mathrm{crit,O64}}$ and $(M/L)_{\mathrm{crit,vir}}$ values for each filament. The $\sigma_\mathrm{tot}$ values used to calculate $M/L_{\mathrm{crit,vir}}$ are listed in Table~\ref{tab:filaments_parameters_kin_1}; they were obtained from the median value of the $\Delta V$, estimated from Gaussian fits (see Appendix~\ref{A:figures}) in different positions along the filaments. We find that $(M/L)_{\mathrm{crit,O64}}$ and $(M/L)_{\mathrm{crit,vir}}$ agree within a factor of $\sim2$, indicating that thermal and non-thermal pressures are similar (see grey circles in Fig.~\ref{fig:Mlin_ratio}, and last columns of Tables~\ref{tab:filaments_parameters_stability_1} and~\ref{tab:filaments_parameters_stability_2}). This is in good agreement with the results of \cite{Pokhrel+2018} work, where the authors present a study of the hierarchical structure in the Perseus molecular cloud at different scales. They show that the thermal motions are least efficient in providing support at larger scales such as the whole cloud ($\sim$10~pc), and most efficient at smaller scales such as the protostellar objects ($\sim$15~AU). Our analysis in \MonR\ corresponds to an intermediate scale between small clumps ($\sim$1~pc) and cores ($\sim$0.05--0.1~pc), in the frontier where the turbulent support starts to be substituted by the thermal support.

In Tables~\ref{tab:filaments_parameters_stability_1} and~\ref{tab:filaments_parameters_stability_2} we compare the observed $M/L$ values for each filament with the critical ones. The masses of the filaments have been calculated using Eq.~\ref{eq:mass} for both $^{13}$CO and C$^{18}$O and for the \textit{Herschel}-derived column density. We find less than a factor of two differences between the masses determined with different tracers. We adopt the mean of these masses for the following analysis and estimate that the uncertainty of the mass is a factor of two. This results in line mass of $M/L$=30--110~\mo~pc$^{-1}$ for the main filaments, which are a factor of 1--4 above the thermally critical values, $(M/L)_{\mathrm{crit,O64}}$=24--30~\mo~pc$^{-1}$. The main filaments are therefore thermally super-critical (see blue circles in Fig.~\ref{fig:Mlin_ratio}). If non-thermal motions are considered, $(M/L)_{\mathrm{crit,vir}}$=30--75~\mo~pc$^{-1}$, most main filaments become trans-critical (see red circles in Fig.~\ref{fig:Mlin_ratio}). For the secondary filaments we obtain $M/L$=12--60~\mo~pc$^{-1}$, which can be compared to $(M/L)_{\mathrm{crit,O64}}$=24--30~\mo~pc$^{-1}$ and $(M/L)_{\mathrm{crit,vir}}$=30--70~\mo~pc$^{-1}$. They are roughly in agreement with the critical line mass regardless of whether non-thermal motions are considered or not. Figure~\ref{fig:Mlin_ratio} shows the results of the line mass comparisons. It is important to mention that for filaments F6, F7, sF5b and sF7a, it is possible to identify more than one velocity component (see Sect.~\ref{sec:kinematics}). This suggests that more than one structure (not resolved with our spatial resolution) may exist in these filaments. In these cases, we may have overestimated the mass of the filaments, leading to too high values of $M/L$. If we assume that the intensities of the two velocity components identified in F6 and F7 are directly proportional to their masses, the two components of F6 would contain 35\% and 65\% of its total mass. The $M/L$ values of these two components would be $\sim20$~\mo~pc$^{-1}$ and $\sim30$~\mo~pc$^{-1}$, similar to the $(M/L)_{\mathrm{crit,O64}}$ value. Following the same procedure, the two components of F7 each contain 50\% of the total filament mass. The two components would be trans-critical under the O64 model but sub-critical under the C51 model. The secondary filaments sF5b and sF7a also show multiple velocity components, but in these cases we can not make a clear separation between them using line intensities.

Figure~\ref{fig:mass_length} presents a comparison of the observed $M/L$ for the main filaments (F1 to F9) with a selection of filaments in other low-mass and high-mass star-forming regions. The main filaments in \MonR\ have line masses similar to the filaments in the Taurus molecular cloud ($M/L$=50~\mo~pc$^{-1}$, \citealt{Palmeirim+2013}), and Serpens ($M/L\sim$70~\mo~pc$^{-1}$, \citealt{Kirk+2013}), and clearly smaller than those found in high-mass star-forming regions such as Orion~A and DR\,21 ($M/L\sim$500~\mo~pc$^{-1}$, \citealt{Bally+1987, Johnstone-Bally1999, Hacar+2018, Stutz-Gould2016, Hennemann+2012}). This is consistent with the fact that the physical conditions measured in the \MonR\ filaments ($T_\mathrm{k}\sim$15--20~K and $n\sim$1--5$\times10^{4}$~cm$^{-3}$, \citealt{Rayner+2017}; see also Tables~\ref{tab:filaments_parameters_stability_3} and~\ref{tab:filaments_parameters_stability_4}) are more similar to those found in low-mass star-forming clouds. Figure~\ref{fig:mass_length} also shows the comparison of the range of $M/L$ values obtained in this work with the range obtained in \citet{Rayner+2017}, which are in agreement within a factor of 1.5.

The dense clumps and cores identified in \textit{Herchel} continuum maps \citep{Rayner+2017} and LMT (Large Millimetre Telescope) continuum maps \citep{Sokol+2019} appear distributed along the filaments of \MonR\ (see Fig.~\ref{fig:large-skeleton}). The clumps and cores identified in both works are consistent, with only a few bound cores in the external regions of the filaments reported only in the work of \citet{Rayner+2017}. In Tables~\ref{tab:filaments_parameters_stability_3} and~\ref{tab:filaments_parameters_stability_4}, we list the ranges of masses separation of the observed clumps/cores in filaments. We compare these values with the predicted masses and separations, which are listed in the Tables and derived following the IM97 and C51 models. The density $n_{\mathrm{c}}$ used to calculate the predicted separations and masses was estimated assuming that the filaments are homogeneous cylinders with $n_{\mathrm{c}}$ being the average density derived from the mass and size of the filament. This value of $n_{\mathrm{c}}$, a few $10^{4}$~cm$^{-3}$, is a lower limit to the density. In order to account for possible density gradients within the filaments, we adopt a value 10 times larger as an upper limit to the central density. The obtained values, a few $10^{5}$~cm$^{-3}$, are similar to those measured by \cite{Berne+2009} and \cite{Ginard+2012} within the central hub (see also \citealt{Rizzo+2003}). Figure~\ref{fig:M_L_clumps} shows a comparison between the observed and predicted clump masses and separations. The observed separations ($\lambda_{\mathrm{cl,obs}}$=0.25--2.00~pc) are in agreement with the predictions of the C51 model ($\lambda_{\mathrm{cl,vir}}$=0.20--1.60~pc), and they are 5--10 times larger than the predictions of the IM97 model ($\lambda_{\mathrm{cl,IM97}}$=0.05--0.25~pc). Similarly, most of the observed masses ($M_{\mathrm{cl,obs}}$=5--35~\mo) are in agreement with the predictions of the C51 model ($M_{\mathrm{cl,vir}}=8$--55~\mo). The observed clump masses are 1--5 times larger than the predictions of the IM97 model ($M_{\mathrm{cl,IM97}}$=1--5~\mo; see Fig.~\ref{fig:M_L_clumps}). In summary, our observations are in good agreement with the C51 model. This indicates that that non-thermal motions are not negligible in the fragmentation and formation of clumps and cores within the filaments of \MonR. Finally, it is worth noting that only 50\% of the mass outside the hub is contained within the filaments (see Table~\ref{tab:hub-masses}), while the rest is distributed in a more extended and diffuse inter-filament medium. This diffuse inter-filament medium is basically devoid of clumps, suggesting that it is non-star-forming gas.

%
\subsection{Filament kinematics\label{sec:kinematics}}

In this section, we study the kinematic properties of the \MonR\ hub-filament system, with special focus on the line shape properties (Sect.~\ref{sec:velocitylinewidths}) and the velocity gradients along the filaments (Sect.~\ref{sec:velocitygradients}) and inside the central hub (Sect.~\ref{sec:hub}).

%
\subsubsection{Velocity components and linewidths\label{sec:velocitylinewidths}}

Most of the main and secondary filaments have a relatively simple velocity structure with one velocity component (see Figs~\ref{fig:spectra_1} to~\ref{fig:spectra_11} in Appendix~\ref{A:figures}). However, few of them show two velocity components (F6, F7, sF5b and sF7a). This is similar to the velocity structure observed towards some filaments in low-mass star forming regions like Taurus, where a number of velocity-coherent, small filaments or `fibers' have been found \citep[\eg][]{Hacar+2013}. However, other authors \citep[\eg][]{Zamora-Aviles+2017, Clarke+2018} suggest that it is not clear that fibers are actual objects. Our low angular resolution ($\sim$25\arcsec, or 0.1~pc), despite resolving the kinematic structure of the filaments, prevents us from searching for `fiber'-like structures in \MonR. Higher angular resolution observations with facilities like ALMA \citep[Atacama Large Milllimeter/Sub-millimeter Array, ][]{ALMA+2015} may help in the search for small-scale sub-structures.

\begin{figure}[t!]
\centering
\includegraphics[width=0.45\textwidth]{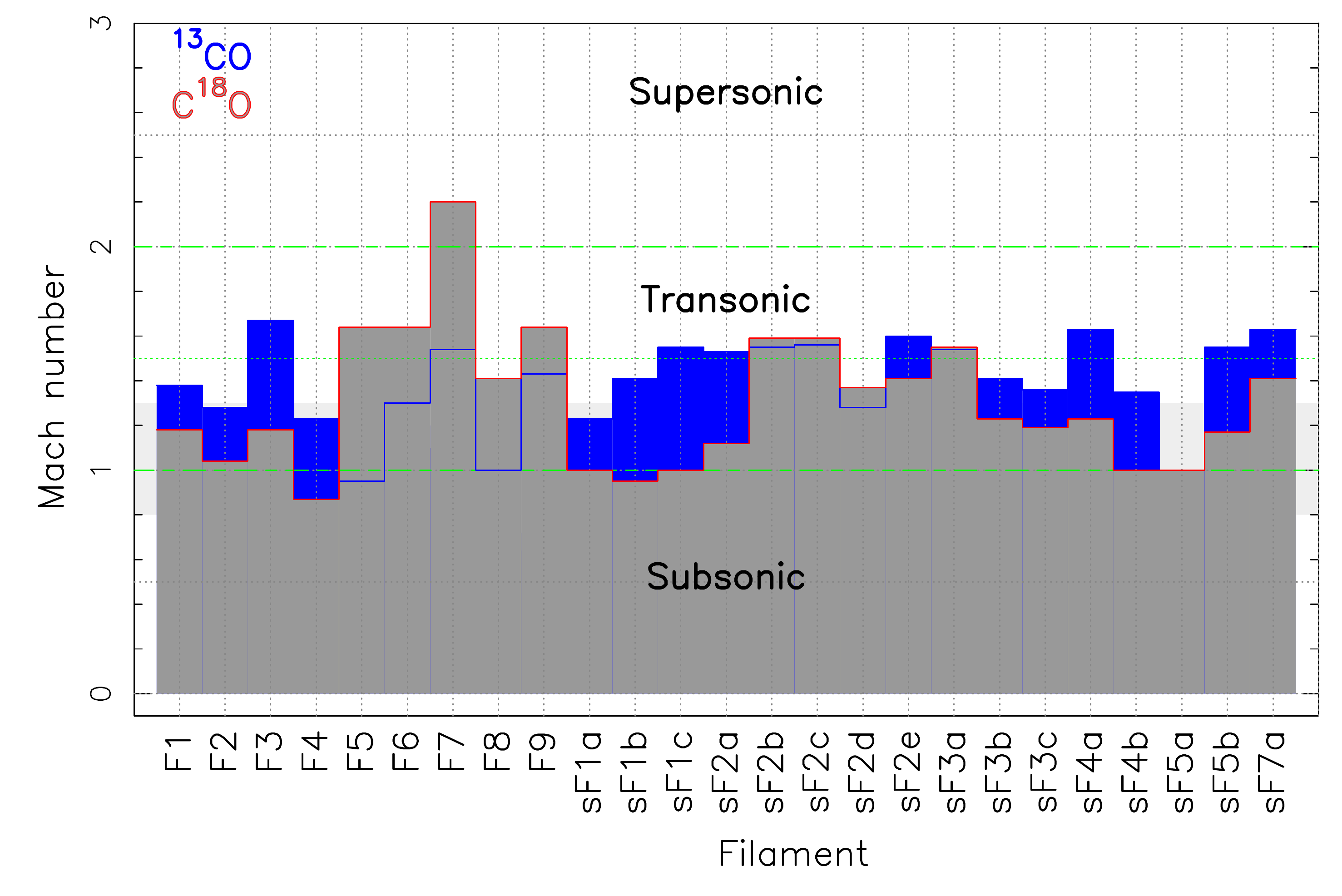}\\
\includegraphics[width=0.45\textwidth]{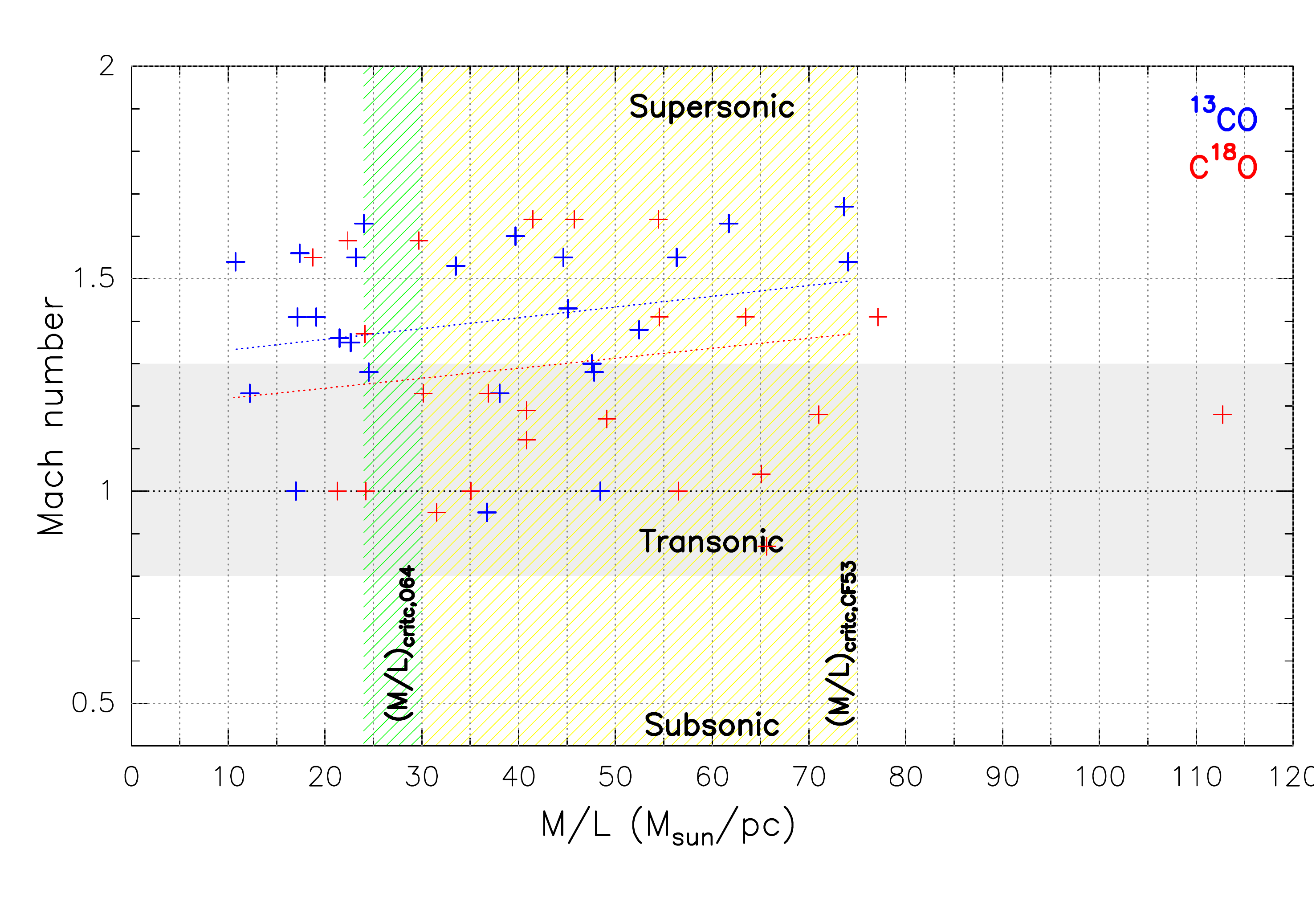}
\caption{\textit{Top:} Distribution of the Mach number calculated from the $^{13}$CO (blue) and C$^{18}$O (red-grey) velocity dispersion. \textit{Bottom:} Relation between the Mach number and the observed $M/L$. The green area indicates the range of the $(M/L)_{\mathrm{crit,O64}}$ values and the yellow one indicates the range of the $(M/L)_{\mathrm{crit,vir}}$ values. A lineal fit is indicated by the blue and red dotted lines.}
\label{fig:Mach_number}
\end{figure}

In order to have a complete image of the kinematical profiles of the filaments, we extracted a number of $^{13}$CO and C$^{18}$O spectra along the filament skeletons. We fit them with Gaussian functions. The whole spectra set and Gaussian fits are shown in Appendix~\ref{A:figures}, while Fig.~\ref{fig:kinematics} presents a summary of the main results. Larger linewidths are observed in the hub, very likely as a consequence of filaments merging together and due to the presence of a hot and expanding UC~\hii\ region \citep[\eg][see also Sect.~\ref{sec:hub}]{Trevino-morales+2016}. The filaments have linewidths of 1--2~\kms\ in $^{13}$CO, and 0.5--1.5~\kms\ in C$^{18}$O. Assuming that the gas and dust are thermalized, $T_\mathrm{k}=T_\mathrm{d}$, the non-thermal velocity dispersion, $\sigma_\mathrm{NT}$, can be determined as
\begin{equation}\label{eq:sigmaNT}
\sigma_\mathrm{NT} =
\left[ \left( \frac{ \Delta V }{ \sqrt{8ln2} }\right)^{2} - \left(\frac{k_\mathrm{B} T_\mathrm{k}}{\mu_{\mathrm{X}}m_{\mathrm{H}}} \right)^{2} \right]^{1/2},
\end{equation}
where $\Delta V$ is the observed full-width at half-maximum, $T_\mathrm{k}$ is the kinetic temperature, m$_{\mathrm{H}}$ is the mass of the hydrogen atom, and $\mu_{\mathrm{X}}$ is the molecular mass of the a specific molecule (\ie\ 29 for $^{13}$CO and 30 for C$^{18}$O). Assuming $T_\mathrm{k} = T_\mathrm{d}$, all the filaments have $T_\mathrm{k}$ between 14 and 18~K (Table~\ref{tab:filaments_parameters_stability_1} and~\ref{tab:filaments_parameters_stability_2}), corresponding to a thermal sound speed\footnote{The thermal sound speed, $c_\mathrm{s}(T_\mathrm{k}) = k_\mathrm{B} T_\mathrm{k}/\mu_{\mathrm{gas}}m_{\mathrm{H_2}}$\, was calculated assuming an average molecular mass of $\mu_{\mathrm{gas}}= 2.3$} $c_\mathrm{s}(T_\mathrm{k})$ of 0.23--0.26~\kms. Using the ratio $\sigma_{\mathrm{NT}}/c_{\mathrm{s}}(T_{\mathrm{K}})$, we calculate the Mach number, $\mathcal{M}$, for the main and secondary filaments (see Table~\ref{tab:filaments_parameters_kin_1}) and look for subsonic ($\mathcal{M}\leq1$), transonic ($1<\mathcal{M}\leq2$) and supersonic ($\mathcal{M}>2$) gas motions along them. For the filaments associated with two velocity components (\eg\ F6, F7), we estimated the Mach number using the most intense velocity component. Figure~\ref{fig:Mach_number}-top presents the distribution of $\mathcal{M}$ of all filaments. There are no significant differences between the main and secondary filaments, with mean (and standard deviation) values of $\mathcal{M}=1.5$($\pm$0.7). Our analysis, therefore, indicates that the main and secondary filaments exhibit transonic non-thermal motions on average. In Fig.~\ref{fig:Mach_number} (bottom panel), we present a comparison of $\mathcal{M}$ with the observed line mass for all the filaments. In the figure it is possible to distinguish a trend suggesting that the filaments that have larger $M/L$ also have larger $\mathcal{M}$ values (see blue and red lines in Fig.~\ref{fig:Mach_number}). 

\begin{figure*}[t!]
\vspace{0.1cm}
\centering  
\includegraphics[angle=0, width=0.8\textwidth]{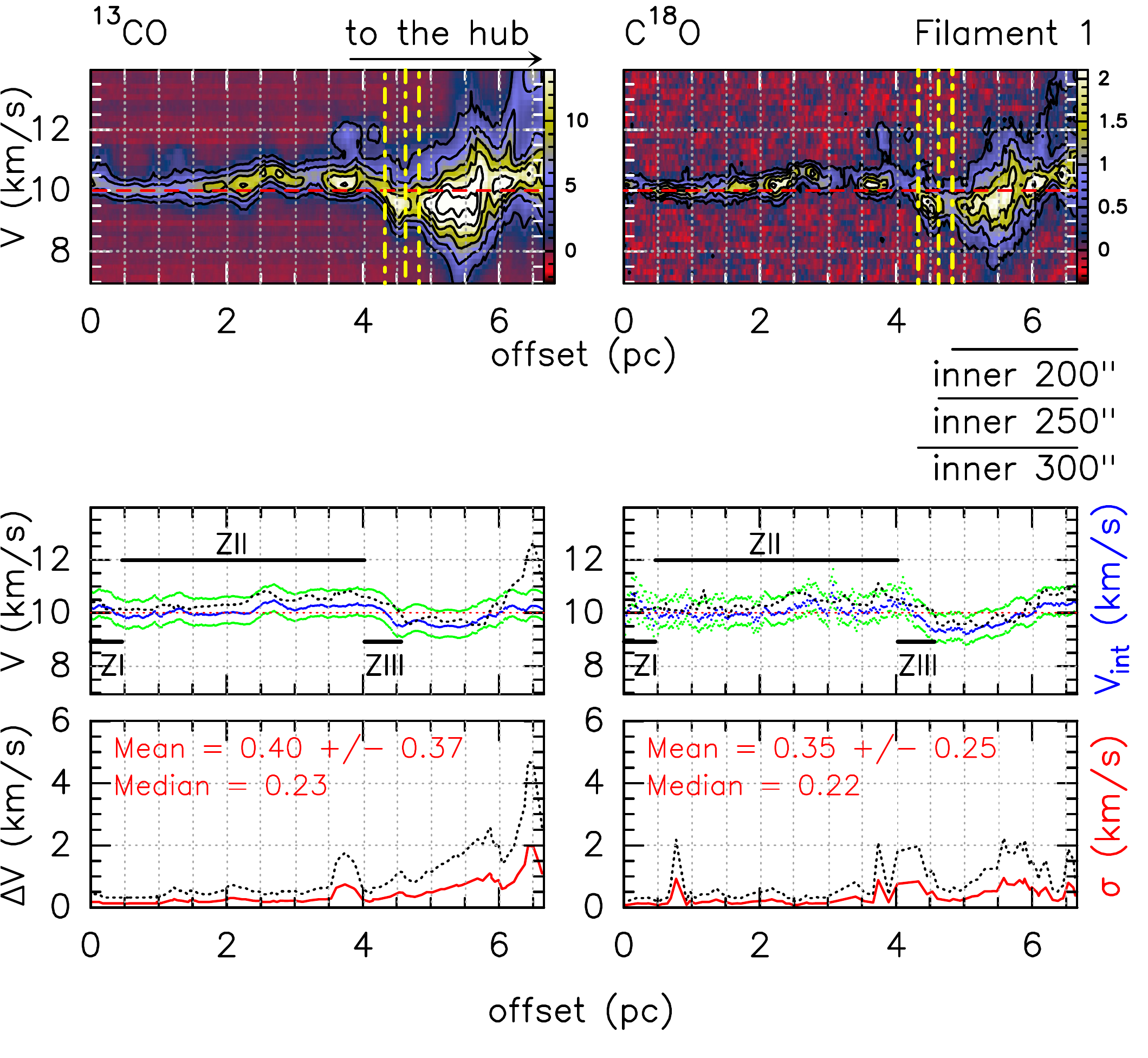} 
\vspace{0.1cm}
\caption{Position-velocity diagrams (\textit{color panels}) along the `skeleton' of filament F1 obtained from the $^{13}$CO (left) and C$^{18}$O (right) data cubes. The vertical yellow (dashed) lines indicate the transition between the hub and the filaments, corresponding to radii 200\arcsec, 250\arcsec\ ($R_\mathrm{hub}$), and 300\arcsec. The \textit{middle} panels show the variation of velocity against the offset along the filament in two different manners: The dotted black line corresponds to the velocity obtained at the central pixel that constitute the skeleton of the filament, while the blue line shows the velocity along the skeleton after averaging over the velocity range shown in the top panels. The green lines indicate the velocity range where most of the emission of the filament resides. The \textit{bottom} panels present the line-width ($\Delta v$) of the skeletons central pixels along the filaments (in black) and the velocity dispersion calculated from $\sigma=\Delta v/\sqrt{8ln2}$ (in red). The text labels show the mean and the median value of the velocity dispersion.}
\label{fig:pv_1}
\end{figure*}

Finally, we study the variation of linewidth (and velocity dispersion, see bottom panels of Figs.~\ref{fig:pv_1} and \ref{A:pv1}). We do not find large variations ($<$0.5~\kms) in the velocity dispersion along the secondary filaments. In contrast, the velocity dispersion increases along the main filaments when approaching and entering the central hub. Inside the central hub ($R_{\mathrm{hub}}<250$\arcsec) the gas has supersonic non-thermal motions on average. It is worth noting that given the moderate spatial resolution of our observations, we cannot exclude the possibility that all our filaments and secondary filaments could contain smaller (subsonic) entities as those observed in other regions (\eg\ Orion~A: \citealt{Hacar+2018}, Perseus: \citealt{Hacar+2017b} and Taurus: \citealt{Hacar+2013}).

%
\subsubsection{Velocity gradients\label{sec:velocitygradients}}

In the following, we study the velocity gradients along the filaments by constructing position-velocity (hereafter PV) diagrams along all the filament skeletons. The PV diagrams were obtained with the python tool \texttt{pvextractor}\footnote{The python package \texttt{pvextractor} is freely available at \url{http://keflavich.gitHub.io/pvextractor}} which generates PV-diagrams along any user-defined path or curved line given its spatial coordinates in a position-position-velocity data set. In the PV diagrams we average over 10~pixels (corresponding to 2~beams, or $\sim$0.2~pc) in the direction perpendicular to the filament skeleton to enhance the signal-to-noise. In this section, we analyze the velocity gradients along the filaments excluding the area located within the hub. The kinematics within the hub are discussed in Sect.~\ref{sec:hub}. 

\begin{figure}[t!]
\vspace{0.1cm}
\centering
\includegraphics[width=0.45\textwidth]{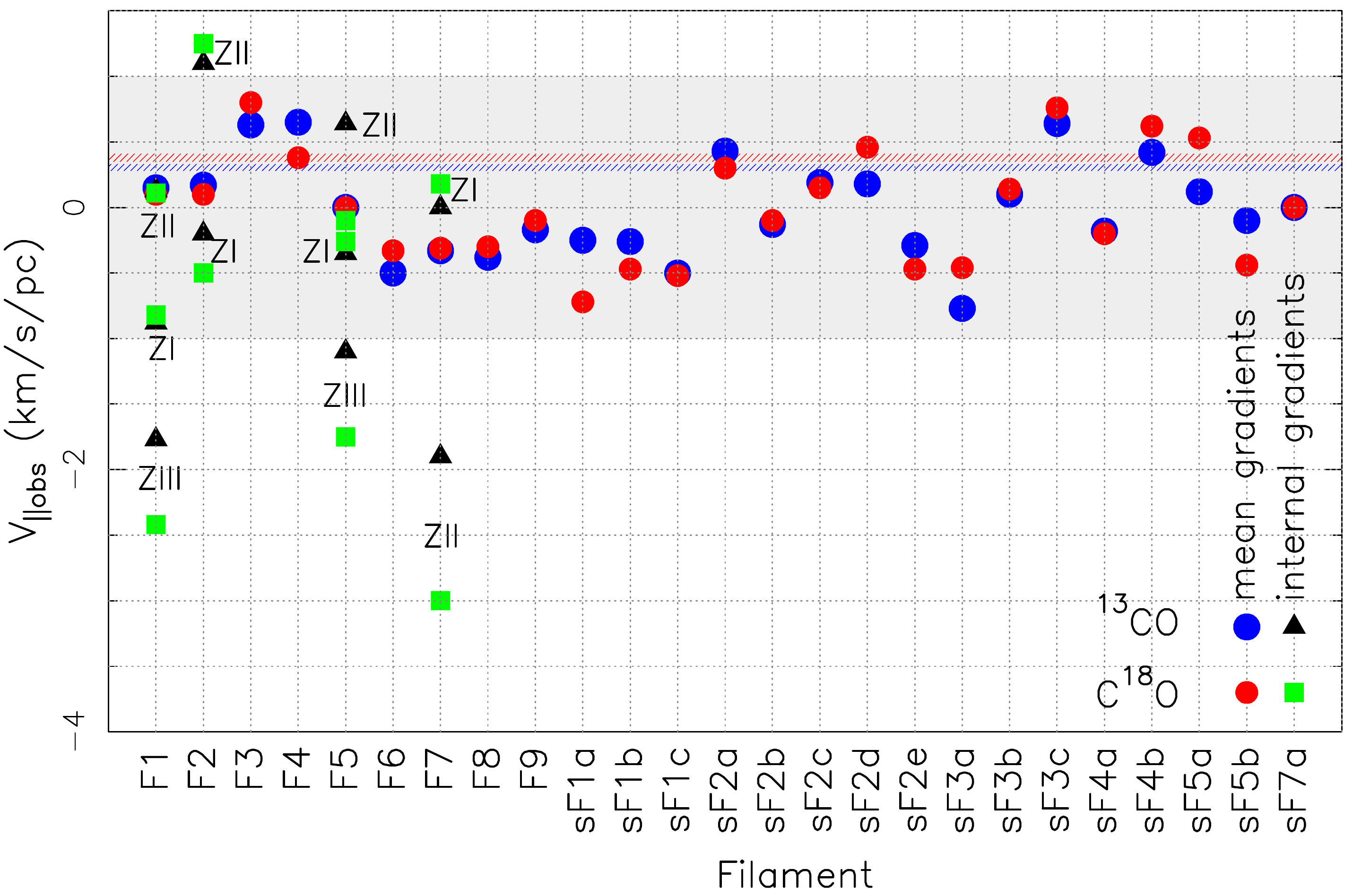}\\
\caption{Distribution of the velocity gradients. The blue dots corresponds to the values calculated from the $^{13}$CO data and the red ones to the values calculated using the C$^{18}$O data. The blue and red dotted lines indicates the average values of the gradients. The black triangles ($^{13}$CO) and green squares (C$^{18}$O) correspond to the different gradients calculated along the filaments F1, F2, F5 and F7. The Zones (ZI, ZII, and ZIII) labeled in those filaments corresponds to the ones indicated in their respective velocity diagrams (\eg\ Fig.~\ref{fig:pv_1}).}
\label{fig:V_grad}
\end{figure}

Figure~\ref{fig:pv_1} (top panels) shows the PV diagrams along the skeleton of the filament F1 for the $^{13}$CO\,(1$\rightarrow$0) and C$^{18}$O\,(1$\rightarrow$0) lines. The PV diagrams for the other filaments are shown in Fig.~\ref{A:pv1}. Most of the filaments show different velocities in the two ends of the filament, \ie\ global velocity gradients. We determine the global velocity gradient of each filament from a linear fit to the velocities along the filament (see middle panels of Figs.~\ref{fig:pv_1} and \ref{A:pv1}) after excluding the region of the filament located inside $R_\mathrm{hub}=250$\arcsec. In Table~\ref{tab:filaments_parameters_kin_1}, we list the velocity gradients derived for each filament, which are in the range 0.0--0.8~\kms~pc$^{-1}$. Figure~\ref{fig:V_grad} shows the distribution of the velocity gradients measured over the entire filaments.

Some main filaments show significant variations or `zig-zag' features in the velocity distribution. In particular, filaments F1, F2, F5 and F7 show different velocity gradients in some segments or zones along the filament. These zones are marked in the PV-diagrams as ZI to ZIII (see \eg\ Fig.~\ref{fig:pv_1}). The velocity gradients seen along the defined zones are in the range 0.2--3.0~\kms~pc$^{-1}$ (see green and black symbols in Fig.~\ref{fig:V_grad}). The larger velocity gradients are found in those regions close to the central hub, suggesting that the gas may be accelerating when approaching the center of the potential well. In contrast to the main filaments, the secondary filaments have smooth and constant velocity gradients along them. These velocity patterns have also been observed in numerical simulations of clouds in global collapse (\eg\ \citealt{Gomez-Vazquez-Semadeni(2014)}).

%
\subsection{Into the hub\label{sec:hub}}

\begin{figure*}[t!]
\centering
\includegraphics[width=0.9\textwidth]{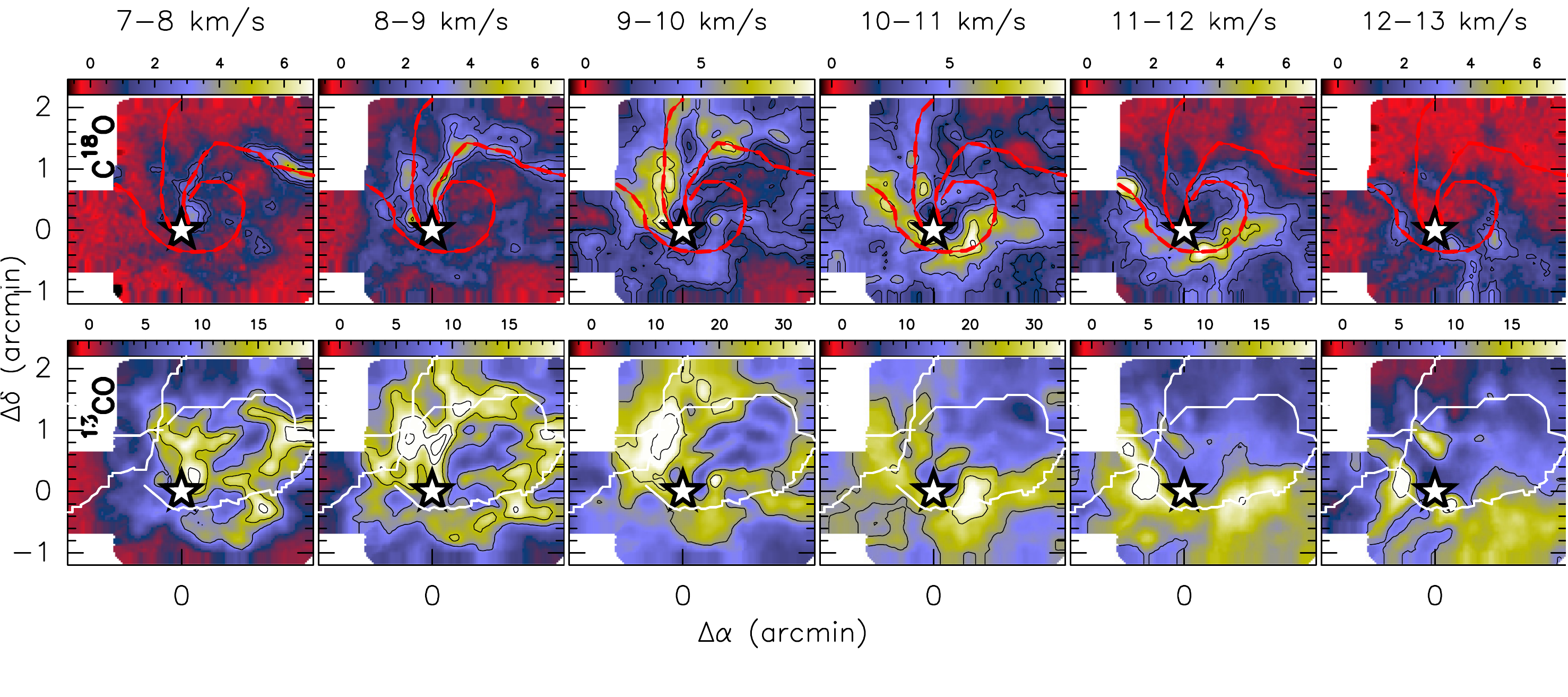} 
\vspace{-0.4cm}
\caption{Integrated emission maps in ranges of 1~\kms\ for the C$^{18}$O\,2$\rightarrow$1 (\emph{top panels}) and $^{13}$CO\,2$\rightarrow$1 (\emph{bottom panels}) lines. The red lines in the top panels depict the brightest features of the C$^{18}$O emission, and mark the possible path that the gas follows to reach the stellar cluster, indicated with a white star. The white lines in the bottom panels mark the 'skeletons' of the filaments as identified by DisPerSE in the $^{13}$CO and C$^{18}$O\,1$\rightarrow$0 maps.}
\label{channels2-1}
\end{figure*} 

As seen in Fig.~\ref{fig:large-skeleton}, the filaments extend into the central hub forming a ring structure traced by the DisPerSE filament skeletons. Several velocity components can be distinguished within the hub suggesting a complex structure that remains unresolved due to the limited angular resolution of the $^{13}$CO and C$^{18}$O\,(1$\rightarrow$0) maps. To explore the morphology and the kinematics of the central hub in more detail, we use the higher-angular resolution maps of the $^{13}$CO and C$^{18}$O\,(2$\rightarrow$1) lines. Figure~\ref{channels2-1} shows, for different velocity ranges, the superposition of the filament skeletons detected with DisPerSE (white contours) with $^{13}$CO and the brightest C$^{18}$O features. The brightest $^{13}$CO\,(2$\rightarrow$1) emission highlights an elliptical structure (hereafter hub-ring) consistent with the skeleton structure identified from the $^{13}$CO\,(1$\rightarrow$0) data. The hub-ring morphology is also observed in the C$^{18}$O\,(2$\rightarrow$1) maps, although it traces an inner layer compared to the $^{13}$CO\,(2$\rightarrow$1) maps. The innermost area of the ring-like structure is, however, devoid of $^{13}$CO and C$^{18}$O emission, suggesting lack of molecular gas, or a lower column density in the very center. This is likely caused by the interaction of the \uchii\ region associated with IRS~1 that affects the dynamics, structure, and chemistry of the gas close to the stellar cluster \citep{Pilleri+2012, Trevino-morales+2016}, creating a cavity devoid of gas.

\begin{figure*}[htp!]
\centering
\includegraphics[width=0.95\textwidth]{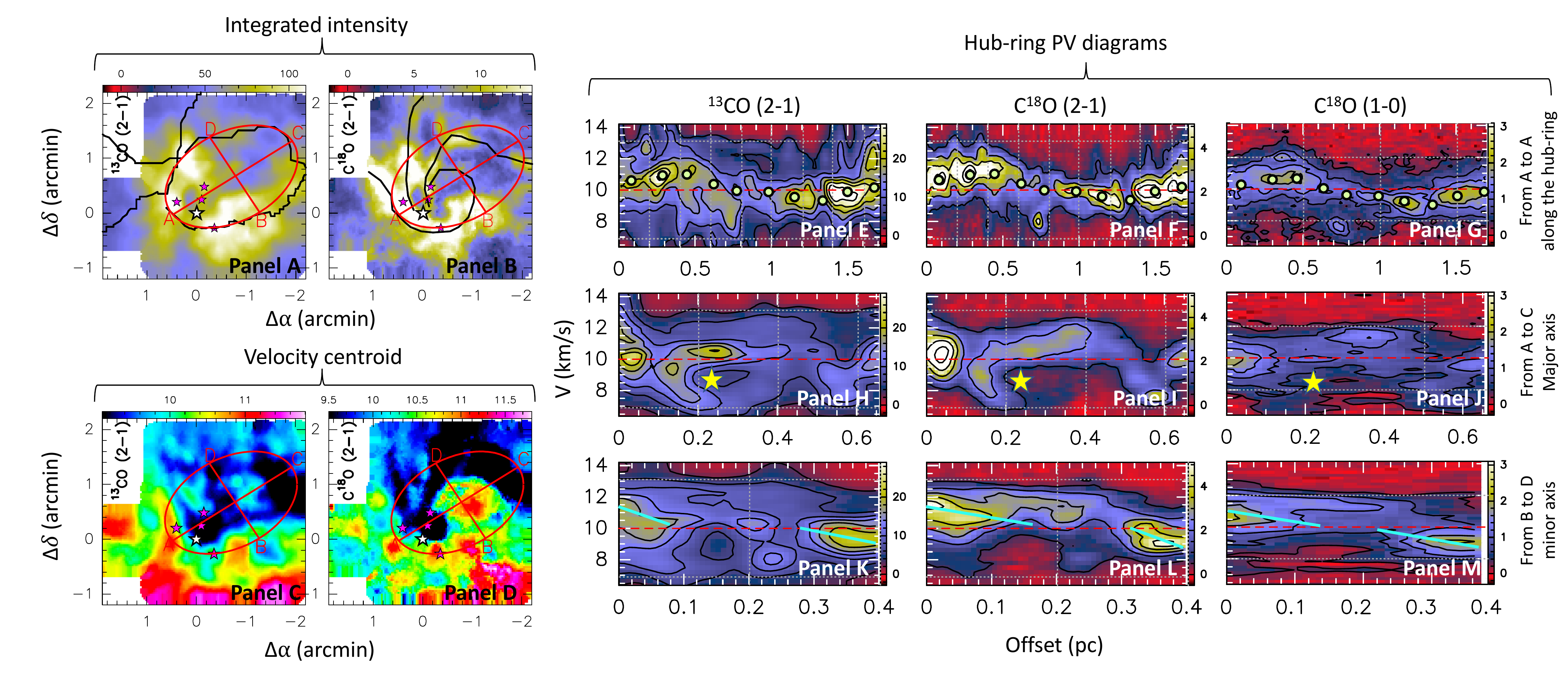}
\caption{Integrated intensity (\textit{Panels~A} and~\textit{B}) and velocity centroid (\textit{Panels~C} and~\textit{D}) maps of the C$^{18}$O and $^{13}$CO\,(2$\rightarrow$1) lines. The red ellipse marks the position of the hub-ring. \textit{Panels~E} to~\textit{G} show the PV-diagrams clockwise along the hub-ring (from point A to point A) for the $^{13}$CO\,(2$\rightarrow$1), C$^{18}$O\,(2$\rightarrow$1) and C$^{18}$O\,(1$\rightarrow$0) lines. \textit{Panels~H} and~\textit{J} show the PV-diagrams along the major axis (from point A to point C). Finally, \textit{Panels~K} and~\textit{M} show the PV-diagrams along the minor axis (from point B to point D). The green dots in \textit{Panels~E} to~\textit{G} indicate the velocities associated with the most intense emission along the ellipse, tracing the sinusoidal pattern. The yellow stars in \textit{Panels~H} to~\textit{J} show the position of the cluster along the major axis. Finally, the cyan lines in \textit{Panels~K} to~\textit{M} mark the strongest velocity gradients along the minor axis.}
\label{spiral_ellipse}
\end{figure*} 

In the following, we describe the kinematics of the gas within the hub-ring. We assume that the gas is falling into the young protostellar cluster while an UC~\hii\ region is developing and breaking out the external cocoon. We make use of PV diagrams to search for possible rotation and infall signatures. The right panels in Fig.~\ref{spiral_ellipse} show the PV diagrams built along the ellipse corresponding to the hub-ring seen in $^{13}$CO (red ellipse in \emph{Panels~A} to~\emph{D} in Fig.~\ref{spiral_ellipse}). \emph{Panels~E} and \emph{G} show the PV diagrams along the hub-ring, while \emph{panels~H} to~\emph{M} show the PV diagrams along the major and minor axis of the ellipse. The gas velocity along the ellipse follows a sinusoidal curve reminiscent of a rotational motion (green dots in \emph{Panels~E} to~\emph{G}). The interpretation of a rotational motion is also supported by the PV-diagrams along the major axis with a velocity gradient of about 4~\kms~pc$^{-1}$ from east to west (see \emph{Panels~H} to~\emph{J} in Fig.~\ref{spiral_ellipse}). However, the PV-diagrams present some features that are not following the rotational patterns. These features are likely the consequence of the interaction of the young stars with the surrounding gas (bipolar outflows and the \uchii\ region, \citealt{Dierickx+2015, Downes+1975, Massi+1985}). A velocity gradient, 1--1.5~\kms\ in 0.1--0.2~pc, is observed along the minor axis which is consistent with the presence of infall (see \emph{Panels~K} and~\emph{M} in Fig.~\ref{spiral_ellipse}). The combination of rotation and infall motions suggest that the molecular gas falls into the stellar cluster following a spiral path as seen in the morphology structure of the C$^{18}$O\,(2$\rightarrow$1) maps. In Fig.~\ref{channels2-1}, it is possible to distinguish three spiral-filament features flowing into the forming cluster. To look for further support for this scenario, it is interesting to compare the velocity gradient measured in the PV diagram with the free-fall velocity in the gravitational potential created by the stellar cluster. The total mass content in the intermediate-mass/massive IRS~1 to IRS~5 cluster is about 48~\mo\ \citep{Carpenter2008}. We need to add the mass of the population of low-mass NIR stars. Following \citet{Carpenter2008}, there are 371 stars within a circle of $R=1.85$~pc. As a first approximation, we can assume that the stellar surface density is uniform, resulting in 154 stars in $R < 0.32$~pc, and a stellar mass of 77~\mo\ assuming an average stellar mass of 0.5~\mo. Finally, we should consider the gas mass. The gas density within the \hii\ region is expected to be $\sim$100 times lower than in the molecular cloud if we assume thermal pressure equilibrium. However, the fully ionized region has a radius of $R_{\mathrm{H{\small II}}} \sim 0.09$~pc, much smaller than our ellipse. On the basis of our molecular data, we estimate a mass of $\sim1600$~\mo\ within $R_\mathrm{hub}=1$~pc. Assuming constant volume density, this would imply 43~\mo\ gas mass in the inner 0.32~pc sphere. In total, we would have a mass of 168~\mo, leading to the free-fall velocity of $\sim2.0$~\kms\ at a distance of 0.32~pc (semi-major axis of the ellipse). This free-fall velocity is consistent with the velocity gradients measured along the minor semi-axis of the hub-ring. It is important to note that ring-hub is not completely edge-on and, thus, the measured infall velocity is a lower limit. However, the mass content also suffers from significant uncertainty. Therefore, we consider that the proposed infall-rotation scenario is consistent with our observational data. Higher angular resolution observations can better resolve the spiral pattern and provide us with more constraints on the kinematics of the gas in the very center of \MonR.

%
\section{Discussion\label{sec:discussion}}

%
\subsection{Mass accretion rate\label{sec:massaccretionrate}}

In the previous sections we presented and analyzed the properties of a filamentary network converging into a dense hub. The kinematic properties can also give us information on the mass of the accretion flow ($\dot{M}_\mathrm{acc}$) along the filaments of \MonR. We calculate $\dot{M}_\mathrm{acc}$ following \citet{Kirk+2013}. We consider that the flilaments are cylinders with mass $M$, length $L$ and radius $r$. They are inclined with respect to the plane of the sky by an angle $\alpha$ and the velocity of the gas along the long axis of the filament is given by $V_{\|}$. The mass accretion rate $\dot{M}_\mathrm{acc}$ is given by
\begin{equation}\label{eq:Macc_1}
\dot{M}_\mathrm{acc} =
\left(\frac{M}{L}\right)\times V_{\|},
\end{equation}
where, due to projection effects, $L_\mathrm{obs} = L\cos(\alpha)$ and 
$V_{\|,\mathrm{obs}} = V_{\|} \sin(\alpha)$. Defining the velocity gradient as $\nabla V_{\|,\mathrm{obs}} = V_{\|,\mathrm{obs}}/L_\mathrm{obs}$, we can write Eq.~\ref{eq:Macc_1} as
\begin{equation}\label{eq:Macc_2}
\dot{M}_\mathrm{acc} =
\frac{\left(\frac{M}{L_\mathrm{obs}}V_{\|, \mathrm{obs}}\right)}{\tan(\alpha)} =
\frac{M \nabla V_{\|,\mathrm{obs}}}{\tan(\alpha)}.
\end{equation}

\begin{figure}[t!]
\centering
\includegraphics[width=0.9\columnwidth]{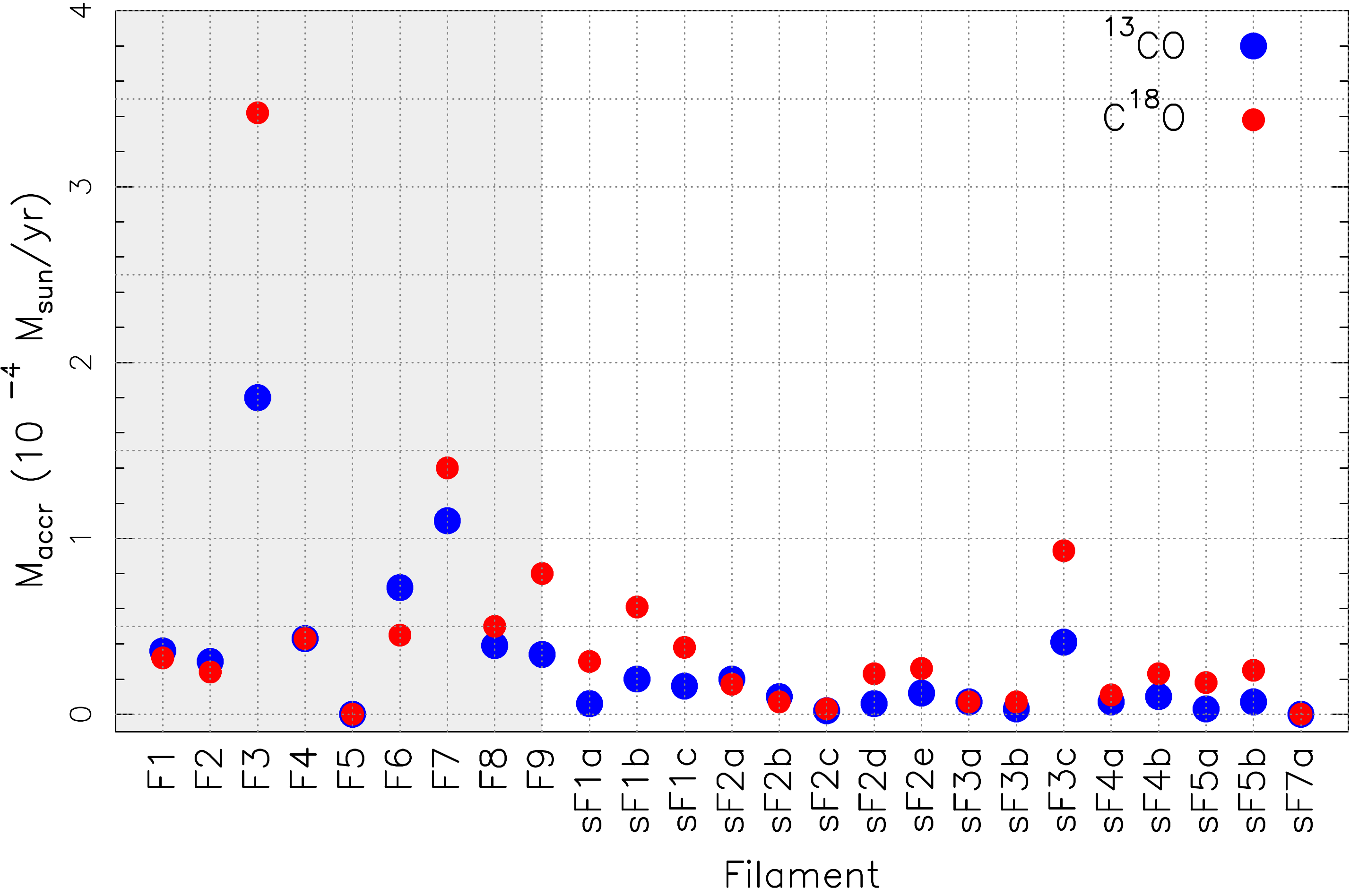} \\
\includegraphics[width=0.9\columnwidth]{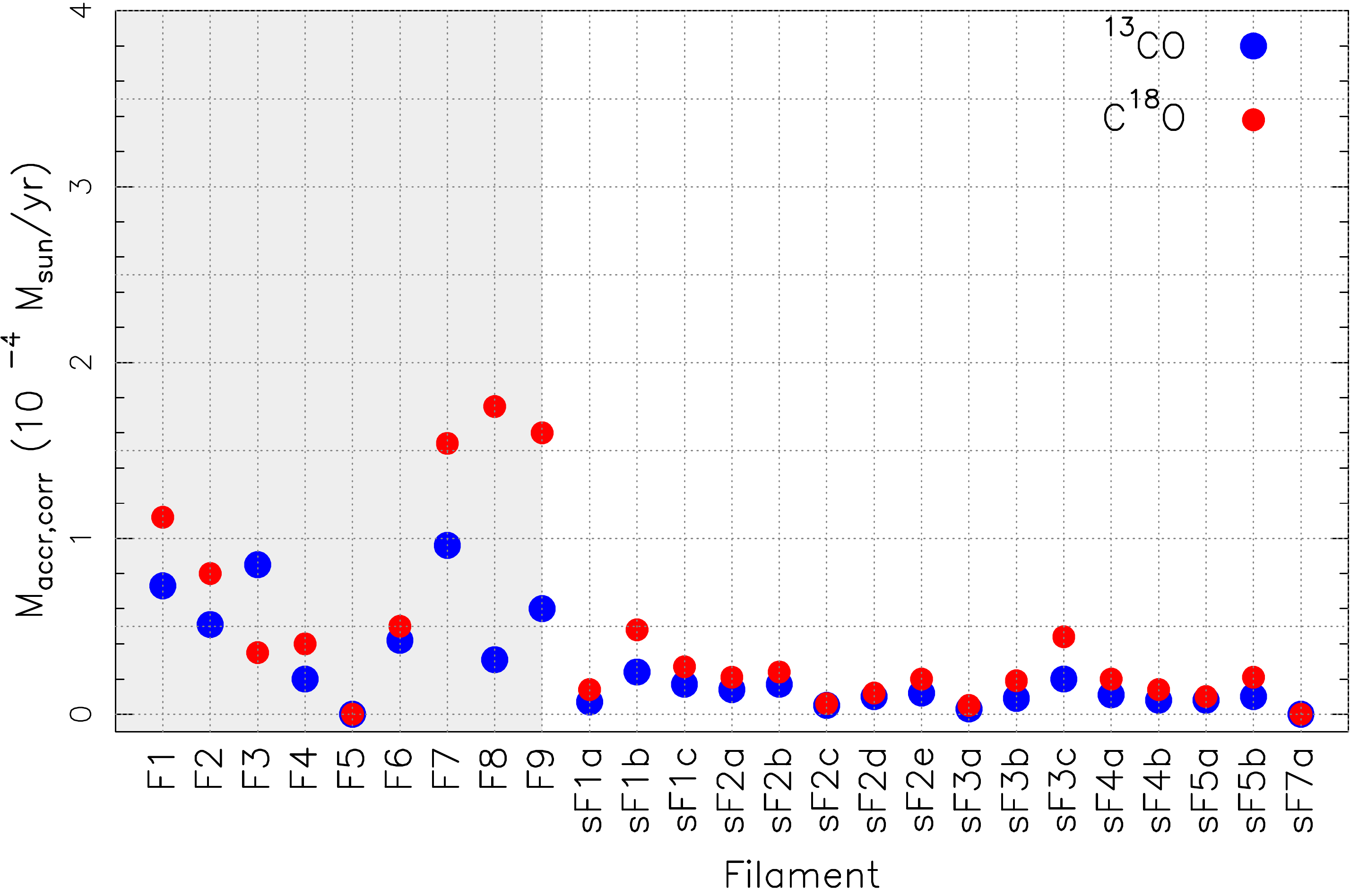} \\
\caption{Mass accretion rate along the main and secondary filament considering an inclination of $\alpha = 45\degr$ (\emph{top}) and the inclination listed in Table~\ref{tab:filaments_parameters_kin_1} (\emph{bottom}). The blue dots correspond to the values calculated from the $^{13}$CO parameters ($M$ and $\nabla V_{\|\mathrm{obs}}$) and the red ones correspond to the values calculated using the C$^{18}$O parameters. The gray zone in the plots indicate the values corresponding to the main filaments.}
\label{fig:mass-accretion-rates}
\end{figure} 

As a first approximation, we assume that all the filaments have an inclination of $\alpha=45^\circ$. In Table~\ref{tab:filaments_parameters_kin_1}, we list, alongside with the velocity gradients, the derived mass accretion rates for the filaments in \MonR\ (see also Fig.~\ref{fig:mass-accretion-rates}). We determine a mean (standard deviation) accretion rate of 0.72($\pm$0.82) $\times10^{-4}$~\mo~yr$^{-1}$ and 0.17($\pm$0.19) $\times10^{-4}$~\mo~yr$^{-1}$ for the main and secondary filaments, respectively. Changing the inclination angle to 30$^\circ$ (60$^\circ$) would increase (reduce) the mass accretion rate by a factor of 1.73. Considering that there is no preferred direction (or inclination angle) for the filaments, the measured mass accretion rates indicate that the secondary filaments transport mass to the main filaments at a rate 4 times lower than the main filaments do to the central hub.

It is important to note that each filament may be distributed around the central core with different inclination angles with respect to the plane of the sky. The angle of the filament can be obtained from
\begin{equation}\label{eq:angle_1}
\frac{V_{\|,\mathrm{obs}}}{L_\mathrm{obs}} =
\frac{V_{\|,\mathrm{real}}}{L_\mathrm{real}} \left(\frac{\sin(\alpha)}{\cos(\alpha)}\right) =
\frac{V_{\|,\mathrm{real}}}{L_\mathrm{real}}\tan(\alpha),
\end{equation}
which results in the inclination angle to be
\begin{equation}\label{eq:angle_2}
\alpha =
\tan^{-1} \frac{\nabla V_{\|,\mathrm{obs}}}{\nabla V_{\|,\mathrm{real}}}.
\end{equation}

Assuming that all the filaments are accreting material onto the hub and have the same velocity gradient, the observed differences can only be due to different inclination angles. Hence, we calculate the average of all the observed velocity gradients to be $\langle \nabla V_{\|} \rangle = 0.30$~\kms~pc$^{-1}$ (for $^{13}$CO; $\langle \nabla V_{\|} \rangle = 0.35$~\kms~pc$^{-1}$ for C$^{18}$O) and consider that this is the velocity gradient at an angle $\alpha=45^\circ$. We then determine the angle of each one of the main filaments as $\alpha=\tan^{-1}(\nabla V_{\|,\mathrm{obs}}/\langle \nabla V_{\|} \rangle$) (see Table~\ref{tab:filaments_parameters_kin_1}). With these angles, we determine the corrected mass accretion rates ($\dot{M}^\mathrm{corr}_\mathrm{acc}$, see Table~\ref{tab:filaments_parameters_kin_1}). Figure~\ref{fig:mass-accretion-rates}-bottom shows the corrected mass accretion rates for all the filaments. We find a mean (standard deviation) accretion rate of 0.70($\pm0.52$)$\times10^{-4}$~\mo~yr$^{-1}$ and 0.20($\pm0.11$)$\times10^{-4}$~\mo~yr$^{-1}$ for the main and secondary filaments, respectively. Considering the eight main filaments that feed the central hub, we determine a total mass accretion rate of 4--$7\times10^{-4}$~\mo~yr$^{-1}$. Using Eqs.~\ref{eq:angle_1} and \ref{eq:angle_2}, it is also possible to determine corrected lengths ($L^\mathrm{corr}$) for the filaments. We find that these values can be larger than the observed $L$ by a factor of 1.2--2.3, which would result in a decrease of about 35\% in the calculated $\lambda_{\mathrm{cl}}$ and $M_{\mathrm{cl}}$ parameters. Moreover, the larger values of $L$ result in a decrease of the observed $M/L$ by a factor of 10--40\%.

\begin{figure}[t!]
\centering
\includegraphics[width=0.85\columnwidth]{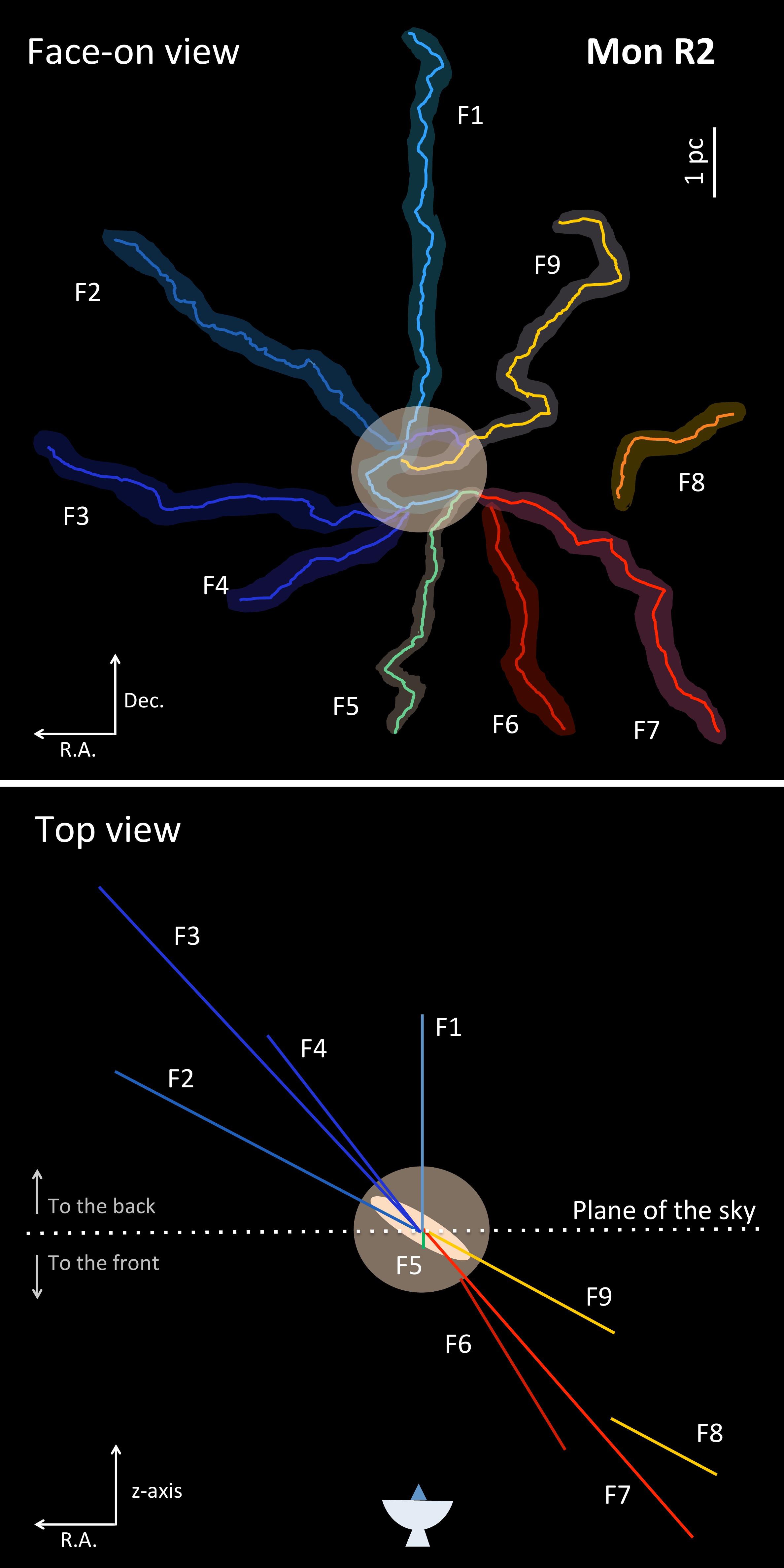}\\
\caption{3-dimensional schematic view of the filamentary structure in \MonR. Top panel shows the face-on view of the filaments, as seen in the plane of the sky. The bottom panel show the top view of the filaments. Filaments F1 to F4 are placed behind the hub (with blue-shifted velocities), while filaments F6 to F9 are placed in front of the hub (with red-shifted velocities).}
\label{fig:3D_view}
\end{figure} 

Compared to other star-forming regions, the mass accretion rates measured along the filaments of \MonR\ ($\sim10^{-4}$~\mo~yr$^{-1}$) are (\textit{i}) similar to those found in Serpens (1--3$\times10^{-4}$~\mo~yr$^{-1}$, \citealt{Kirk+2013}) Perseus (0.1--0.4$\times10^{-4}$~\mo~yr$^{-1}$, \citealt{Hacar+2017b}), and Orion ($\sim0.6\times10^{-4}$~\mo~yr$^{-1}$, \citealt{Rodriguez-Franco+1992, Hacar+2017}), (\textit{ii}) smaller by one order of magntiude than those measured in the DR\,21 ridge ($\sim10^{-3}$~\mo~yr$^{-1}$, \citealt{Schneider+2010}), and (\textit{iii}) larger than those seen in Taurus (0.1--0.9$\times10^{-5}$~\mo~yr$^{-1}$, \citealt{Hacar+2013}) and SDC\,13 (2--5$\times10^{-5}$\mo~yr$^{-1}$, \citealt{Peretto+2014}). 

It is important to note that $V_{\|,\mathrm{obs}}$ was calculated as an average velocity gradient along the filament. However, it is possible to distinguish changes in the velocity gradients along the filaments F1, F2, F5 and F7. The velocity gradients seen in the different zones (see Figs.~\ref{fig:pv_1} and \ref{A:pv1}) are in the range 0.2--3.0~\kms~pc$^{-1}$ (see green and black markers in Fig.~\ref{fig:V_grad}), and correspond to $\dot{M}_\mathrm{acc}$ of 0.3--3.5~\mo~yr$^{-1}$. The largest velocity gradients are found in the vicinity of the hub, \ie\ when the filaments reach and enter the hub. This is due to the larger masses (main filaments are gathering mass in their trajectories to the hub) and the acceleration of the material when approaching the hub. The behavior seen in filaments F1, F2, F5 and F7 is reminiscent to a gravitational collapse, where a rapid acceleration is expected in the proximity of the potential well, with the velocity varying as $R^{-0.5}$. In this expression, $R$ is the distance to the center of the potential well which is related to the distance measured in our maps, $R_{\mathrm{hub}}$, by $R=R_{\mathrm{hub}}/sin(\alpha)$ with $\alpha$ being the inclination angle relative to the plane of sky. In a rotating cloud, because of the conservation of the angular momentum, the trajectories of the infalling material change from a large-scale radial infall to a rotating flattened structure around the potential well. The rotation within the hub can produce the `zig-zag' variations seen in the PV-diagrams. In contrast with the main filaments, the velocity gradients along the secondary filaments show a constant gradient with no significant variations.

We make use of the velocity gradients and the angles derived for each filament to build a 3-dimensional vision of the filamentary network in \MonR. Figure~\ref{fig:3D_view} shows a sketch in which we assign a color to each filament depending on its location. We find that the north (F1) and eastern filaments (F2 to F4) are placed behind the hub (blue shifted in velocity), while the western filaments (F6 to F9) are placed in front of the hub (red-shifted velocities), with the ones in the north-south direction being less shifted and most likely located close to the plane of the sky. This suggests that the main filaments are located in a extended 2D sheet with an angle of 30\degr\ with respect to the plane of the sky, \ie\ the easter side being located behing the plane, and the western side in front of it.

%
\subsection{Timing a global collapse\label{sec:comparison}}

In the context of a hub-filamentary system presenting a global non-isotropic collapse, the gas flows through the filaments to form the central hub. We determine a mass-doubling time of 4--7.5~Myr to build-up the current mass of the hub ($\sim3000$~\mo) considering the total mass accretion rate of the main filaments (4--$7\times10^{-4}$~\mo~yr$^{-1}$). A slightly smaller mass-doubling time ($\sim2.5$~Myr) is obtained if we consider the larger mass accretion rates measured in the vicinity of the central hub ($\sim12\times10^{-4}$~\mo~yr$^{-1}$, see Sect.~\ref{sec:massaccretionrate}). This last value is comparable with the velocity gradients and timescale presented by \citet{Rayner+2017} when analyzing only the inner part of the filaments in \MonR. The mass-doubling time derived from the velocity gradients seen in the filaments is one order of magnitude larger than the free-fall\footnote{Considering the gas density of $\sim10^4$~cm$^{-3}$ for \MonR, the free-fall time ($t_\mathrm{ff}=\left(3\pi/(32G\rho)\right)^{-(1/2)}$) is $\sim3\times10^5$~yr} time in \MonR, suggesting a dynamically old region. If the initial density of the cloud was lower, and in the order of $\sim5\times10^2$~cm$^{-3}$, the free-fall time is in agreement with the mass-doubling time suggesting a dynamically young region.

In general, hub-filament systems are likely to be very common in massive collapsing regions as a consequence of the interaction between turbulence and gravitational instabilities. The similarity between observed hub-filament systems with numerical simulations is striking \citep[see \eg][]{Smith+2009, Gomez-Vazquez-Semadeni(2014), Vazquez-Semadeni+2017, Ballesteros-Paredes+2018, Lee-Hennebelle2016, Lee-Hennebelle2018}. \citet{Lee-Hennebelle2018} present simulations of a collapsing molecular cloud and summarize the main features of the process in: (\textit{i}) a global collapse forming a central stellar cluster, (\textit{ii}) prominent filamentary structures, and (\textit{iii}) stars forming along the radial filaments that feed the central cluster. The presence of radial filamentary structures like the one seen in \MonR\ is more prominent in simulations with a low initial density. In this situation (case~A of \citealt{Lee-Hennebelle2018}) the global collapse precedes the formation of most of the stars. Contrary to that, for initially denser clouds (see case~C of \citealt{Lee-Hennebelle2018}), star formation activity is more widespread and the global collapse is less efficient, resulting in a web-like cloud instead of a radially filamentary cloud. A different interpretation for the generation of a radial filamentary structures in a molecular cloud, is presented in \cite{Ballesteros-Paredes+2015}, where the turbulent crossing time is $\sim6$--7 times larger than the sound crossing time (consistent with the obtained in the case-A of \citealt{Lee-Hennebelle2018}). For turbulent crossing times much larger or smaller, the morphology can be substantially different (case C of \citealt{Lee-Hennebelle2018}, \citealt{Ballesteros-Paredes+2015}). 

In a recent work, \cite{Motte+2018} present an evolutionary scheme for the formation of high-mass stars (see their Fig.~8) that follows an empirical scenario qualitatively recalling the global hierarchical collapse and clump-feed accretion scenarios \citep[see][]{Vazquez-Semadeni+2009, Vazquez-Semadeni+2017, Smith+2009}. In this scenario, parsec-scale massive clumps/clouds such as ridges (\eg\ DR\,21) and hub-filament systems (\eg\ \MonR) are the preferred sites for high-mass star formation, and their physical characteristics (velocity, density and structure) favor a global controlled collapse. The Motte \et\ scheme (adapted from \citealt{Tige+2017}) represents a molecular cloud complex containing a hub/ridge filamentary system with gas flowing through the filaments to the central hub, where a number of massive dense cores/clumps (MDCs, in a 0.1~pc scale) form. During the starless phase ($\sim10^{4}$~yr), MDCs only harbor low-mass prestellar cores. The MDCs become protostellar when hosting a stellar embryo of low mass ($\sim3\times10^{5}$~yr). Then, the protostellar envelopes feed from the gravitationally-driven inflows and lead to the formation of high-mass protostars. High-mass protostars become IR-bright for stellar embryos with masses larger than 8~\mo. Finally, the main accretion phase terminates when the stellar UV radiation ionizes the envelope and generates an \hii\ region (in a time of few $10^{5}$--$10^{6}$~yr). The properties of the \MonR\ hub-filament system agree with the morphological description of the scheme presented in \cite{Motte+2018}. Adapting this evolutionary scheme for the case of \MonR, we consider that it was necessary a low initial collapsing mass (dense structure) to reach the current physical and morphological properties of the hub-filament system after $\sim1$--2~Myr. Moreover, massive star formation exist in the central hub of \MonR\ for about $10^{5}$~yr, as determined on basis of the UC~\hii\ region and surrounding PDRs \citep[see][]{Trevino-morales+2014, Didelon+2015}.

Thus far, very few massive star forming regions have been studied with a detailed similar to that presented in this paper (among them: Orion and DR\,21, \citealt{Stutz-Gould2016, Hacar+2018, Suri+2019}). Even though this group is not numerous, it is clear that giant molecular clouds may undergo different types of collapse, related more likely to their initial physical conditions. \MonR\ shows differentiated dynamical properties from the others. While DR\,21 and Orion have massive super-critical ridges with high star formation rates, \MonR\ is formed by a network of filaments resembling those in low-mass star-forming regions which converge in a single well-defined gravitational well where a cluster of massive stars are forming. The formation of the hub and radial filamentary structure has taken more than one million of years. Up to our knowledge, this is the first massive cloud with these characteristics and thus essential to compare with 3D magneto-hydrodynamic simulations to better understand the star formation process. With its simple geometry and located at only 830~pc from the Sun, \MonR\ appears as an ideal candidate to study the global collapse of a massive cloud. 

%
\section{Summary and Conclusions\label{sec:summary}}
  
In this paper, we have studied the stability and the kinematic/dynamic properties of the cluster-forming hub-filament system in th Monoceros~R2 molecular cloud. We have used large-scale maps of different molecular tracers obtained with the IRAM-30m telescope, as well as H$_2$ column density map derived from \textit{Herschel} observations. Our main results can be summarized as follows:

\begin{itemize}
\item The large scale emission seen in $^{13}$CO, C$^{18}$O, HNC and N$_2$H$^+$ correlates with the \textit{Herschel}-derived H$_2$ column density. All tracers reveal a hub-filament system in \MonR. 

\item We identify nine main filaments and 16 secondary filaments in the position-position-velocity datasets. The main filaments converge to the central hub for which we determine a radius $R_\mathrm{hub}\approx1$~pc, while the secondary filaments merge into main filaments.

\item We study the stability of the filaments by determining their line mass ($M/L$) and comparing it with the critical line masses of a thermally-supported filament and a filament supported by non-thermal motions. Both critical line masses are similar suggesting that thermal pressure and turbulence have similar contributions to the stability of the filaments. The line mass for the main filaments is 30--100~\mo~pc$^{-1}$, and is lower for the secondary filaments (12--60~\mo~pc$^{-1}$). The main filaments are slightly super-critical, while the secondary filaments are trans-critical.

\item We study the fragmentation of the filaments by comparing the masses and separations of clumps located within the filaments, with the estimates of a fragmenting filament as predicted in two different models: a filament regulated by thermal motions, and a filament with non-thermal support. The observed clump masses ($M_{\mathrm{cl,obs}}$=5--35~\mo) and separations ($\lambda_{\mathrm{cl,obs}}$=0.25--2.00~pc) are in agreement with a fragmenting properties of a filament if the non-thermal motions are considered.

\item We study the kinematic properties of the filaments by inspecting the velocity and linewidth along them. Most of the filaments have a simple velocity structure with one velocity component, and linewidths $\sim$~0.5--1.5~\kms. The linewidth increases inside the hub, likely due to the filaments merging together and the presence of a hot and expanding \uchii\ region. We find sub-sonic non-thermal motions along the filaments, which become super-sonic inside the hub.

\item We measure velocity gradients $\approx0.4$~\kms~pc$^{-1}$ in the filaments of \MonR, and derive mass accretion rates of $\approx0.7\times10^{-4}$~\mo~yr$^{-1}$ and $\approx0.2\times10^{-4}$~\mo~yr$^{-1}$ for the main and secondary filaments, respectively. We find significant variations in the velocity of some main filaments, in particular when approaching or entering the hub. The velocity gradients and mass accretion rates of these filaments increase by a factor of a few in the vicinity of the central hub, likely due to an acceleration of the accretion flow when approaching the center of the potential well.

\item Most of the main filaments extend into the central hub forming a ring structure. The kinematics of the hub-ring reveal signs of rotation and infall motions with gas flowing from the external filaments to the central massive cluster following a spiral-like pattern.
 
\item We construct a 3D schematic view of the filamentary structure in \MonR. Filaments F1 to F4 (located to the north and east) are placed behind the hub. Filaments F6 to F9 (located in the south and west) are placed in front of the hub. This scheme suggests that the filaments in \MonR\ may be distributed in a 2D plane with an angle of about 30\degr\ with respect to the plane of the sky.

\item Considering that the velocity gradients seen in the main filaments converging to the central hub correspond to infall, we estimate a timescale of about $\sim2.5$~Myr as the necessary time to gather the current mass in the central hub ($\sim3000$~\mo).

\end{itemize}

Overall, the properties of \MonR\ are in agreement with a scenario of a massive star-forming region that has been formed by a global non-isotropic collapse. The main filaments converge in the central hub from different directions feeding it at an accretion rate of $10^{-3}$--$10^{-4}$~\mo~yr$^{-1}$. The mass accretion rates increase along the filaments when approaching or entering in the hub, which may be due to an acceleration of the gas when entering the hub. In a similar way, secondary filaments feed the main filaments at smaller mass accretion rates. The main filaments extend into the central hub forming a ring structure. Within the hub, it is possible to distinguish several velocity components suggesting a complex structure that remains unresolved. The kinematics inside the hub show signs of rotation and infall motions with the gas converging in to the stellar cluster following a spiral like pattern, while the central \uchii\ region is expanding and breaking out the surrounding envelope. Thanks to its simple geometry and nearby distance (830~pc), \MonR\ is an ideal candidate to study the global collapse of a massive cloud and the formation process of high-mass stars, combining both high-spatial resolution observations and numerical simulations.

%
%
\begin{acknowledgements}
SPTM and JK acknowledge to the European Union's Horizon 2020 research and innovation program for funding support given under grant agreement No~639459 (PROMISE). AF thanks the Spanish MINECO for funding support from grants AYA2016-75066-C2-2-P, and ERC under ERC-2013-SyG, G.A.\ 610256 NANOCOSMOS. ASM and SS thank the Deutsche Forschungsgemeinschaft (DFG) for funding support via the collaborative research grant SFB\,956, projects A6 and A4. PP acknowledges financial support from the Center National de Etudes Spatiales (CNES). NS acknowledges support by the French ANR and the German DFG through the project "GENESIS" (ANR-16-CE92-0035-01/DFG1591/2-1). SS acknowledges support from the European Research Council under the Horizon 2020 Framework Program via the ERC Consolidator Grant CSF-648505. SG is funded by the European Research Council under Grant Agreement no. 339177 (STARLIGHT) of the European Community's Seventh Framework Programme (FP7/2007-2013). JRG thanks the Spanish MICIU for funding support from grant AYA2017-85111-P. SPTM acknowledges to J. Orkisz for useful discussions. We thank the anonymous referee for her/his constructive comments. 
\end{acknowledgements}
%
%

%
\begin{appendix}

%
\section{Mass and column density\label{A:ColDens}}

The mass $M$ of the filaments is given by
\begin{equation}\label{Aeq1}
M=N(\mathrm{H}_2)\,A(2.8m_\mathrm{H}),
\end{equation}
where $N(\mathrm{H}_2)$ is the total column density of the H$_2$ molecule, $A$ is the surface area of the filament and $m_\mathrm{H}$ is the hydrogen mass. When the mass is determined from a molecular tracer different to H$_2$, \eg\ $^{13}$CO, C$^{18}$O, Eq.~\ref{Aeq1} is written as
\begin{equation}\label{Aeq2}
M=\frac{N}{X}A(2.8m_\mathrm{H}),
\end{equation}
where $N$ is the total molecular column density and $X$ its abundance with respect to H$_2$.

The molecular column density can be determined from observations of a molecular transition from level $u$ (upper) to level $l$ (lower). In particular \citep[see \eg][]{Estalella-Anglada1999, Sanchez-Monge2011}, the column density of molecules in the $u$ level ($N_\mathrm{u}$) is related to the optical depth as a function of velocity ($\tau_\mathrm{v}$) by
\begin{equation}\label{Aeq3}
\tau_{\mathrm{v}}=\frac{c^{3}A_{\mathrm{ul}}}{8\pi\nu^{3}}\,N_{\mathrm{u}}\,\left(e^{ \frac{h\nu}{kT_{\mathrm{ex}}}} -1\right)\phi_{\mathrm{v}}\ (v),
\end{equation}
where $c$ is the speed of light, $A_{\mathrm{ul}}$ is the Einstein spontaneous emission coefficient, $\nu$ is the frequency of the transition, $T_{\mathrm{ex}}$ is the excitation temperature, $h$ is the Planck constant, $k$ is the Boltzmann constant and $\phi_{\mathrm{v}}\ (v)$ is the line profile function. $N_{\mathrm{u}}$ corresponds to the number of molecules in the energy level $u$ (integrated over the pathlength $dx$). The optical depth ($\tau_{\mathrm{v}}$) can be rewriten in terms of the maximum optical depth ($\tau_{\mathrm{0}}$ at the center of the line) and the linewdith ($\Delta v$) using
\begin{equation}\label{Aeq3b}
\tau_{\mathrm{v}}=\tau_{\mathrm{0}}\Delta v \phi_{\mathrm{v}}\ (v). 
\end{equation}
Inserting Eq.~\ref{Aeq3b} in Eq.~\ref{Aeq3} and normalizing the line profile to 1, $\int \phi_{\mathrm{v}}\ (v) =1$, we obtain
\begin{equation}\label{Aeq4}
\centering
\tau_{\mathrm{0}}\Delta v=\frac{c^{3}}{8\pi k\nu^{3}}\,A_{\mathrm{ul}}N_{\mathrm{u}},\left(e^{ \frac{h\nu}{kT_{\mathrm{ex}}}} -1\right).
\end{equation}
In the Rayleigh-Jeans aproximation ($h\nu \ll kT_{\mathrm{ex}}$), Eq.~\ref{Aeq4} can be written as
\begin{equation}\label{Aeq5}
\centering
\tau_{\mathrm{0}}\Delta v=\frac{c^{3}}{8\pi k\nu^{2}}\,\frac{A_{\mathrm{ul}}N_{\mathrm{u}}}{T_{\mathrm{ex}}}.
\end{equation}
The number of molecules in the energy level $u$ ($N_\mathrm{u}$) is related to the total number of molecules ($N$) by
\begin{equation}\label{Aeq6} 
\centering
N_{\mathrm{u}}=N\, \frac{g_{\mathrm{u}}}{Q(T_{\mathrm{ex}})}e^{\left(\frac{-E_{\mathrm{u}}}{T_{\mathrm{ex}}}\right)},
\end{equation}
where $g_{\mathrm{u}}$ and $E_{\mathrm{u}}$ are the upper state degeneracy and energy, respectively, and $Q(T_{\mathrm{ex}})$ is the partition function defined as the sum over all the posible energy levels. Substituting Eq.~\ref{Aeq6} in Eq.~\ref{Aeq5}, we have 
\begin{equation}\label{Aeq7}
\centering
T_{\mathrm{ex}}\tau_{\mathrm{0}}\Delta v=\frac{c^{3}A_{\mathrm{ul}}}{8\pi k\nu^{2}}\, N\, \frac{g_{\mathrm{u}}}{Q(T_{\mathrm{ex}})}\,e^{\left(\frac{-E_{\mathrm{u}}}{T_{\mathrm{ex}}}\right)}. 
\end{equation}

The opacity term in Eq.~\ref{Aeq7} can be written \citep[see \eg][]{Palau+2006} as
\begin{equation}\label{Aeq8}
\centering
\tau_{0}\Delta v=\int \tau(v)~dv=\frac{1}{J_{v}(T_{\mathrm{ex}})-J_v(T_{\mathrm{bg}})}\frac{\tau_0}{1-e^{-\tau_0}}\int T_\mathrm{L}(v)~dv,
\end{equation}
where $J_{v}(T)$ is defined as
\begin{equation}
J_nu(T) = \frac{h\nu/k}{\exp\left(\frac{h\nu}{kT}\right)-1}.
\end{equation}
If $J_{v}(T_{\mathrm{ex}}) \gg T_{\mathrm{bg}}$, where $T_\mathrm{bg}$ is the background temperature, Eq.~\ref{Aeq8} can be written as
\begin{equation}\label{Aeq9}
\centering
\tau_{0}\Delta v=\frac{1}{T_{\mathrm{ex}}}\left(\frac{\tau_0}{1-e^{-\tau_0}}\right)\int T_\mathrm{L}(v)~dv.
\end{equation}
Combining Eqs.~\ref{Aeq7} and \ref{Aeq9}, the total molecular column density $N$ can be written as 
\begin{equation}\label{Aeq10}
\centering
N=\frac{8\pi k\nu^2}{hc^3A_{\mathrm{ul}}}\left(\frac{Q(T_{\mathrm{ex}})}{g_{\mathrm{u}}}\right)\left(\frac{\tau_{0}}{1-e^{\tau_{0}}}\right)e^{\frac{E_{\mathrm{u}}}{T_{\mathrm{ex}}}}\int T_{\mathrm{L}}\,(v)dv,
\end{equation}
which simplifies to
\begin{equation}\label{Aeq11}
\centering
N=\frac{8\pi k\nu^2}{hc^3A_{\mathrm{ul}}}\left(\frac{Q(T_{\mathrm{ex}})}{g_{\mathrm{u}}}\right)e^{\frac{E_{\mathrm{u}}}{T_{\mathrm{ex}}}}\int T_{\mathrm{L}}\,(v)dv.
\end{equation}
in the optically thin scenario ($\tau \ll 1$).

The partition funtion is $Q(T_{\mathrm{ex}})=\sum g_{\mathrm{u}}e^{\left(\frac{-E_{\mathrm{u}}}{k T_{\mathrm{ex}}}\right)}$. For linear molecules like CO, the degeneracy and energy of a rotational transition going from level $u$ (described by the quantum number $J$) to a lower level $l$ (described by the quantum number $J-1$) are given by $g_{\mathrm{u}}=(2J+1)$ and $E_{\mathrm{u}}=J(J+1)hB_0$, where $B_0=h/(8\pi^{2}I)$ is the rotational constant of the molecule and $I$ its moment of inertia. Then, the partition function can be written as
\begin{equation}\label{Aeq12} 
Q(T_{\mathrm{ex}}) \simeq \frac{kT_{\mathrm{ex}}}{hB_{\mathrm{0}}} + \frac{1}{3} + \frac{1}{15}\frac{hB_{\mathrm{0}}}{kT_{\mathrm{ex}}} + ... \simeq \frac{kT_{\mathrm{ex}}}{hB_{\mathrm{0}}}. 
\end{equation}
For $^{13}$CO and C$^{18}$O, $B_\mathrm{0}$ is $55101.012$~MHz and $54891.421$~MHz, respectively, and the partition function can be calculated as $Q(T_{\mathrm{ex}})= T_{\mathrm{ex}}/2.644416~\mathrm{K}$ for $^{13}$CO, and $Q(T_{\mathrm{ex}})= T_{\mathrm{ex}}/2.634358~\mathrm{K}$ for C$^{18}$O. Applying this to Eq.~\ref{Aeq11}, we find that the column density $N$ is
\begin{equation}\label{Aeq13}
\left[\frac{N(^{13}\mathrm{CO})}{\mathrm{cm}^{-2}}\right]=4.69\times10^{13}\left[T_{\mathrm{ex}}\right]e^{\left(\frac{5.3}{T_{\mathrm{ex}}}\right)}\left[\frac{\int T(v)~dv}{\mathrm{K~km~s}^{-1}}\right],
\end{equation}
for the $^{13}$CO\,(1$\rightarrow$0) line, and
\begin{equation}\label{Aeq14}
\left[\frac{N(\mathrm{C^{18}O})}{\mathrm{cm}^{-2}}\right]=4.723\times10^{13}\left[T_{\mathrm{ex}}\right]e^{\left(\frac{5.28}{T_{\mathrm{ex}}}\right)}\left[\frac{\int T(v)~dv}{\mathrm{K~km~s}^{-1}}\right]
\end{equation}
for the C$^{18}$O\,(1$\rightarrow$0) line\footnote{We use $\nu_{10}=110.2013541$~GHz, $A_{\mathrm{ul}}=6.338\times 10^{-8}$~s$^{-1}$, $g_{\mathrm{u}}=3$ for $^{13}$CO\,(1$\rightarrow$0), and $\nu_{10}= 109.7821734$~GHz, $A_{\mathrm{ul}}=6.266\times 10^{-8}$~s$^{-1}$, $g_{\mathrm{u}}=3$ for C$^{18}$O\,(1$\rightarrow$0). Values reported in the Cologne Database for Molecular Spectroscopy (CDMS, \url{http://www.astro.uni-koeln.de/cdms/entries}).}. 

%
\section{Additional figures and tables\label{A:figures}}

\begin{figure}[t!]
\centering  
\includegraphics[width=0.9\columnwidth]{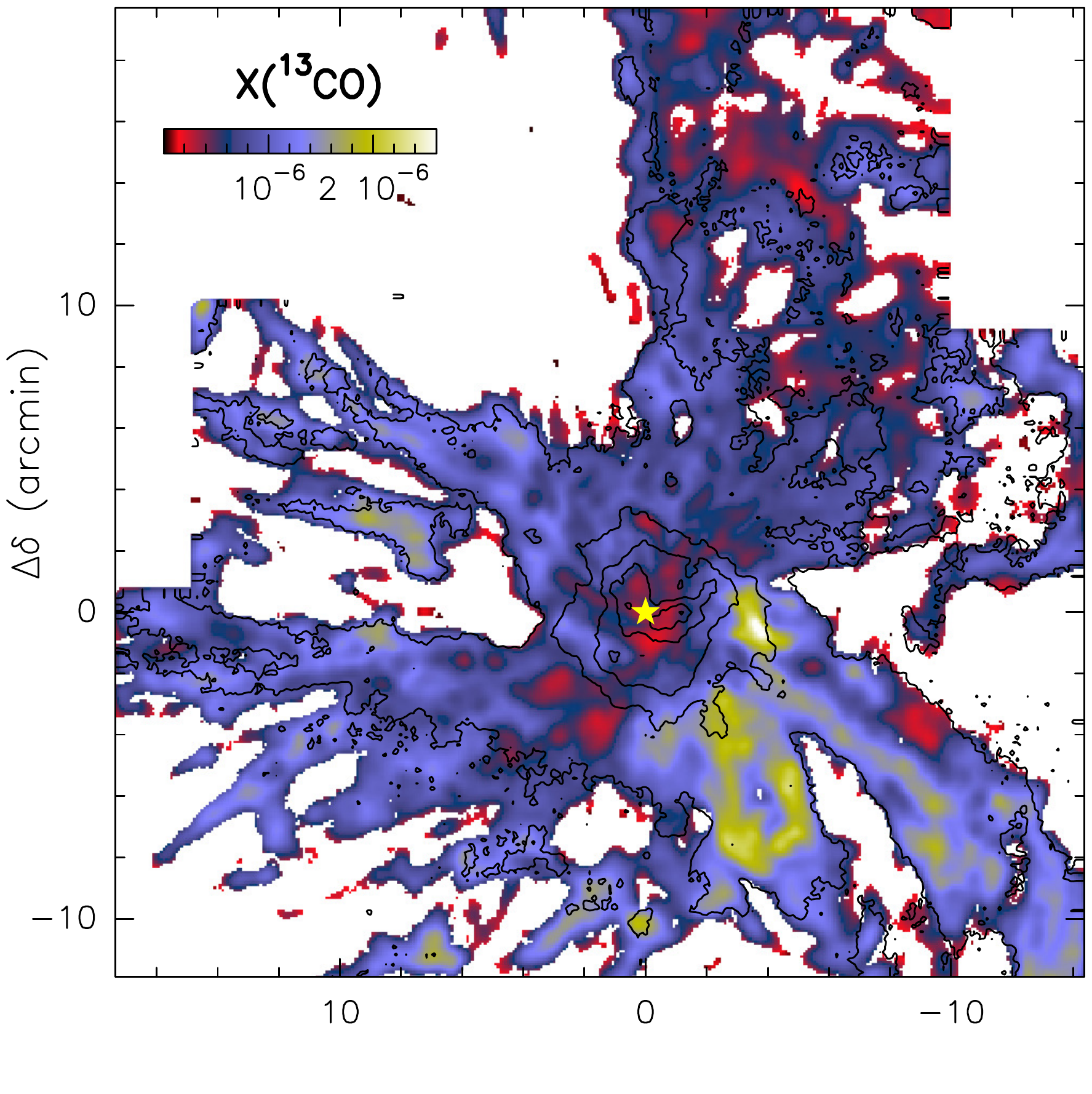}\\ 
\includegraphics[width=0.9\columnwidth]{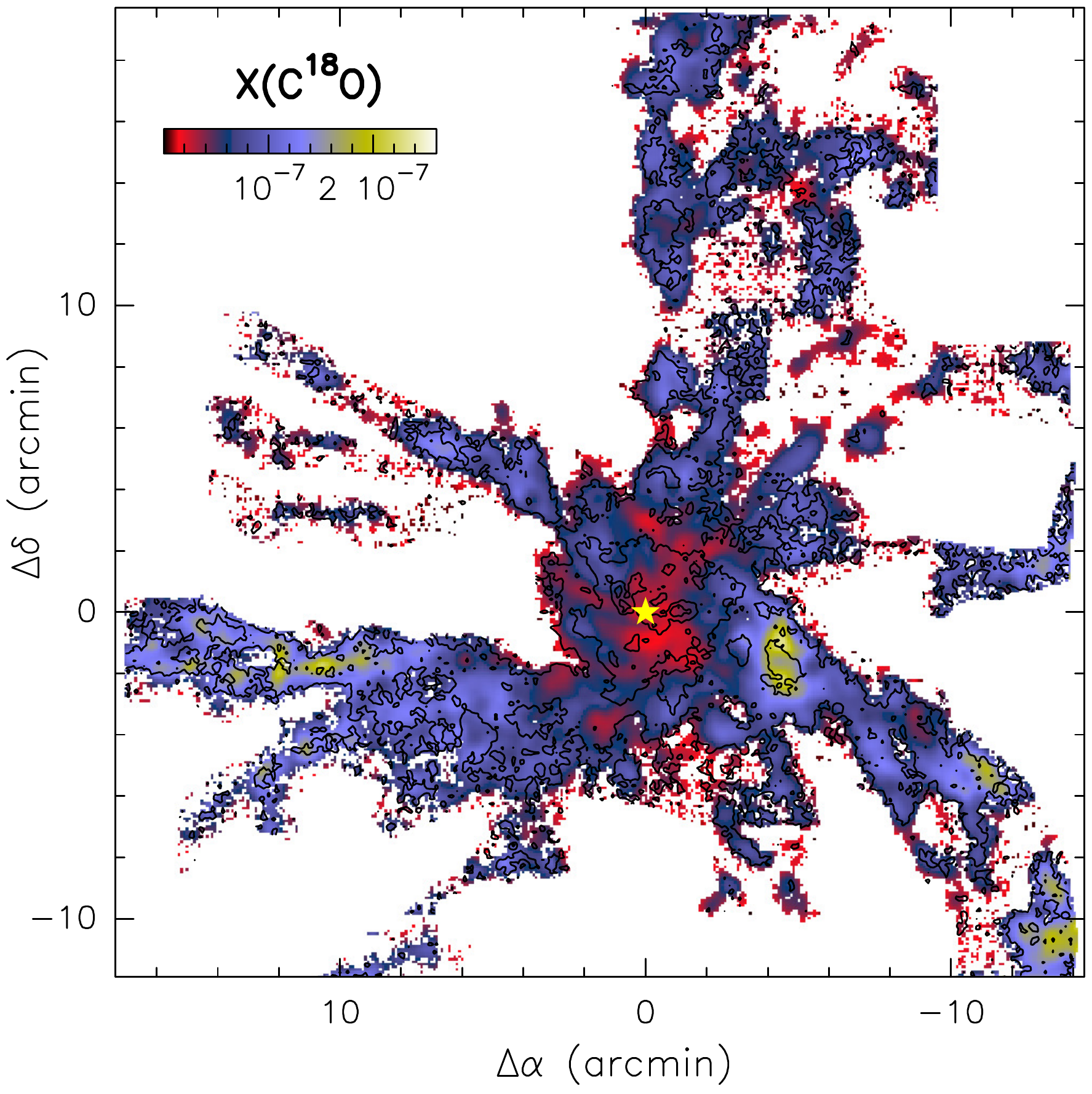}\\
\caption{$^{13}$CO and C$^{18}$O abundance maps towards \MonR. The yellow star at offset (0\arcsec, 0\arcsec) marks the position of IRS~1.}
\label{A:abundances}
\end{figure} 

\begin{figure}[hp!]
\centering
\includegraphics[width=0.9\columnwidth]{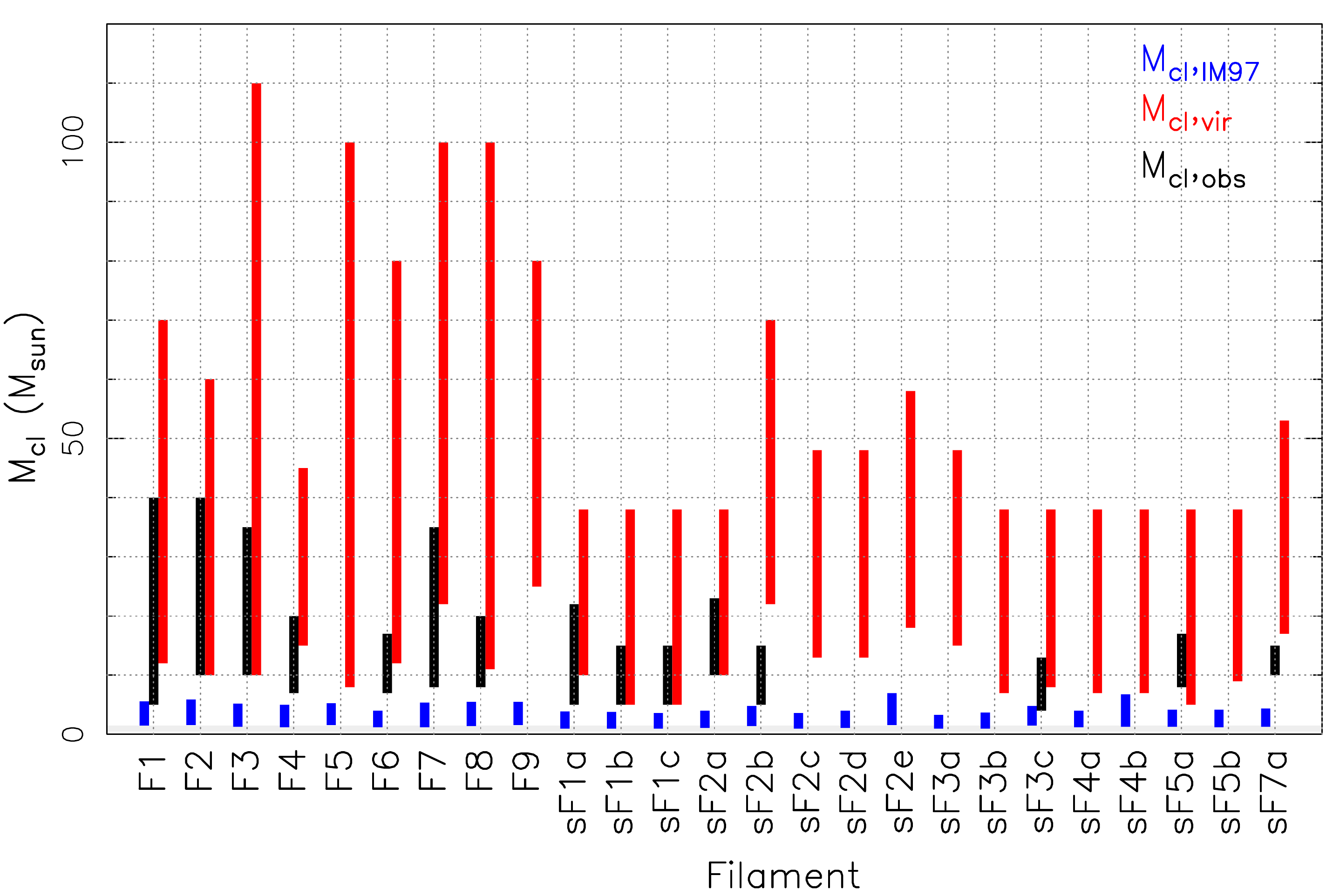} \\
\includegraphics[width=0.9\columnwidth]{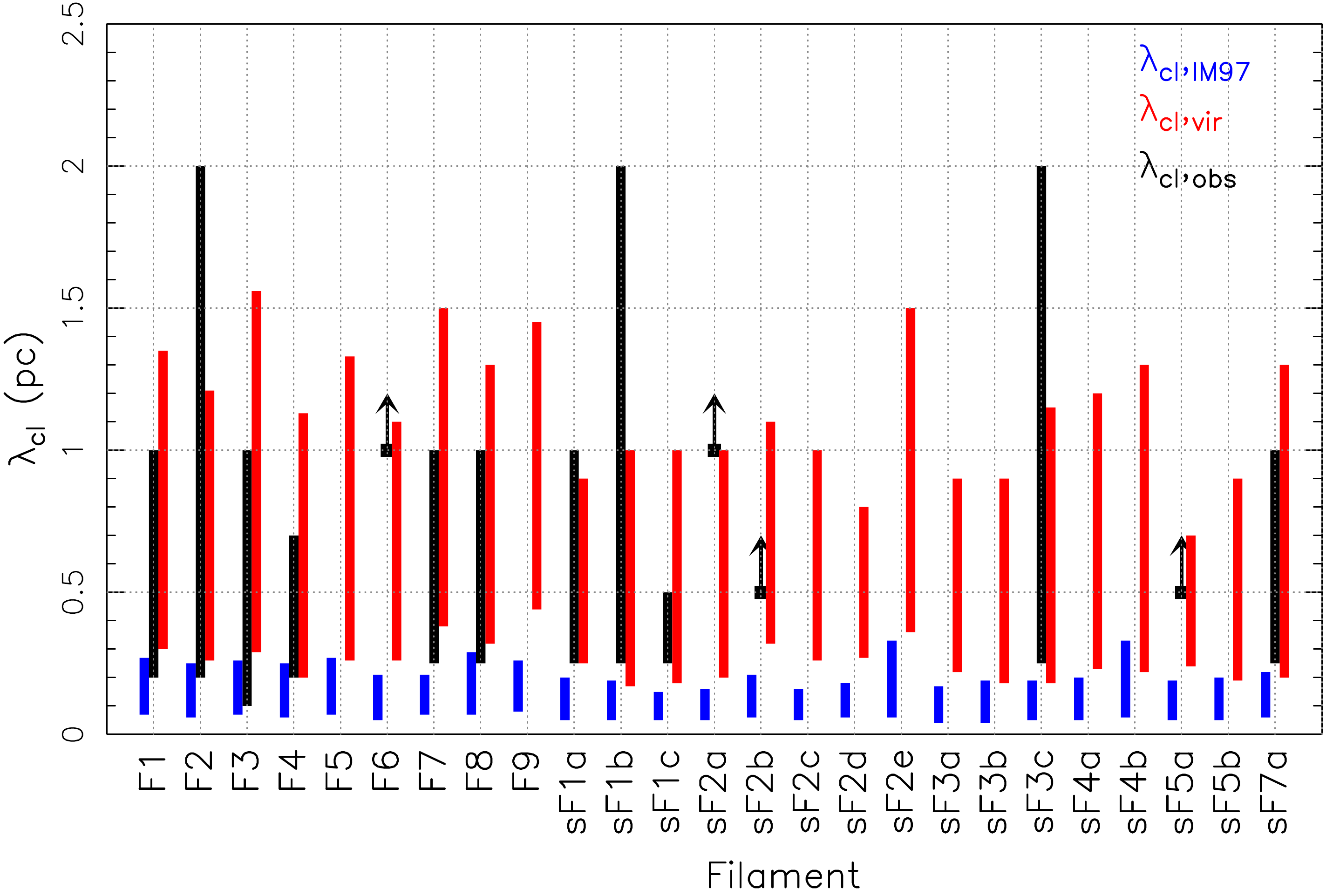} \\
\caption{Masses (top panel) and separation (bottom panels) of clumps within main and secondary filaments. The observed values are marked in black, and the ranges are set from the minimum and maximum values obtained from the different molecular tracers (see Tables~\ref{tab:filaments_parameters_stability_1} to \ref{tab:filaments_parameters_stability_4}). Theoretical values are marked in blue (following the O64 model) and in red (following the C51 model), see Sect.~\ref{sec:stability}.}
\label{fig:M_L_clumps}
\end{figure} 

In this section, we present additional Figures and Tables listing the main parameters of the filaments identified in \MonR.

Figure~\ref{A:abundances} shows an abundance map of the main analyzed species $^{13}$CO and C$^{18}$O. These maps have been obtained from the molecular column density maps obtained pixel by pixel using Eqs.~\ref{Aeq13} and \ref{Aeq14}, and the H$_2$ column density map obtained from \textit{Herschel} \citep{Didelon+2015}.

In Fig.~\ref{fig:M_L_clumps}, we show the ranges of the observed and predicted masses (\emph{top}) and separations (\emph{bottom}) between the clumps obtained from the stability analysis presented in Sect.~\ref{sec:stability}. Tables~\ref{tab:filaments_parameters_stability_1} to \ref{tab:filaments_parameters_stability_4} list the stability parameters of the main and secondary filaments identified in \MonR. All the parameters are calculated on basis of the $^{13}$CO, C$^{18}$O, and H$_2$ (derived from dust) molecular emission, following the analysis presented in Appendix~\ref{A:ColDens}.

Table~\ref{tab:filaments_parameters_kin_1} lists the kinematic parameters of the filaments derived from the $^{13}$CO and C$^{18}$O emission maps. Figure~\ref{A:pv1} shows the position-velocity diagrams along the `skeletons' of the main and secondary filaments for both $^{13}$CO and C$^{18}$O. Figures~\ref{fig:spectra_1} and~\ref{fig:spectra_11} show $^{13}$CO (black) and C$^{18}$O (red) spectra along the main filaments. The green solid lines correspond to Gaussian fits. The parameters of the Gaussian functions are listed in Table~\ref{tab:gauss_parameters_1}.

\clearpage
\begin{table*}[t!]
\centering
\caption{Physical parameters of the main filaments}
\begin{tabular}{l c c c c c c c c c}
\hline
\noalign{\smallskip}
&$L$\tablefootmark{a}
&Area\tablefootmark{b}
&$\int{T_\mathrm{MB}\,dv}$\tablefootmark{b}
&$T$\tablefootmark{c}
&$N$\tablefootmark{d}
&$M$\tablefootmark{e}
&$M/L$
&($M/L$)$_{\mathrm{crit,O64}}$\tablefootmark{f}
&($M/L$)$_{\mathrm{crit,vir}}$\tablefootmark{g}
\\
Filament
&(pc)
&(pc$^{2}$)
&(K~\kms)
&(K)
&($\times10^{16}$cm$^{-2}$)
&(\mo)
&(\mo/pc)
&(\mo/pc)
&(\mo/pc)
\\
\hline
\noalign{\smallskip}
\multicolumn{10}{l}{Derived from $^{13}$CO} \\
\hline
\noalign{\smallskip}
F1    &4.6   &1.45  &\phn9.54 &15.80  &\phn1.26  &241.22    &\phn52.44  &27  &54 \\
F2    &3.6   &1.00  &\phn9.35 &17.27  &\phn1.28  &172.04    &\phn47.80  &29  &50 \\
F3    &3.8   &1.43  &11.61    &14.97  &\phn1.48  &279.86    &\phn73.65  &25  &70 \\
F4    &1.7   &0.45  &\phn8.50 &14.97  &\phn1.10  &\phn64.67 &\phn38.04  &25  &40 \\
F5    &2.6   &0.61  &\phn8.90 &16.50  &\phn1.20  &\phn95.50 &\phn36.73  &28  &29 \\
F6    &3.0   &0.60  &13.54    &16.32  &\phn1.82  &142.77    &\phn47.59  &27  &60 \\
F7    &4.3   &1.36  &13.85    &14.73  &\phn1.76  &318.37    &\phn74.04  &25  &55 \\
F8    &2.1   &0.70  &\phn8.86 &14.61  &\phn1.11  &101.74    &\phn48.45  &24  &55 \\
F9    &4.4   &1.29  &\phn8.90 &15.66  &\phn1.16  &198.54    &\phn45.12  &26  &56 \\
\hline
\noalign{\smallskip}
\multicolumn{10}{l}{Derived from C$^{18}$O} \\
\hline
\noalign{\smallskip}
F1    &4.6   &1.45  &\phn1.29 &15.80  &\phn1.70  &326.60    &\phn71.00  &27  &40 \\
F2    &3.6   &1.00  &\phn1.27 &17.27  &\phn1.75  &234.31    &\phn65.08  &29  &34 \\
F3    &3.8   &1.43  &\phn1.78 &14.97  &\phn2.27  &428.28    &112.71     &25  &37 \\
F4    &1.7   &0.45  &\phn1.46 &14.97  &\phn1.87  &111.62    &\phn65.66  &25  &25 \\
F5    &2.6   &0.61  &\phn1.00 &16.50  &\phn1.34  &107.78    &\phn41.46  &28  &74 \\
F6    &3.0   &0.60  &\phn1.29 &16.32  &\phn1.75  &137.23    &\phn45.74  &27  &74 \\
F7    &4.3   &1.36  &\phn1.83 &14.73  &\phn2.35  &424.57    &\phn98.74  &25  &50 \\
F8    &2.1   &0.70  &\phn1.40 &14.61  &\phn1.77  &161.96    &\phn77.12  &24  &50 \\
F9    &4.4   &1.29  &\phn1.06 &15.66  &\phn1.40  &239.60    &\phn54.45  &26  &74 \\
\hline
\noalign{\smallskip}
\multicolumn{10}{l}{Derived from the H$_2$ \textit{Herschel}-derived column density maps} \\
\hline
\noalign{\smallskip}
F1    &4.6   &1.45  &\ldots   &15.80  &10.23     &334.18    &\phn72.65  &27  &\ldots \\
F2    &3.6   &1.00  &\ldots   &17.27  &\phn7.90  &179.98    &\phn50.00  &29  &\ldots \\
F3    &3.8   &1.43  &\ldots   &14.97  &10.83     &347.37    &\phn91.41  &25  &\ldots \\
F4    &1.7   &0.45  &\ldots   &14.97  &10.67     &108.37    &\phn63.75  &25  &\ldots \\
F5    &2.6   &0.61  &\ldots   &16.50  &\phn5.56  &\phn75.91 &\phn29.20  &28  &\ldots \\
F6    &3.0   &0.60  &\ldots   &16.32  &\phn7.91  &105.30    &\phn35.10  &27  &\ldots \\
F7    &4.3   &1.36  &\ldots   &14.73  &11.17     &342.98    &\phn79.76  &25  &\ldots \\
F8    &2.1   &0.70  &\ldots   &14.61  &\phn6.34  &\phn98.46 &\phn46.89  &24  &\ldots \\
F9    &4.4   &1.29  &\ldots   &15.66  &\phn7.34  &213.21    &\phn48.46  &26  &\ldots \\
\hline
\end{tabular}  
\tablefoot{
\tablefoottext{a}{The lengths are calculated from the PV-diagrams (see Figures~\ref{fig:pv_1} and \ref{A:pv1}).}
\tablefoottext{b}{From a polygon defined from the emission (over 5$\sigma$) around each filament skeleton.}
\tablefoottext{c}{From a polygon defined in the $T_{\mathrm{dust}}$ map (Figure~\ref{fig:large-skeleton}) around each filament skeleton.}
\tablefoottext{d}{From a polygon defined in the $N(^{13}\mathrm{CO})$ map (Figure~\ref{fig:large-skeleton}) around each filament skeleton.}
\tablefoottext{e}{Mass of the filament derived from Eq.~\ref{eq:mass}.}
\tablefoottext{f}{Calculated from Eq.~\ref{eq:M_L_ther}.}
\tablefoottext{g}{Calculated from Eq.~\ref{eq:M_L_non_ther}. For filaments F6 and F7, the values are calculated considering the velocity dispersion of the velocity component with larger $T_{\mathrm{MB}}$.}}
\label{tab:filaments_parameters_stability_1} 
\end{table*}

\begin{table*}[t!]
\centering
\caption{Physical parameters of the secondary filaments}
\begin{tabular}{l c c c c c c c c c}
\hline
\noalign{\smallskip}
&$L$\tablefootmark{a}
&Area\tablefootmark{b}
&$\int{T_\mathrm{MB}\,dv}$\tablefootmark{b}
&$T$\tablefootmark{c}
&$N(^{13}\mathrm{CO})$\tablefootmark{d}
&$M$\tablefootmark{e}
&$M/L$
&($M/L$)$_{\mathrm{crit,O64}}$\tablefootmark{f}
&($M/L$)$_{\mathrm{crit,vir}}$\tablefootmark{g}
\\
Filament
&(pc)
&(pc$^{2}$)
&(K~\kms)
&(K)
&($\times10^{16}$cm$^{-2}$)
&(\mo)
&(\mo/pc)
&(\mo/pc)
&(\mo/pc)
\\
\hline
\noalign{\smallskip}
\multicolumn{10}{l}{Derived from $^{13}$CO} \\
\hline
\noalign{\smallskip}
sF1a  &1.90  &0.22  &\phn6.29 &14.22  &\phn0.78  &\phn23.24 &\phn12.23  &24  &39 \\
sF1b  &4.10  &0.60  &\phn7.87 &14.63  &\phn1.00  &\phn78.32 &\phn19.10  &24  &50 \\
sF1c  &1.30  &0.25  &13.90    &15.60  &\phn1.80  &\phn58.03 &\phn45.50  &26  &67 \\
sF2a  &1.40  &0.22  &11.90    &17.37  &\phn1.63  &\phn46.93 &\phn33.52  &29  &70 \\
sF2b  &2.25  &0.35  &\phn8.25 &16.70  &\phn1.11  &\phn52.21 &\phn23.20  &29  &67 \\
sF2c  &0.80  &0.09  &\phn9.25 &15.98  &\phn1.22  &\phn13.93 &\phn17.41  &27  &67 \\
sF2d  &1.30  &0.17  &10.22    &17.80  &\phn1.43  &\phn31.89 &\phn24.53  &30  &50 \\
sF2e  &1.00  &0.28  &\phn7.90 &16.73  &\phn1.10  &\phn39.68 &\phn39.68  &28  &70 \\
sF3a  &0.80  &0.07  &\phn7.50 &15.20  &\phn1.00  &\phnn8.61 &\phn10.76  &25  &60 \\
sF3b  &1.70  &0.21  &\phn8.18 &14.60  &\phn1.00  &\phn29.17 &\phn17.16  &24  &50 \\
sF3c  &3.00  &0.55  &\phn6.89 &15.00  &\phn0.90  &\phn64.47 &\phn21.50  &25  &48 \\
sF4a  &1.55  &0.26  &\phn8.37 &15.00  &\phn1.10  &\phn37.20 &\phn24.00  &25  &67 \\
sF4b  &1.10  &0.22  &\phn6.50 &15.70  &\phn0.85  &\phn24.91 &\phn22.65  &26  &50 \\
sF5a  &1.40  &0.18  &\phn6.62 &15.90  &\phn0.87  &\phn24.91 &\phn17.80  &27  &30 \\
sF5b  &1.15  &0.24  &15.58    &16.00  &\phn2.00  &\phn64.81 &\phn56.35  &27  &60 \\
sF7a  &2.20  &0.62  &13.09    &14.80  &\phn1.66  &135.73    &\phn61.70  &25  &60 \\
\hline
\noalign{\smallskip}
\multicolumn{10}{l}{Derived from C$^{18}$O} \\
\hline
\noalign{\smallskip}
sF1a  &1.90  &0.22  &\phn1.08 &14.22  &\phn1.40  &\phn40.47 &\phn21.30  &24  &30 \\
sF1b  &4.10  &0.60  &\phn1.30 &14.63  &\phn1.64  &129.32    &\phn31.54  &24  &25 \\
sF1c  &1.30  &0.25  &\phn1.73 &15.60  &\phn2.28  &\phn73.47 &\phn56.52  &26  &30 \\
sF2a  &1.40  &0.22  &\phn1.44 &17.37  &\phn2.00  &\phn57.22 &\phn40.87  &29  &39 \\
sF2b  &2.25  &0.35  &\phn1.00 &16.70  &\phn1.42  &\phn66.82 &\phn29.70  &28  &70 \\
sF2c  &0.80  &0.09  &\phn1.17 &15.98  &\phn1.60  &\phn17.88 &\phn22.35  &27  &70 \\
sF2d  &1.30  &0.17  &\phn1.00 &17.80  &\phn1.40  &\phn31.35 &\phn24.12  &30  &57 \\
sF2e  &1.00  &0.28  &\phn1.10 &16.73  &\phn1.50  &\phn54.56 &\phn54.56  &28  &50 \\
sF3a  &0.80  &0.07  &\phn1.30 &15.20  &\phn1.70  &\phn15.01 &\phn18.76  &25  &67 \\
sF3b  &1.70  &0.21  &\phn1.42 &14.60  &\phn1.80  &\phn51.24 &\phn30.14  &24  &39 \\
sF3c  &3.00  &0.55  &\phn1.30 &15.00  &\phn1.67  &122.40    &\phn40.80  &25  &37 \\
sF4a  &1.55  &0.26  &\phn1.27 &15.00  &\phn1.64  &\phn57.20 &\phn36.90  &25  &39 \\
sF4b  &1.10  &0.22  &\phn1.00 &15.70  &\phn1.32  &\phn38.55 &\phn35.05  &26  &30 \\
sF5a  &1.40  &0.18  &\phn1.13 &15.90  &\phn1.50  &\phn33.91 &\phn24.22  &27  &30 \\
sF5b  &1.15  &0.24  &\phn1.35 &16.00  &\phn1.80  &\phn56.54 &\phn49.17  &27  &39 \\
sF7b  &2.20  &0.62  &\phn1.34 &14.80  &\phn1.70  &139.59    &\phn63.45  &25  &50 \\
\hline
\noalign{\smallskip}
\multicolumn{10}{l}{Derived from the H$_2$ \textit{Herschel}-derived column density maps} \\
\hline
\noalign{\smallskip}
sF1a  &1.90  &0.22  &\ldots   &14.22  &\phn5.93  &\phn29.99 &\phn15.78  &24  &\ldots \\
sF1b  &4.10  &0.60  &\ldots   &14.63  &\phn7.86  &105.47    &\phn25.72  &24  &\ldots \\
sF1c  &1.30  &0.25  &\ldots   &15.60  &13.72     &\phn75.11 &\phn57.78  &26  &\ldots \\
sF2a  &1.40  &0.22  &\ldots   &17.37  &11.81     &\phn57.84 &\phn41.31  &29  &\ldots \\
sF2b  &2.25  &0.35  &\ldots   &16.70  &\phn5.84  &\phn46.70 &\phn20.75  &29  &\ldots \\
sF2c  &0.80  &0.09  &\ldots   &15.98  &\phn7.34  &\phn14.26 &\phn17.83  &27  &\ldots \\
sF2d  &1.30  &0.17  &\ldots   &17.80  &\phn7.51  &\phn28.57 &\phn21.98  &30  &\ldots \\
sF2e  &1.00  &0.28  &\ldots   &16.73  &\phn4.93  &\phn31.10 &\phn31.10  &28  &\ldots \\
sF3a  &0.80  &0.07  &\ldots   &15.20  &\phn4.73  &\phnn7.20 &\phnn9.00  &25  &\ldots \\
sF3b  &1.70  &0.21  &\ldots   &14.60  &\phn5.23  &\phn25.21 &\phn14.83  &24  &\ldots \\
sF3c  &3.00  &0.55  &\ldots   &15.00  &\phn5.08  &\phn63.16 &\phn21.05  &25  &\ldots \\
sF4a  &1.55  &0.26  &\ldots   &15.00  &\phn7.40  &\phn43.63 &\phn28.15  &25  &\ldots \\
sF4b  &1.10  &0.22  &\ldots   &15.70  &\phn2.99  &\phn14.84 &\phn13.49  &26  &\ldots \\
sF5a  &1.40  &0.18  &\ldots   &15.90  &\phn5.82  &\phn22.47 &\phn16.05  &27  &\ldots \\
sF5b  &1.15  &0.24  &\ldots   &16.00  &\phn8.64  &\phn46.38 &\phn40.33  &27  &\ldots \\
sF7a  &2.20  &0.62  &\ldots   &14.80  &\phn8.69  &121.25    &\phn55.11  &25  &\ldots \\
\hline
\end{tabular} 
\tablefoot{
\tablefoottext{a}{The lengths are calculated from the PV-diagrams (see Figures~\ref{fig:pv_1} and \ref{A:pv1}).}
\tablefoottext{b}{From a polygon defined from the emission (over 5$\sigma$) around each filament skeleton.}
\tablefoottext{c}{From a polygon defined in the $T_{\mathrm{dust}}$ map (Figure~\ref{fig:large-skeleton}) around each filament skeleton.}
\tablefoottext{d}{From a polygon defined in the $N(^{13}\mathrm{CO})$ map (Figure~\ref{fig:large-skeleton}) around each filament skeleton.}
\tablefoottext{e}{Mass of the filament derived from Eq.~\ref{eq:mass}.}
\tablefoottext{f}{Calculated from Eq.~\ref{eq:M_L_ther}.}
\tablefoottext{g}{Calculated from Eq.~\ref{eq:M_L_non_ther}.}
\tablefoottext{*}{These values were calculated considering the velocity dispersion of the velocity component presenting the largest $T_{\mathrm{MB}}$.}
}
\label{tab:filaments_parameters_stability_2}
\end{table*}

\begin{table*}[t!]
\centering
\caption{Clumps properties of the main filaments}
\begin{tabular}{l c c c c c c c} 
\hline
\noalign{\smallskip}
&$n_{\mathrm{c}}$\tablefootmark{a}
&$\lambda_{\mathrm{cl,obs}}$\tablefootmark{b}
&$M_{\mathrm{cl,obs}}$\tablefootmark{b}
&$\lambda_{\mathrm{cl,IM97}}$\tablefootmark{c}
&$M_{\mathrm{cl,IM97}}$\tablefootmark{d}
&$\lambda_{\mathrm{cl,vir}}$\tablefootmark{e}
&$M_{\mathrm{cl,vir}}$\tablefootmark{f}  
\\
Filament
&($10^{4}$~cm$^{-3}$)
&(pc)
&(\mo)
&(pc)
&(\mo)
&(pc)
&(\mo)
\\
\hline
\noalign{\smallskip}
\multicolumn{8}{l}{Derived from $^{13}$CO} \\
\hline
\noalign{\smallskip}
F1   &0.97 & 0.20--1.00  & \phn5--35  &0.09--0.27  &1.8--5.6 &0.43--1.35 &20--\phn70    \\ 
F2   &1.13 & 0.20--2.00  & 10--25     &0.08--0.25  &1.9--5.9 &0.38--1.21 &20--\phn60    \\
F3   &0.95 & 0.10--1.00  & 10--15     &0.08--0.26  &1.6--5.2 &0.49--1.56 &35--110       \\ 
F4   &1.00 & 0.20--0.70  & \phn7--10  &0.08--0.25  &1.6--5.0 &0.35--1.13 &15--\phn45    \\
F5   &1.23 & \ldots      & \ldots     &0.07--0.24  &1.7--5.3 &0.26--0.85 &\phn8--\phn22 \\ 
F6   &2.19 & $>1.00$     & 12--17     &0.05--0.18  &1.2--3.9 &0.26--0.85 &12--\phn40    \\
F7   &1.36 & 0.25--1.00  & \phn8--25  &0.07--0.21  &1.3--4.3 &0.38--1.20 &22--\phn70    \\ 
F8   &0.80 & 0.25--1.00  & \phn8--11  &0.09--0.28  &1.7--5.5 &0.32--1.00 &11--\phn35    \\
F9   &0.96 & \ldots      & \ldots     &0.08--0.26  &1.7--5.5 &0.44--1.40 &25--\phn80    \\
\hline
\noalign{\smallskip}
\multicolumn{8}{l}{Derived from C$^{18}$O} \\
\hline
\noalign{\smallskip}
F1   &1.31 & 0.20--1.00  & 10--30     &0.07--0.22  &1.5--4.8 &0.30--1.00 &12--\phn38    \\ 
F2   &1.55 & 0.20--2.00  & 15--30     &0.06--0.22  &1.6--5.0 &0.26--0.85 &10--\phn28    \\
F3   &1.46 & 0.10--1.00  & 15--35     &0.07--0.21  &1.3--4.2 &0.29--0.90 &10--\phn33    \\ 
F4   &1.73 & 0.20--0.70  & 10--20     &0.06--0.19  &1.2--3.8 &0.19--0.63 &\phn5--\phn13 \\
F5   &1.38 & \ldots      & \ldots     &0.07--0.22  &1.6--5.0 &0.42--1.33 &30--100       \\ 
F6   &2.10 & $>1.00$     & 10--15     &0.06--0.18  &1.3--3.8 &0.34--1.10 &25--\phn80    \\
F7   &1.81 & 0.25--1.00  & 12--35     &0.06--0.18  &1.2--3.7 &0.46--1.50 &55--100       \\ 
F8   &1.27 & 0.25--1.00  & 15--20     &0.07--0.22  &1.4--4.3 &0.40--1.30 &18--100       \\
F9   &1.16 & \ldots      & \ldots     &0.07--0.25  &1.6--5.0 &0.45--1.45 &23--\phn35    \\
\hline
\noalign{\smallskip}
\multicolumn{8}{l}{Derived from the H$_2$ \textit{Herschel}-derived column density maps} \\
\hline
\noalign{\smallskip}
F1   &1.34 & 0.20--1.00  & 10--40     &0.07--0.23  &1.5--4.8 &\ldots     &\ldots        \\
F2   &1.19 & 0.20--2.00  & 10--40     &0.08--0.25  &1.8--5.8 &\ldots     &\ldots        \\
F3   &1.18 & 0.10--1.00  & 10--25     &0.07--0.23  &1.5--4.6 &\ldots     &\ldots        \\
F4   &1.66 & 0.20--0.70  & \phn7--17  &0.06--0.19  &1.3--4.0 &\ldots     &\ldots        \\
F5   &0.97 & \ldots      & \ldots     &0.09--0.27  &1.9--4.5 &\ldots     &\ldots        \\
F6   &1.61 & $>1.00$     & \phn7--12  &0.07--0.21  &1.5--4.0 &\ldots     &\ldots        \\
F7   &1.46 & 0.25--1.00  & 10--30     &0.07--0.21  &1.3--5.4 &\ldots     &\ldots        \\
F8   &0.78 & 0.25--1.00  & 10--17     &0.09--0.29  &1.8--3.2 &\ldots     &\ldots        \\
F9   &1.03 & \ldots      & \ldots     &0.08--0.25  &1.7--2.3 &\ldots     &\ldots        \\
\hline
\end{tabular} 
\tablefoot{
\tablefoottext{a}{The density, $n_{\mathrm{c}}$, was estimated considering that the filaments are homogeneous cylinder.}
\tablefoottext{b}{Minimum (left) and maximum (right) values of the masses and distances between clumps. These values were measured from the $^{13}$CO, C$^{18}$O and H$_2$ maps. For this, we set a polygon around the clumps and protostar presented in Fig~\ref{fig:large-skeleton}.}
\tablefoottext{c}{Calculated from the Eq. $\lambda_{\mathrm{cl,O64}}=0.066~\mathrm{pc}~(T/10~\mathrm{K})^{1/2}~(n_\mathrm{c}/10^{5}~\mathrm{cm}^{-3})^{-1/2}$}
\tablefoottext{d}{Calculated from the Eq. $M_{\mathrm{cl,O64}}=0.877$~\mo$~(T/10~\mathrm{K})^{3/2}~(n_\mathrm{c}/10^{5}~\mathrm{cm}^{-3})^{-1/2}$}
\tablefoottext{e}{Calculated from the Eq. $\lambda_{\mathrm{cl,IM97}}=1.24~\mathrm{pc}~(\sigma_{\mathrm{tot}}/1~\mathrm{km~s}^{-1})(n_\mathrm{c}/10^{5}~\mathrm{cm}^{-3})^{-1/2}$}
\tablefoottext{f}{Calculated from the Eq. $M_{\mathrm{cl,IM97}}=575.3$~\mo~$(\sigma_{\mathrm{tot}}/1~\mathrm{km~s}^{-1})^{3}~(n_\mathrm{c}/10^{5}~\mathrm{cm}^{-3})^{-1/2}$. The values at the right of the columns 5 to 8 were calculated using the $n_\mathrm{c}$ listed in column 2. The values at the left of the columns 5 to 8 were calculated using the $n_\mathrm{c}$ as an order of magnitude larger that the values listed in column 2.}
\tablefootmark{*}{The filaments that are not associated with any clump or protostar (see Fig.~\ref{fig:large-skeleton}) were filled with the '--' mark.}
}
\label{tab:filaments_parameters_stability_3} 
\end{table*}

\begin{table*}[t!]
\centering
\caption{Clumps properties of the secondary filaments}
\begin{tabular}{l c c c c c c c} 
\hline
\noalign{\smallskip}
&$n_{\mathrm{c}}$\tablefootmark{a}
&$\lambda_{\mathrm{cl,obs}}$\tablefootmark{b}
&$M_{\mathrm{cl,obs}}$\tablefootmark{b}
&$\lambda_{\mathrm{cl,IM97}}$\tablefootmark{c}
&$M_{\mathrm{cl,IM97}}$\tablefootmark{d}
&$\lambda_{\mathrm{cl,vir}}$\tablefootmark{e}
&$M_{\mathrm{cl,vir}}$\tablefootmark{f}  
\\
Filament
&($10^{4}$~cm$^{-3}$)
&(pc)
&(\mo) 
&(pc)
&(\mo) 
&(pc)
&(\mo) 
\\
\hline
\noalign{\smallskip}
\multicolumn{8}{l}{Derived from $^{13}$CO} \\
\hline
\noalign{\smallskip}
sF1a &1.66 & 0.25--1.00  & \phn7--15  &0.05--0.19   &1.2--3.5 &0.29--0.90 &10--23     \\ 
sF1b &1.63 & 0.25--2.00  & \phn5--10  &0.05--0.19   &1.2--3.8 &0.32--1.00 &15--13     \\ 
sF1c &2.22 & 0.25--0.50  & \phn5--10  &0.05--0.17   &1.1--3.6 &0.31--1.00 &16--20     \\ 
sF2a &2.50 & $>1$        & 10--15     &0.05--0.17   &1.3--4.0 &0.30--1.00 &22--25     \\
sF2b &1.75 & $>0.5$      & \phn5--10  &0.07--0.20   &1.4--4.5 &0.35--1.10 &24--70     \\
sF2c &2.54 & \ldots      & \ldots     &0.05--0.16   &1.1--3.5 &0.35--1.00 &20--59     \\
sF2d &2.64 & \ldots      & \ldots     &0.06--0.17   &1.3--4.0 &0.30--0.80 &13--48     \\
sF2e &0.94 & \ldots      & \ldots     &0.10--0.27   &1.9--6.2 &0.25--1.50 &35--58     \\
sF3a &2.70 & \ldots      & \ldots     &0.05--0.15   &1.0--3.2 &0.50--0.90 &15--48     \\ 
sF3b &2.13 & \ldots      & \ldots     &0.06--0.17   &1.0--3.4 &0.27--0.90 &15--23     \\ 
sF3c &1.19 & 0.25--2.00  & \phn4--\phn8 &0.08--0.23 &1.5--4.7 &0.28--1.15 &17--27     \\ 
sF4a &1.56 & \ldots      & \ldots     &0.06--0.20   &1.5--4.0 &0.36--1.20 &25--28     \\
sF4b &1.04 & \ldots      & \ldots     &0.10--0.25   &1.3--5.4 &0.38--1.30 &20--22     \\
sF5a &1.99 & $>0.5$      & 10--15     &0.06--0.19   &1.7--3.9 &0.40--0.70 &\phn5--18  \\
sF5b &2.39 & \ldots      & \ldots     &0.06--0.17   &1.2--3.6 &0.22--0.90 &20--30     \\
sF7a &1.43 & 0.25--1.00  & 10--15     &0.07--0.21   &1.3--4.2 &0.30--1.30 &25--53     \\ 
\hline
\noalign{\smallskip}
\multicolumn{8}{l}{Derived from C$^{18}$O} \\
\hline
\noalign{\smallskip}
sF1a &1.43 & 0.25--1.00  & 10--20     &0.05--0.20  &1.0--3.9 &0.25--0.82 &10--38     \\ 
sF1b &2.72 & 0.25--2.00  & \phn8--15  &0.05--0.15  &1.0--3.0 &0.17--0.55 &\phn5--38  \\ 
sF1c &2.83 & 0.25--0.50  & \phn8--15  &0.05--0.15  &1.0--3.2 &0.18--0.82 &\phn5--38  \\ 
sF2a &3.03 & $>1.0$      & 15--20     &0.05--0.15  &1.1--3.6 &0.20--0.58 &10--38     \\
sF2b &2.26 & $>0.5$      & 10--15     &0.06--0.17  &1.5--4.0 &0.32--0.65 &22--38     \\
sF2c &3.27 & \ldots      & \ldots     &0.05--0.15  &1.0--3.0 &0.26--1.00 &20--38     \\
sF2d &2.56 & \ldots      & \ldots     &0.06--0.17  &1.3--4.1 &0.27--0.85 &15--38     \\
sF2e &1.29 & \ldots      & \ldots     &0.07--0.23  &2.2--5.3 &0.36--1.10 &18--38     \\
sF3a &4.50 & \ldots      & \ldots     &0.04--0.12  &1.0--2.5 &0.22--0.70 &15--38     \\ 
sF3b &3.68 & \ldots      & \ldots     &0.04--0.13  &1.2--2.5 &0.18--0.60 &\phn7--38  \\ 
sF3c &2.22 & 0.25--2.00  & \phn8--13  &0.05--0.17  &1.5--3.4 &0.23--0.74 &\phn8--38  \\ 
sF4a &2.44 & \ldots      & \ldots     &0.05--0.16  &1.2--3.3 &0.22--0.73 &\phn7--38  \\
sF4b &1.63 & \ldots      & \ldots     &0.07--0.20  &2.2--4.3 &0.24--0.77 &\phn7--38  \\
sF5a &2.70 & $>0.5$      & 13--17     &0.05--0.16  &1.3--3.4 &0.19--0.60 &\phn5--38  \\
sF5b &2.09 & \ldots      & \ldots     &0.05--0.18  &1.4--4.0 &0.20--0.60 &\phn9--38  \\
sF7a &1.47 & 0.25--1.00  & 10--15     &0.06--0.20  &1.4--4.1 &0.25--0.80 &17--38     \\ 
\hline
\noalign{\smallskip}
\multicolumn{8}{l}{Derived from the H$_2$ \textit{Herschel}-derived column density maps} \\
\hline
\noalign{\smallskip}
sF1a &2.17 & 0.25--1.00  & 12--22     &0.06--0.17   &1.6--3.2 &\ldots     &\ldots     \\
sF1b &2.20 & 0.25--2.00  & \phn7--10  &0.06--0.17   &1.6--3.3 &\ldots     &\ldots     \\
sF1c &2.87 & 0.25--0.50  & \phn5--10  &0.06--0.15   &1.6--3.2 &\ldots     &\ldots     \\
sF2a &3.09 & $>1.0$      & 16--23     &0.06--0.16   &1.6--3.6 &\ldots     &\ldots     \\
sF2b &1.59 & $>0.5$      & \phn5--\phn8 &0.06--0.21 &1.6--4.8 &\ldots     &\ldots     \\
sF2c &2.72 & \ldots      & \ldots     &0.06--0.16   &1.6--3.4 &\ldots     &\ldots     \\
sF2d &2.40 & \ldots      & \ldots     &0.06--0.18   &1.6--4.3 &\ldots     &\ldots     \\
sF2e &0.72 & \ldots      & \ldots     &0.06--0.31   &1.6--7.0 &\ldots     &\ldots     \\
sF3a &2.40 & \ldots      & \ldots     &0.06--0.17   &1.6--3.3 &\ldots     &\ldots     \\
sF3b &1.77 & \ldots      & \ldots     &0.06--0.19   &1.6--3.7 &\ldots     &\ldots     \\
sF3c &1.15 & 0.25--2.00  & \phn4--\phn6 &0.06--0.23 &1.6--4.8 &\ldots     &\ldots     \\
sF4a &1.85 & \ldots      & \ldots     &0.06--0.19   &1.6--3.8 &\ldots     &\ldots     \\
sF4b &0.62 & \ldots      & \ldots     &0.06--0.33   &1.6--6.8 &\ldots     &\ldots     \\
sF5a &1.74 & $>0.5$      & \phn8--11  &0.06--0.19   &1.6--4.2 &\ldots     &\ldots     \\
sF5b &1.72 & \ldots      & \ldots     &0.06--0.20   &1.6--4.2 &\ldots     &\ldots     \\
sF7a &1.26 & 0.25--1.00  & 12--15     &0.06--0.22   &1.6--4.4 &\ldots     &\ldots     \\
\hline
\end{tabular}
\label{tab:filaments_parameters_stability_4} 
\end{table*}

\begin{table*}[t!]
\centering
\caption{Kinematical parameters of the main and secondary filaments}
\begin{tabular}{l c c c c c c c c c c} 
\hline
\noalign{\smallskip}
&$M$
&$\sigma_{\mathrm{NT}}$\tablefootmark{a}
&$\sigma_{\mathrm{tot}}$\tablefootmark{b}
&$\mathcal{M}$\tablefootmark{c}
&$\nabla V_{\|\mathrm{obs}}$ 
&$\dot{M}_\mathrm{accr}$\tablefootmark{d}
&$\alpha$\tablefootmark{e}
&$\dot{M}_\mathrm{accr}^{\mathrm{corr,}}$\tablefootmark{f}
\\
Filament
&(\mo)
&(\kms) 
&(\kms) 
&
&(\kms~pc$^{-1}$)
&($10^{-4}$~\mo~yr$^{-1}$)
&(degrees)
&($10^{-4}$~\mo~yr$^{-1}$)
\\
\hline
\noalign{\smallskip}
\multicolumn{8}{l}{Derived from $^{13}$CO} \\
\hline
\noalign{\smallskip}
F1       &241.22    &0.33  &0.34  &1.38  &$+$0.15  &0.36  &$+$26    &0.73  \\  
F2       &172.04    &0.32  &0.33  &1.28  &$+$0.17  &0.30  &$+$30    &0.51  \\  
F3       &279.86    &0.38  &0.39  &1.67  &$+$0.63  &1.80  &$+$64    &0.85  \\ 
F4       &\phn64.67 &0.28  &0.29  &1.23  &$+$0.65  &0.43  &$+$65    &0.20  \\ 
F5       &\phn95.50 &0.23  &0.24  &0.95  &$+$0.00  &0.00  &\phn$+$0 &0.00  \\  
F6       &142.77    &0.31  &0.32  &1.30  &$-$0.50  &0.72  &$-$60    &0.42  \\
F7       &318.37    &0.35  &0.36  &1.54  &$-$0.33  &1.10  &$-$48    &0.96  \\
F8       &101.74    &0.25  &0.26  &1.00  &$-$0.38  &0.39  &$-$28    &0.31  \\ 
F9       &198.54    &0.34  &0.35  &1.43  &$-$0.17  &0.34  &$-$30    &0.60  \\
sF1a    &\phn23.24  &0.28  &0.29  &1.23  &$-$0.25  &0.06  &$-$41    &0.07  \\
sF1b    &\phn78.32  &0.32  &0.33  &1.41  &$-$0.26  &0.20  &$-$42    &0.21  \\
sF1c    &\phn58.03  &0.37  &0.38  &1.55  &$-$0.50  &0.16  &$-$60    &0.17  \\
sF2a    &\phn46.93  &0.38  &0.39  &1.53  &$+$0.43  &0.20  &$+$56    &0.13  \\
sF2b    &\phn52.21  &0.37  &0.38  &1.55  &$-$0.13  &0.10  &$-$25    &0.15  \\
sF2c    &\phn13.93  &0.37  &0.38  &1.56  &$+$0.19  &0.02  &$+$34    &0.03  \\
sF2d    &\phn31.89  &0.32  &0.33  &1.28  &$+$0.18  &0.06  &$+$33    &0.10  \\
sF2e    &\phn39.68  &0.38  &0.39  &1.60  &$-$0.29  &0.12  &$-$46    &0.11  \\ 
sF3a    &\phnn8.61  &0.35  &0.36  &1.54  &$-$0.77  &0.07  &$-$70    &0.03  \\
sF3b    &\phn29.17  &0.32  &0.33  &1.41  &$+$0.10  &0.03  &$+$19    &0.09  \\
sF3c    &\phn64.47  &0.31  &0.32  &1.36  &$+$0.64  &0.41  &$+$65    &0.20  \\
sF4a    &\phn37.20  &0.38  &0.38  &1.63  &$-$0.18  &0.07  &$-$33    &0.10  \\
sF4b    &\phn24.91  &0.32  &0.33  &1.35  &$+$0.42  &0.10  &$+$55    &0.07  \\ 
sF5a    &\phn24.91  &0.24  &0.25  &1.00  &$+$0.12  &0.03  &$+$23    &0.07  \\  
sF5b    &\phn64.81  &0.37  &0.38  &1.55  &$-$0.10  &0.07  &$+$20    &0.18  \\
sF7a    &135.73     &0.37  &0.38  &1.63  &$+$0.00  &0.00  &\phn$+$0 &0.00  \\
\hline
\noalign{\smallskip}
\multicolumn{8}{l}{Derived from C$^{18}$O} \\
\hline
\noalign{\smallskip}
F1       &241.22    &0.33  &0.34  &1.38  &$+$0.15  &0.36  &$+$26    &0.73  \\  
F1       &326.60    &0.28  &0.29  &1.18  &$+$0.10  &0.32  &$+$16    &1.12  \\ 
F2       &234.31    &0.26  &0.27  &1.04  &$+$0.10  &0.24  &$+$16    &0.80  \\ 
F3       &428.28    &0.27  &0.28  &1.18  &$+$0.80  &3.42  &$+$84    &0.35  \\
F4       &111.62    &0.20  &0.21  &0.87  &$+$0.38  &0.43  &$+$46    &0.40  \\ 
F5       &107.78    &0.39  &0.40  &1.64  &$+$0.00  &0.00  &\phn$+$0 &0.00  \\
F6       &137.23    &0.39  &0.40  &1.64  &$-$0.33  &0.45  &$-$42    &0.50  \\ 
F7       &424.57    &0.50  &0.51  &2.20  &$-$0.31  &1.40  &$-$41    &1.54  \\ 
F8       &161.96    &0.32  &0.33  &1.41  &$-$0.30  &0.50  &$-$40    &1.75  \\ 
F9       &239.60    &0.39  &0.40  &1.64  &$-$0.10  &0.80  &$-$15    &1.60  \\  
sF1a     &\phn40.47 &0.24  &0.25  &1.00  &$-$0.72  &0.30  &$-$60    &0.17  \\
sF1b     &129.32    &0.22  &0.23  &0.95  &$-$0.47  &0.61  &$-$49    &0.53  \\
sF1c     &\phn73.47 &0.24  &0.25  &1.00  &$-$0.52  &0.38  &$-$52    &0.30  \\
sF2a     &\phn57.22 &0.28  &0.29  &1.12  &$+$0.30  &0.17  &$+$55    &0.12  \\
sF2b     &\phn66.82 &0.38  &0.39  &1.59  &$-$0.10  &0.07  &$-$14    &0.27  \\
sF2c     &\phn17.88 &0.38  &0.39  &1.59  &$+$0.15  &0.03  &$+$20    &0.07  \\
sF2d     &\phn31.35 &0.34  &0.35  &1.37  &$+$0.73  &0.23  &$+$60    &0.13  \\
sF2e     &\phn54.56 &0.32  &0.33  &1.41  &$-$0.47  &0.26  &$-$48    &0.23  \\ 
sF3a     &\phn15.01 &0.37  &0.38  &1.55  &$-$0.46  &0.07  &$+$48    &0.06  \\
sF3b     &\phn51.24 &0.28  &0.29  &1.23  &$+$0.14  &0.07  &$+$19    &0.21  \\
sF3c     &122.40    &0.27  &0.28  &1.19  &$+$0.76  &0.93  &$-$62    &0.49  \\
sF4a     &\phn57.20 &0.28  &0.29  &1.23  &$-$0.20  &0.11  &$-$26    &0.24  \\
sF4b     &\phn38.55 &0.24  &0.25  &1.00  &$+$0.62  &0.23  &$+$56    &0.16  \\ 
sF5a     &\phn33.91 &0.24  &0.25  &1.00  &$+$0.53  &0.18  &$-$50    &0.15  \\  
sF5b     &\phn56.54 &0.28  &0.29  &1.17  &$-$0.44  &0.25  &$+$47    &0.23  \\
sF7b      &139.59   &0.32  &0.33  &1.41  &$+$0.00  &0.00  &\phn$+$0 &0.00  \\
\hline
\end{tabular} 
\tablefoot{
\tablefoottext{a}{Calculated with the Eq. $\sigma_{\mathrm{NT}} = [(\Delta V/\sqrt{8ln2})^{2} - (k_\mathrm{B}T_\mathrm{k}/\mu_{\mathrm{X}}m_{\mathrm{H}})^{2}]^{1/2}$}%
\tablefoottext{b}{velocity dispersion calculated from $\sigma_{\mathrm{tot}} = \Delta v/\sqrt{8ln2}$}
\tablefoottext{c}{Mach number calculated from $\sigma_{\mathrm{NT}} /c_{\mathrm{s}}(T_{\mathrm{K}})$.}
\tablefoottext{d}{Calculated with the Eq. $\dot{M}_{\mathrm{accr}}=M\,\nabla V_{\|\mathrm{obs}}$.}
\tablefoottext{e}{Calculated with the Eq. $\alpha=\tan^{-1}\left(\frac{\nabla V_{\|\mathrm{obs}}}{\nabla V_{\|\mathrm{real}}}\right)$.}
\tablefoottext{f}{Calculated with $\dot{M}_\mathrm{accr}^\mathrm{corr}=\frac{M\,\nabla V_{\|\mathrm{obs}}}{\tan\alpha}$.}
}
\label{tab:filaments_parameters_kin_1}
\end{table*}

\begin{figure*}[t!]
\centering 
\subfloat{
\begin{tabular}{c c}
\includegraphics[width=0.9\columnwidth]{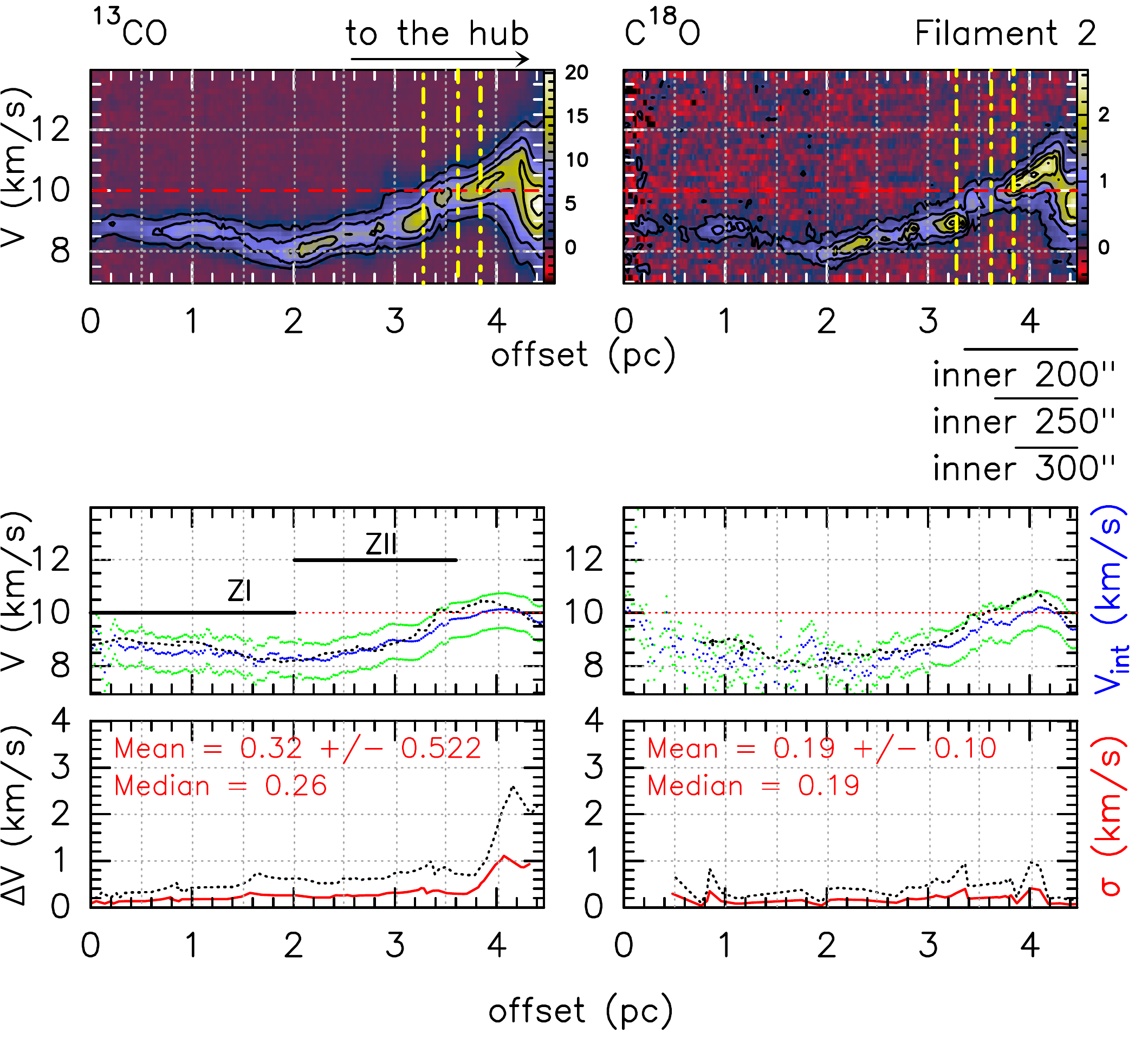} &
\includegraphics[width=0.9\columnwidth]{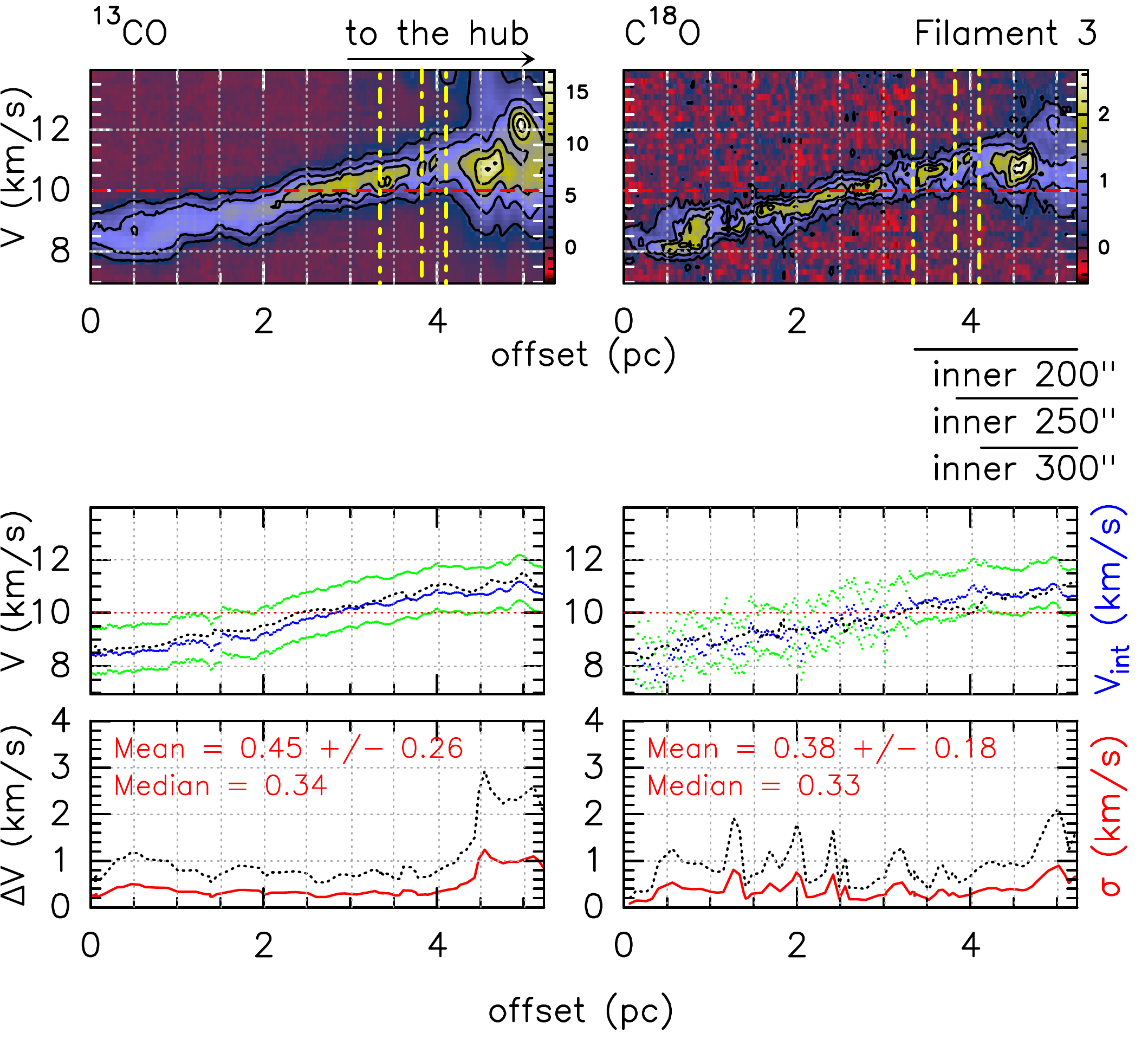} \\
\includegraphics[width=0.9\columnwidth]{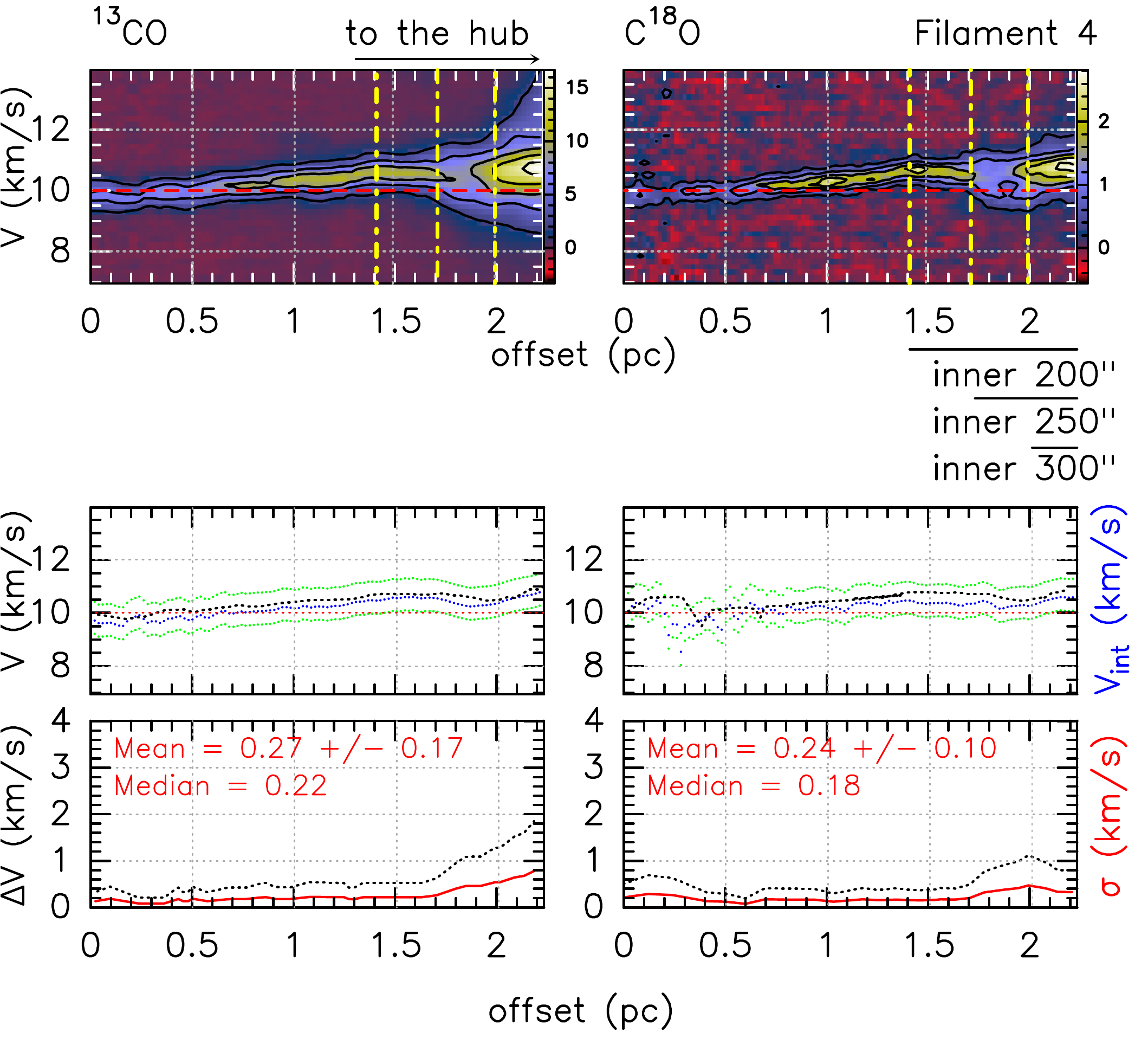} &
\includegraphics[width=0.9\columnwidth]{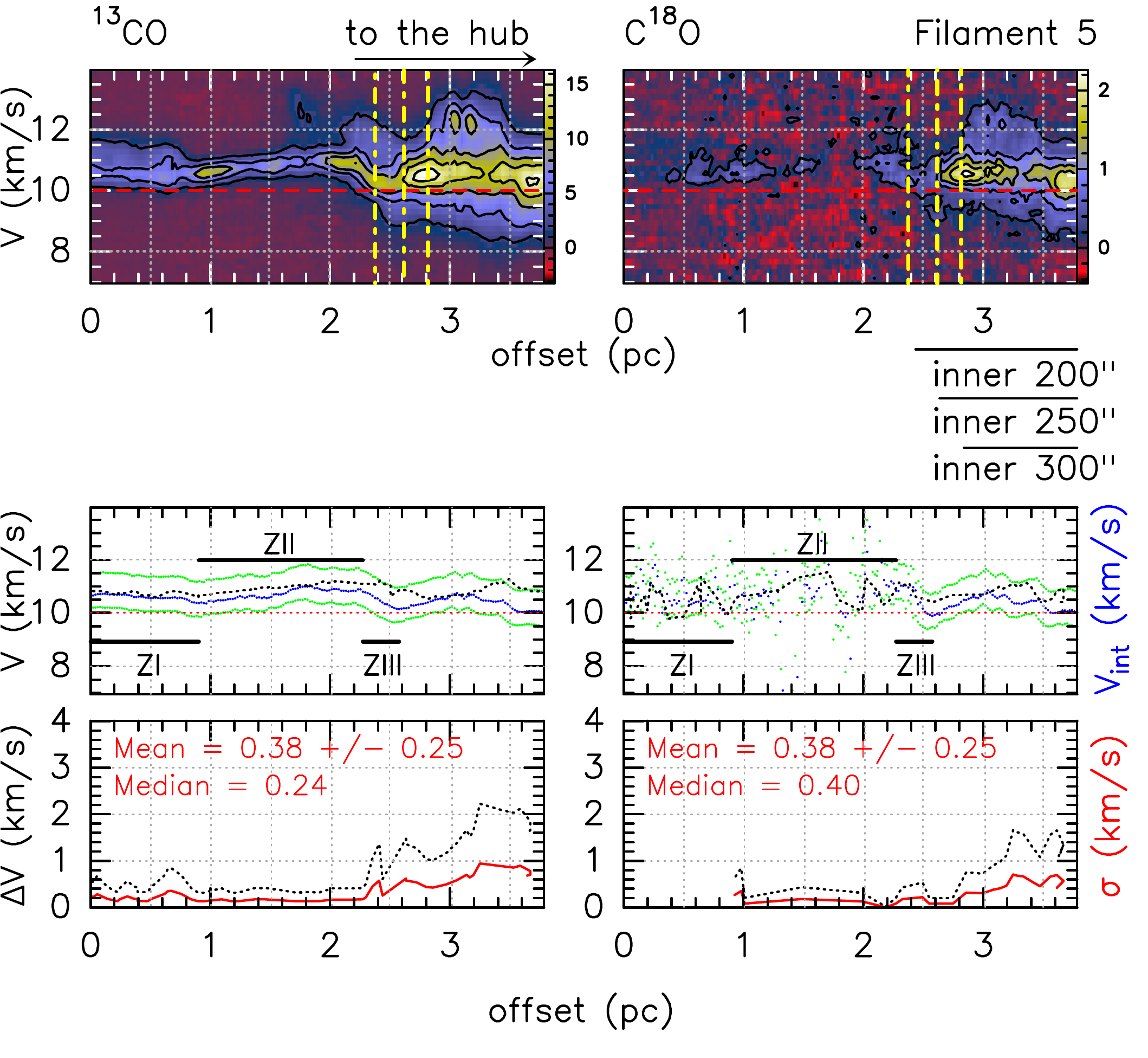} \\ 
\includegraphics[width=0.9\columnwidth]{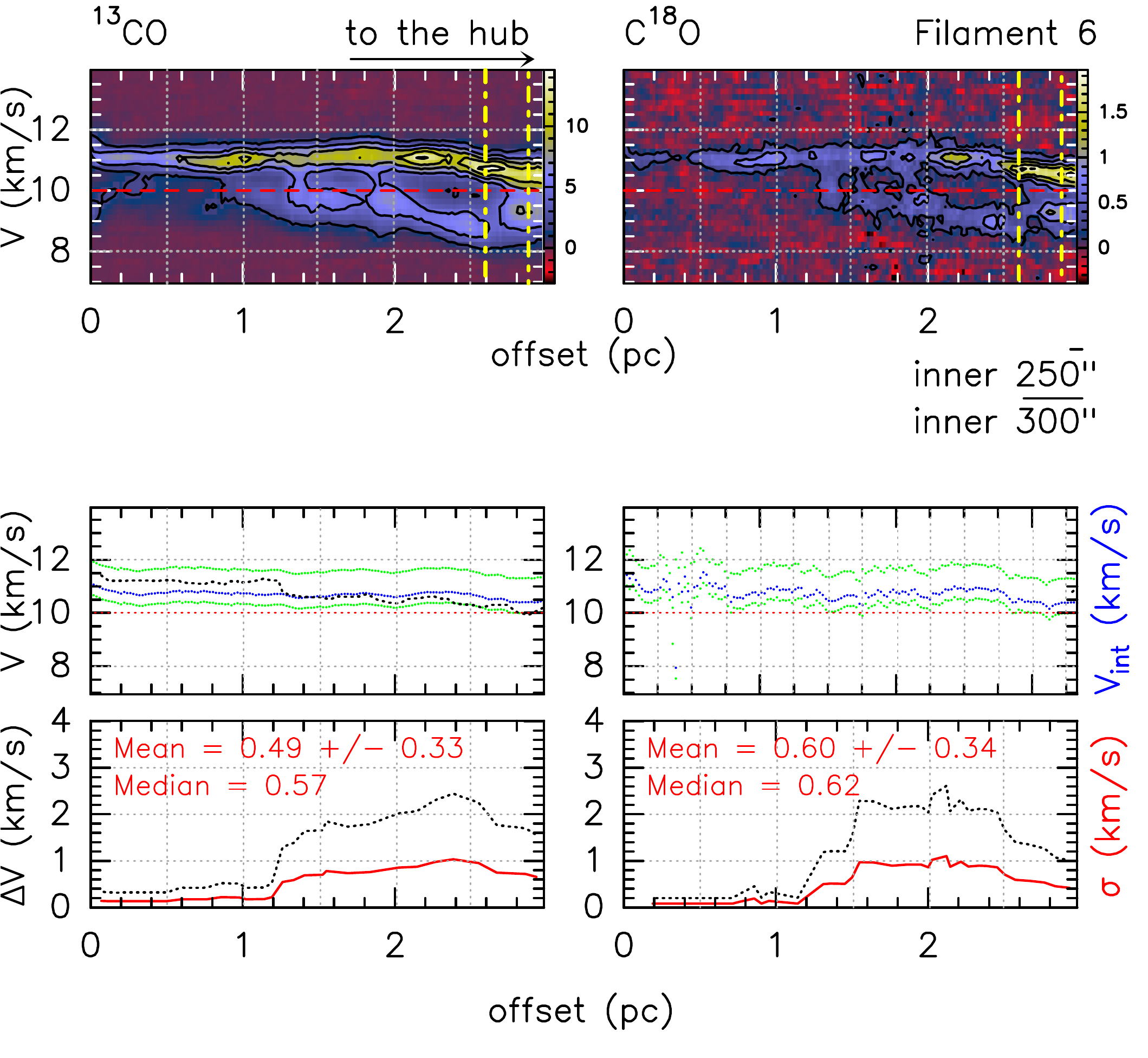} &
\includegraphics[width=0.9\columnwidth]{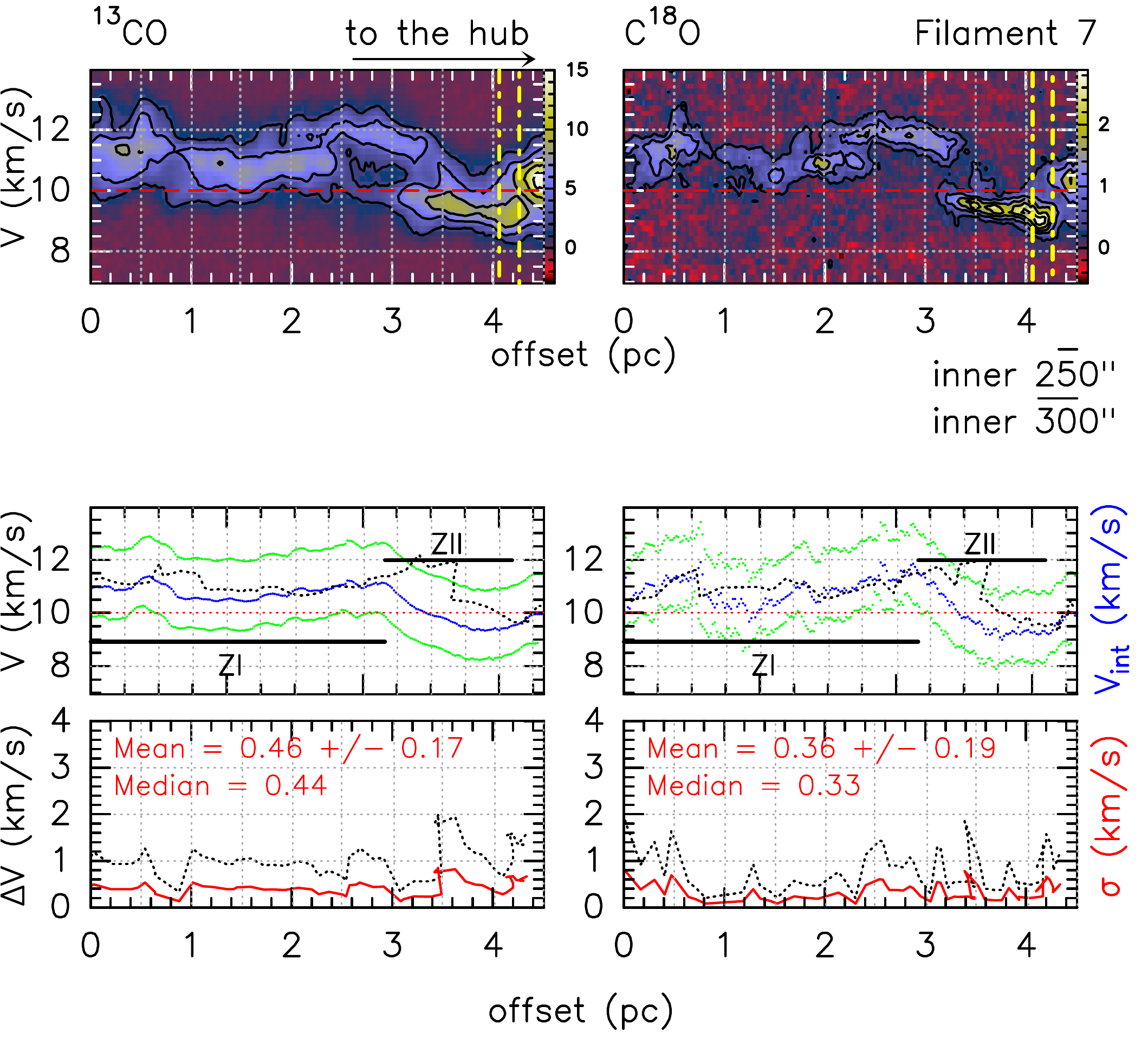} \\
\end{tabular}
}
\caption{Position-velocity diagrams along the `skeletons' of the main and secondary filaments. For each filament there is a set of six plots showing the results for $^{13}$CO (left) and C$^{18}$O (right). The description of the panels and symbols can be found in Fig.~\ref{fig:pv_1}.}
\label{A:pv1}
\end{figure*} 
\begin{figure*}[t!]
\ContinuedFloat
\centering 
\subfloat{
\begin{tabular}{c c}
\includegraphics[width=0.9\columnwidth]{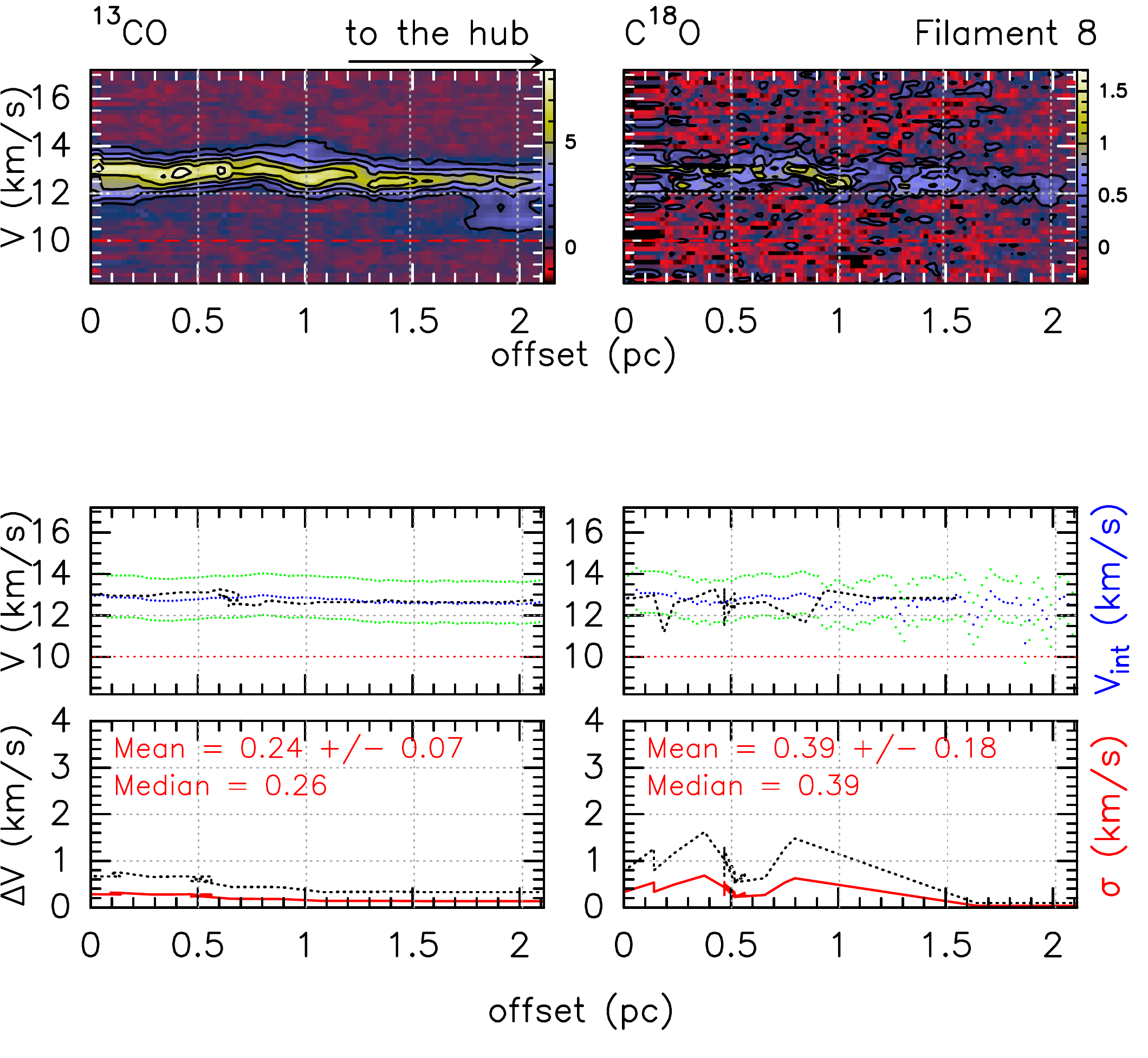} &
\includegraphics[width=0.9\columnwidth]{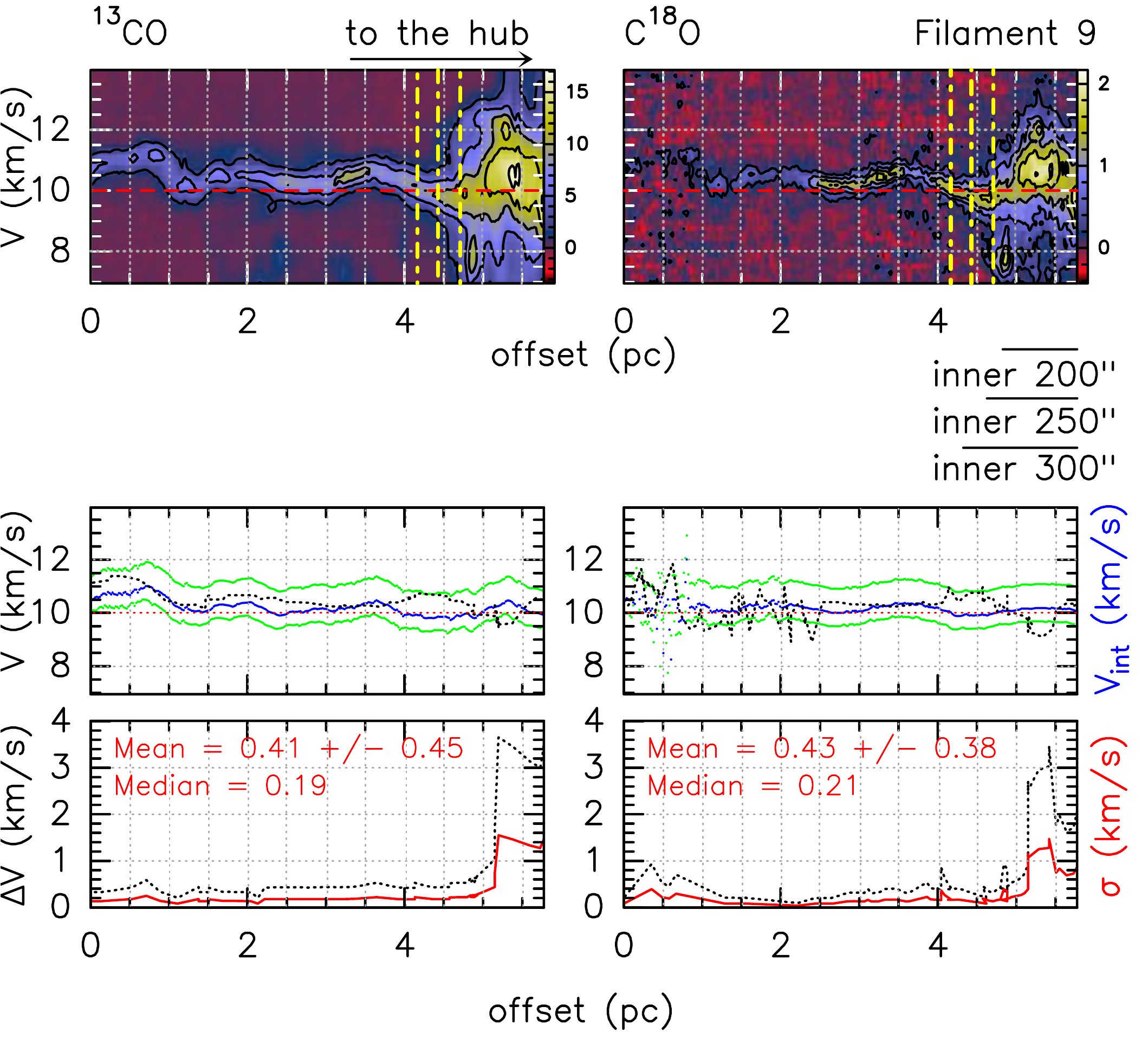} \\
\includegraphics[width=0.9\columnwidth]{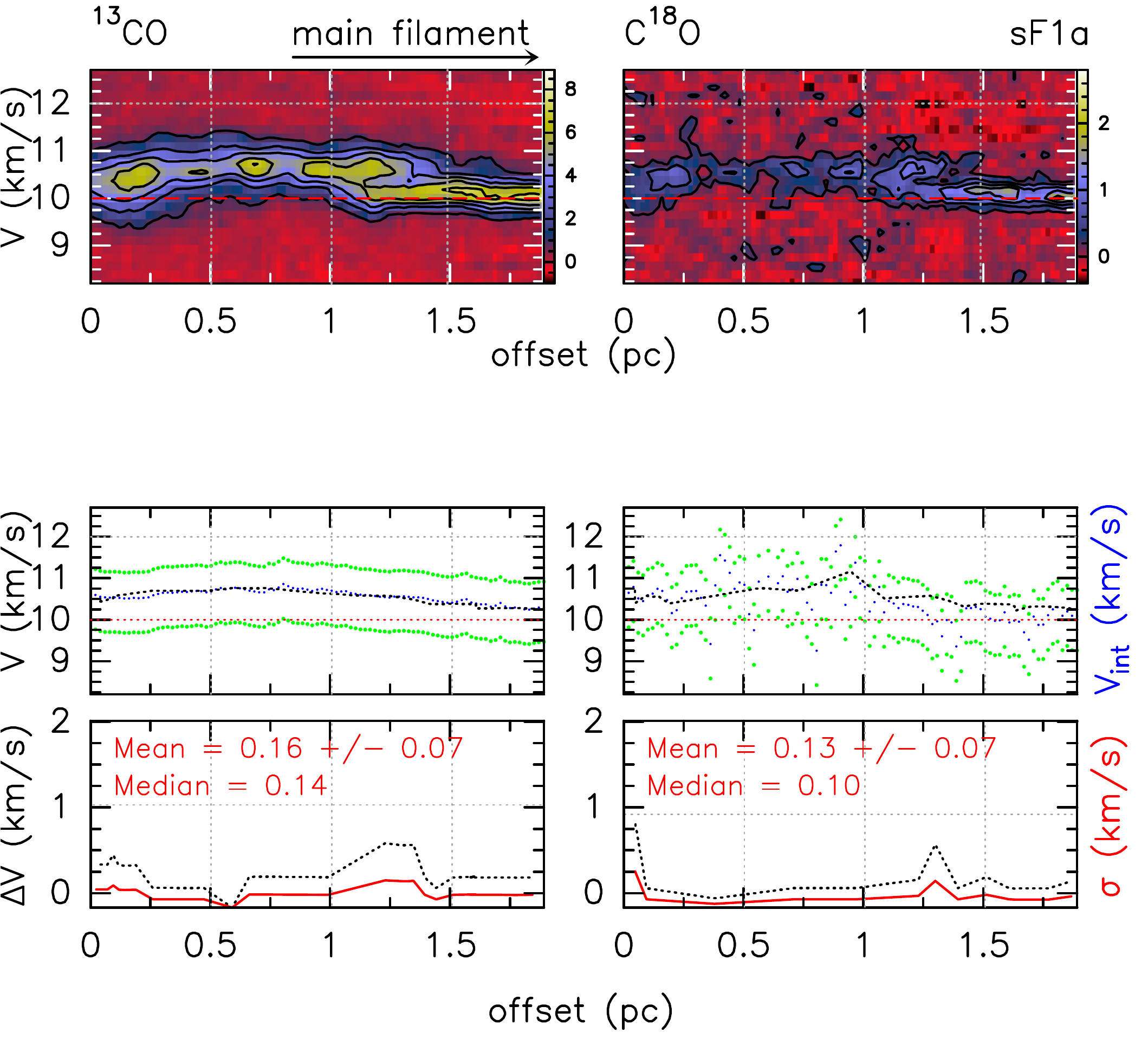} & 
\includegraphics[width=0.9\columnwidth]{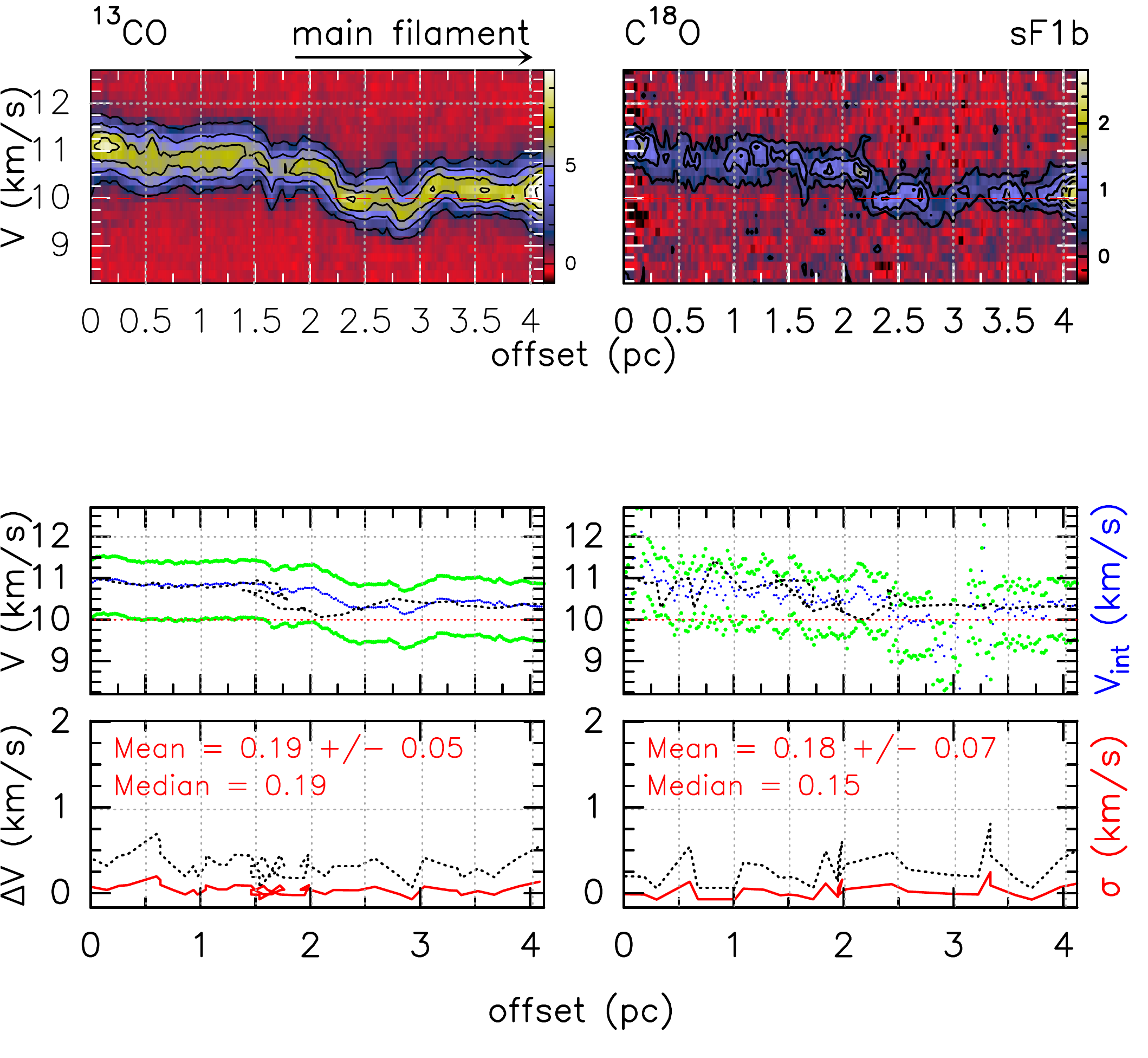} \\
\includegraphics[width=0.9\columnwidth]{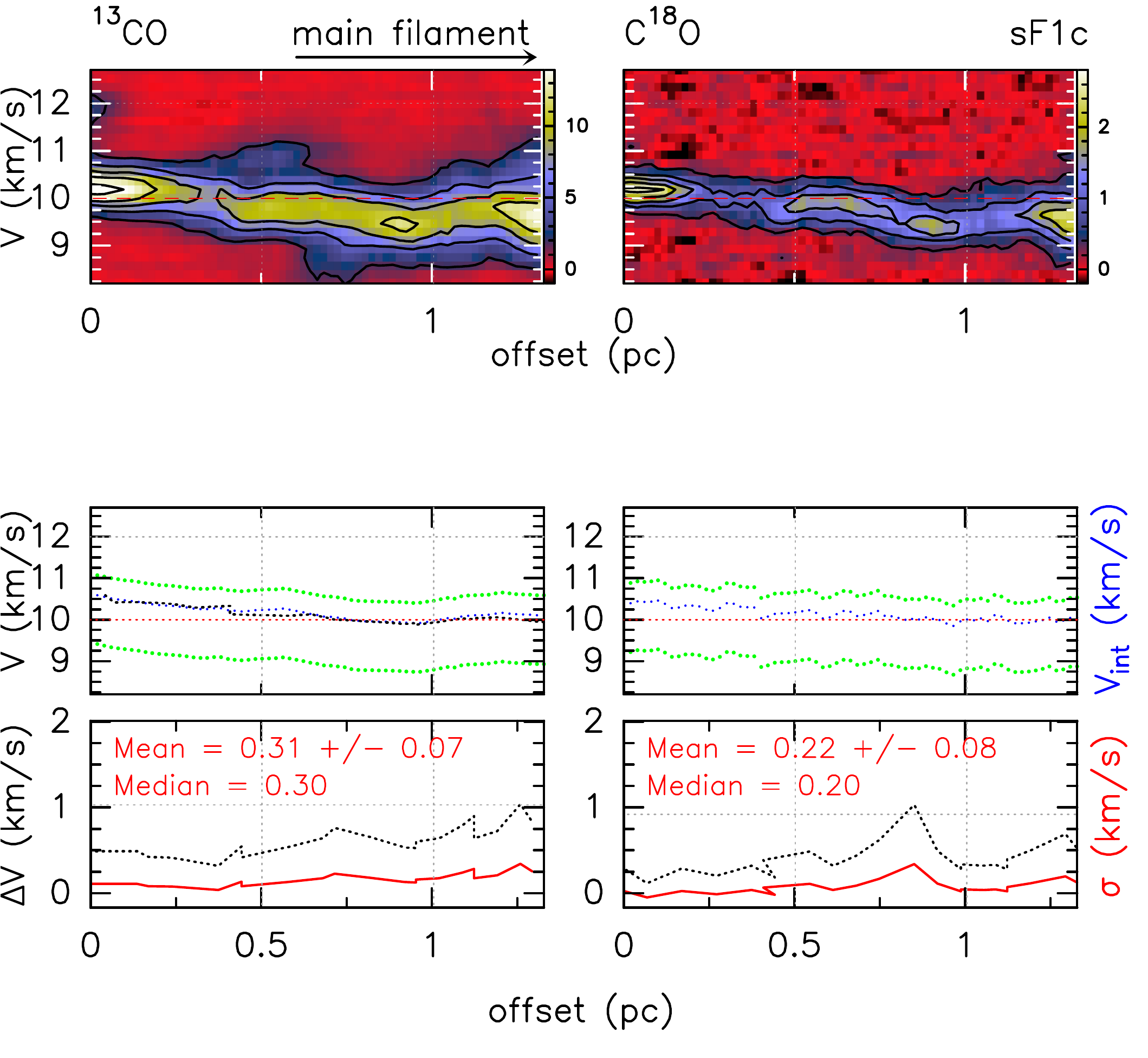} &
\includegraphics[width=0.9\columnwidth]{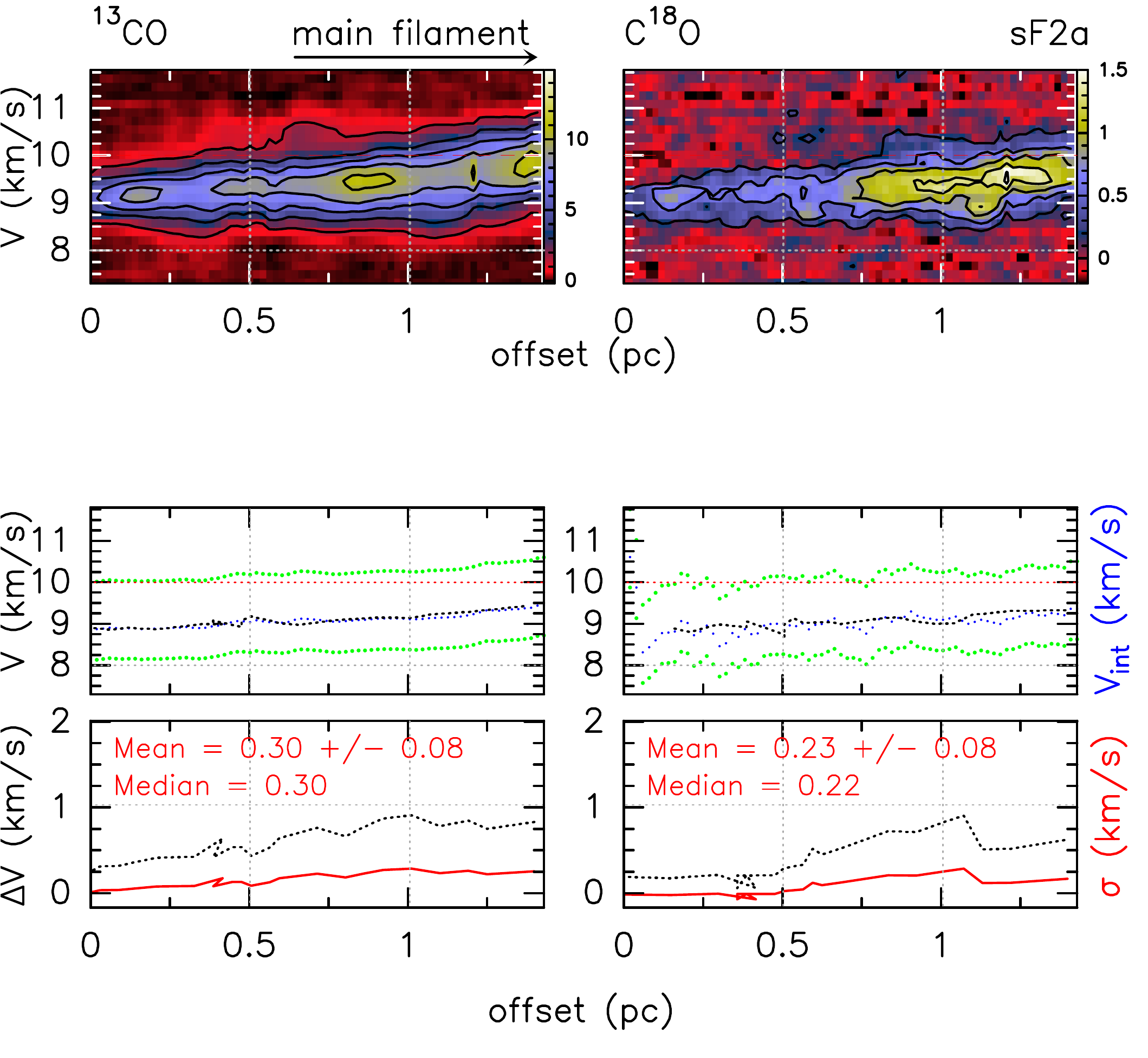} \\
\end{tabular}
}
\caption[]{continued.}
\label{A:pv1}
\end{figure*} 
\begin{figure*}[t!]
\ContinuedFloat
\centering 
\subfloat{
\begin{tabular}{c c}
\includegraphics[width=0.9\columnwidth]{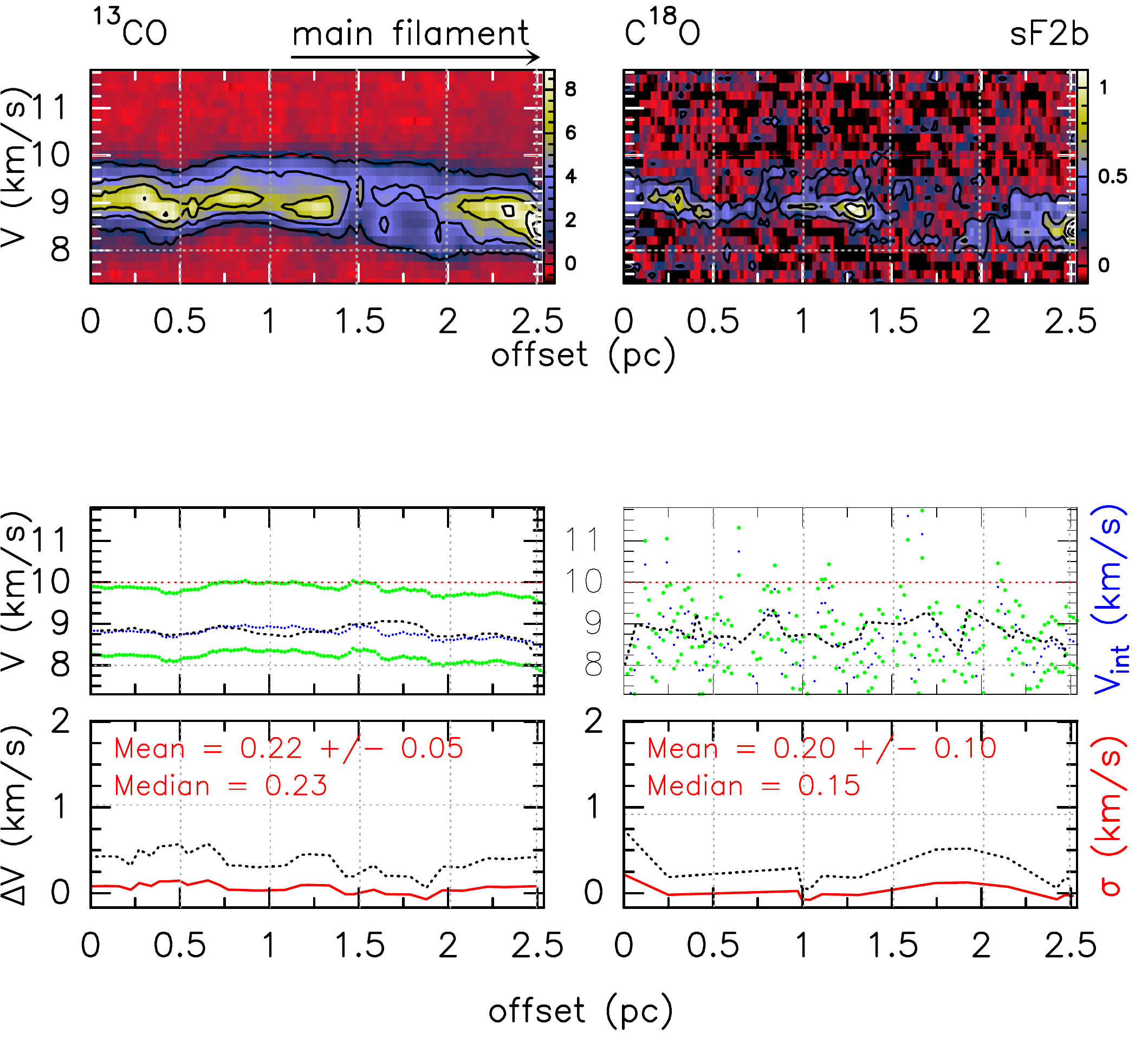} &
\includegraphics[width=0.9\columnwidth]{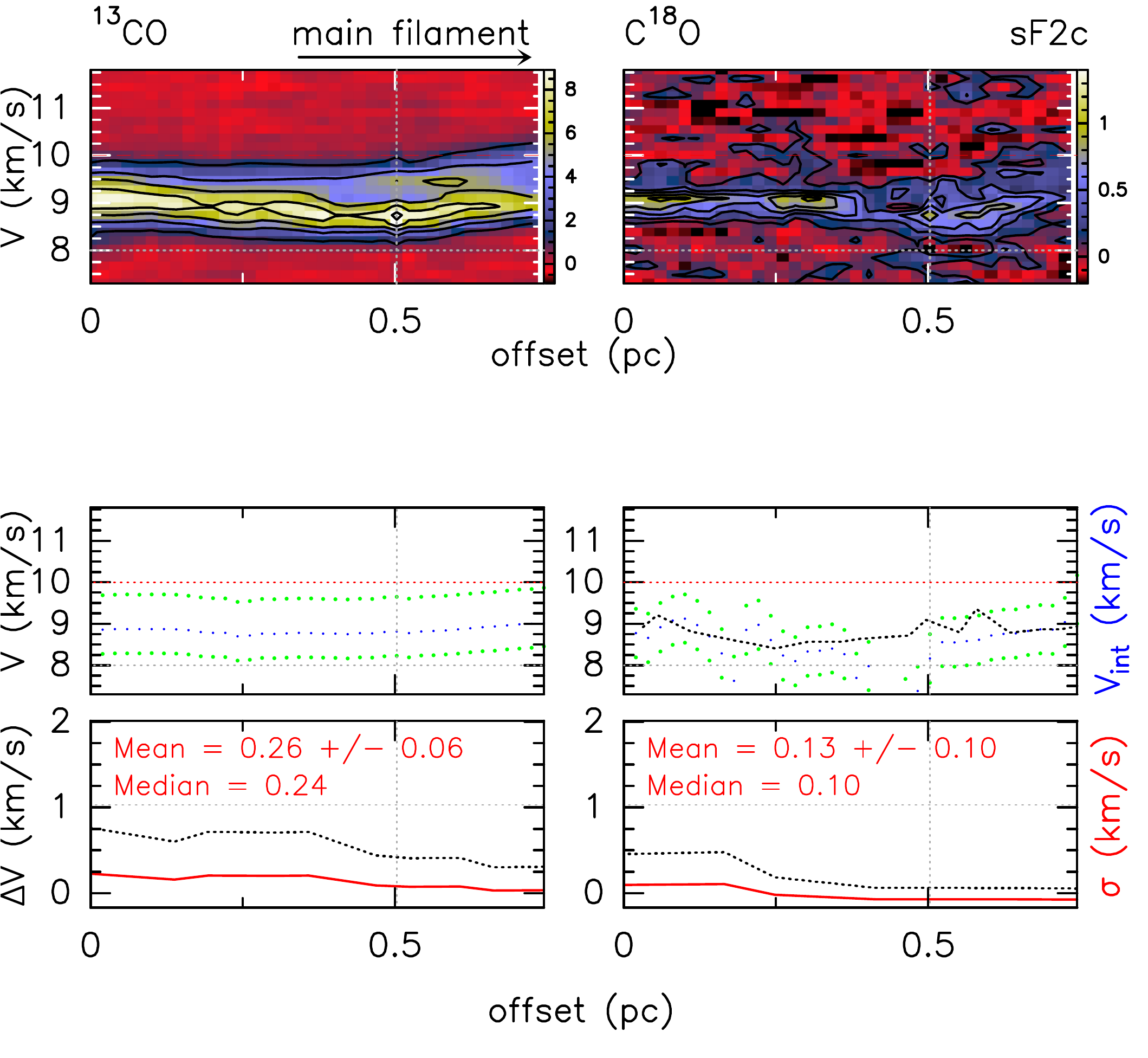} \\
\includegraphics[width=0.9\columnwidth]{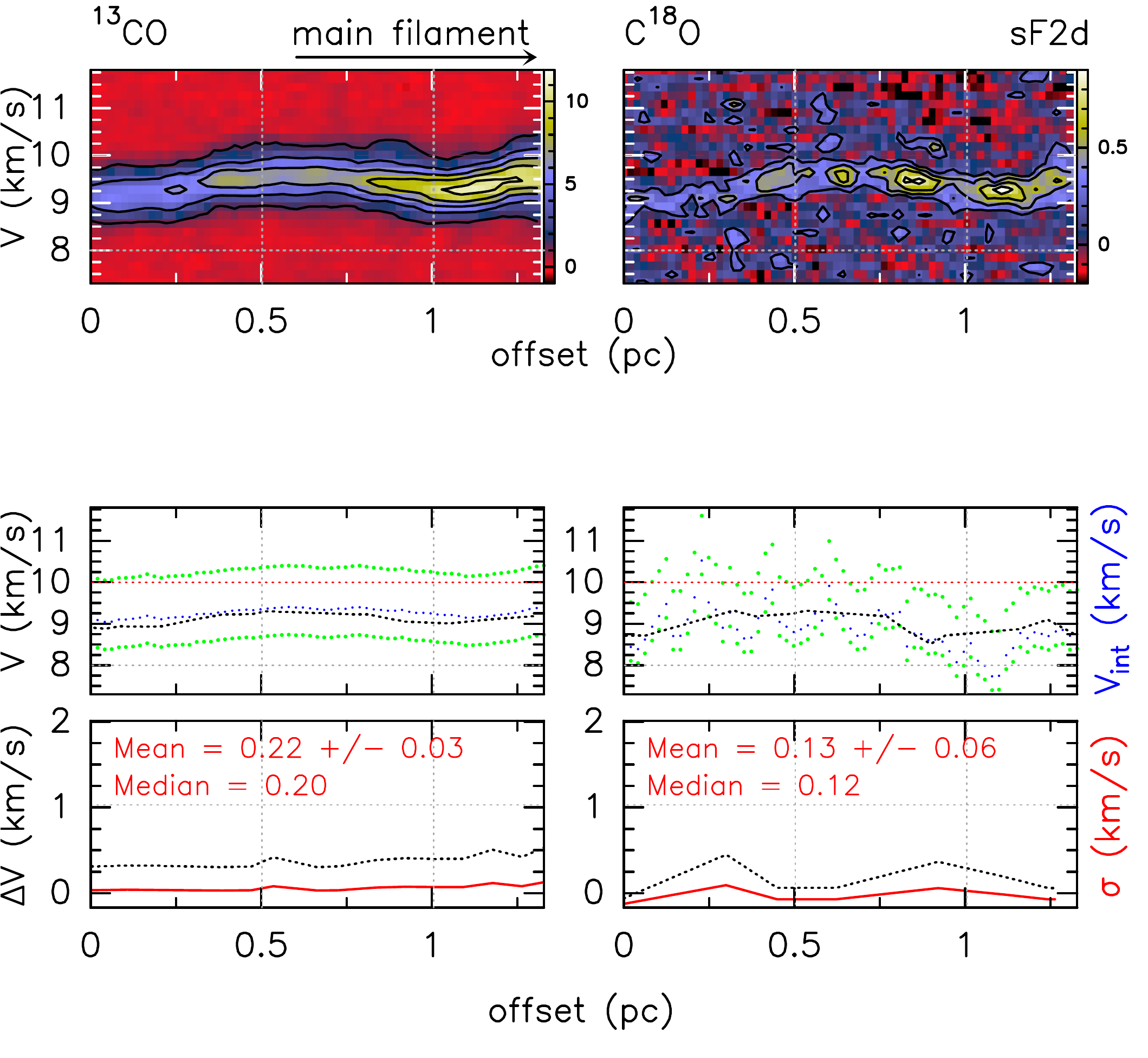} &
\includegraphics[width=0.9\columnwidth]{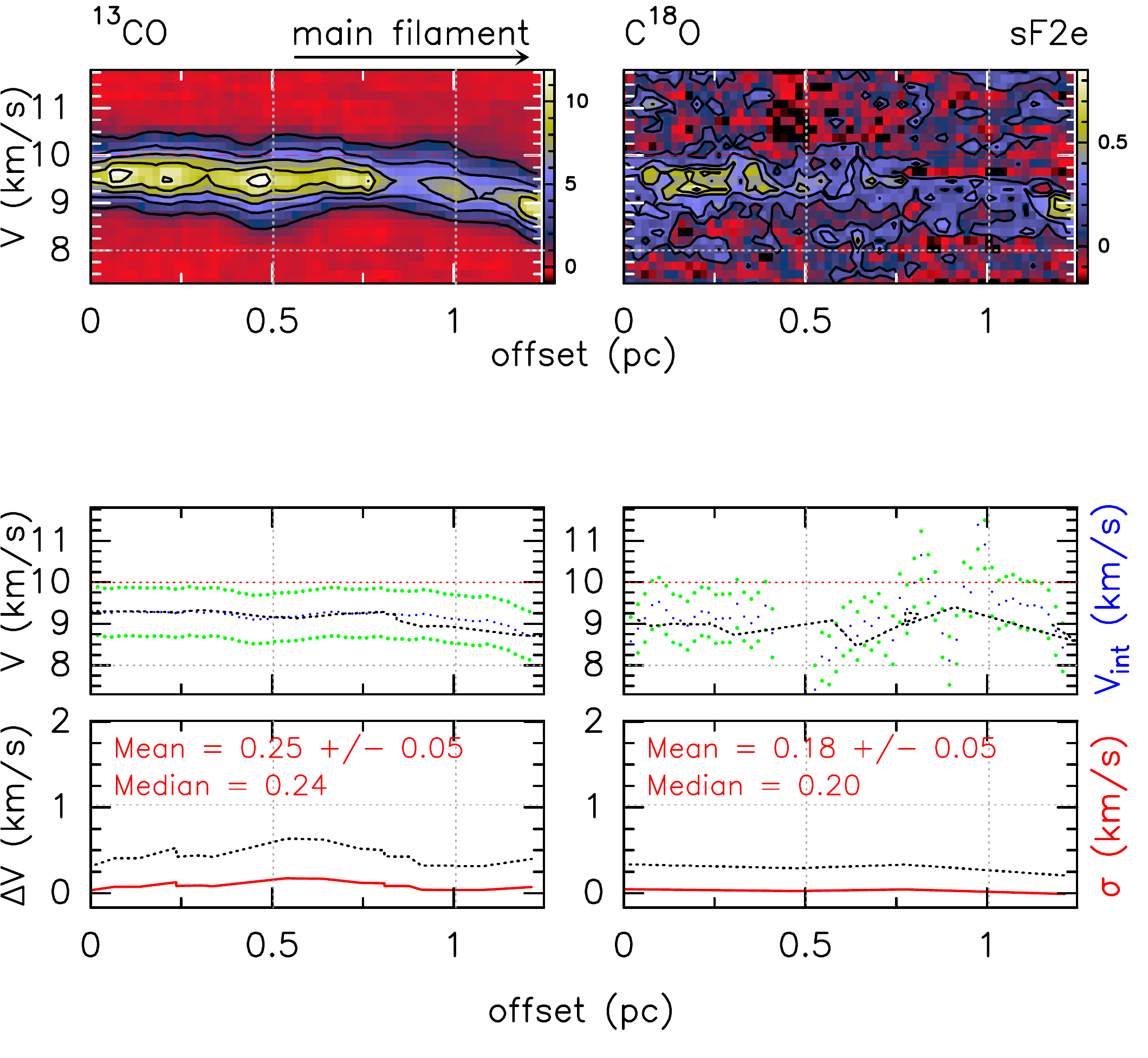} \\
\includegraphics[width=0.9\columnwidth]{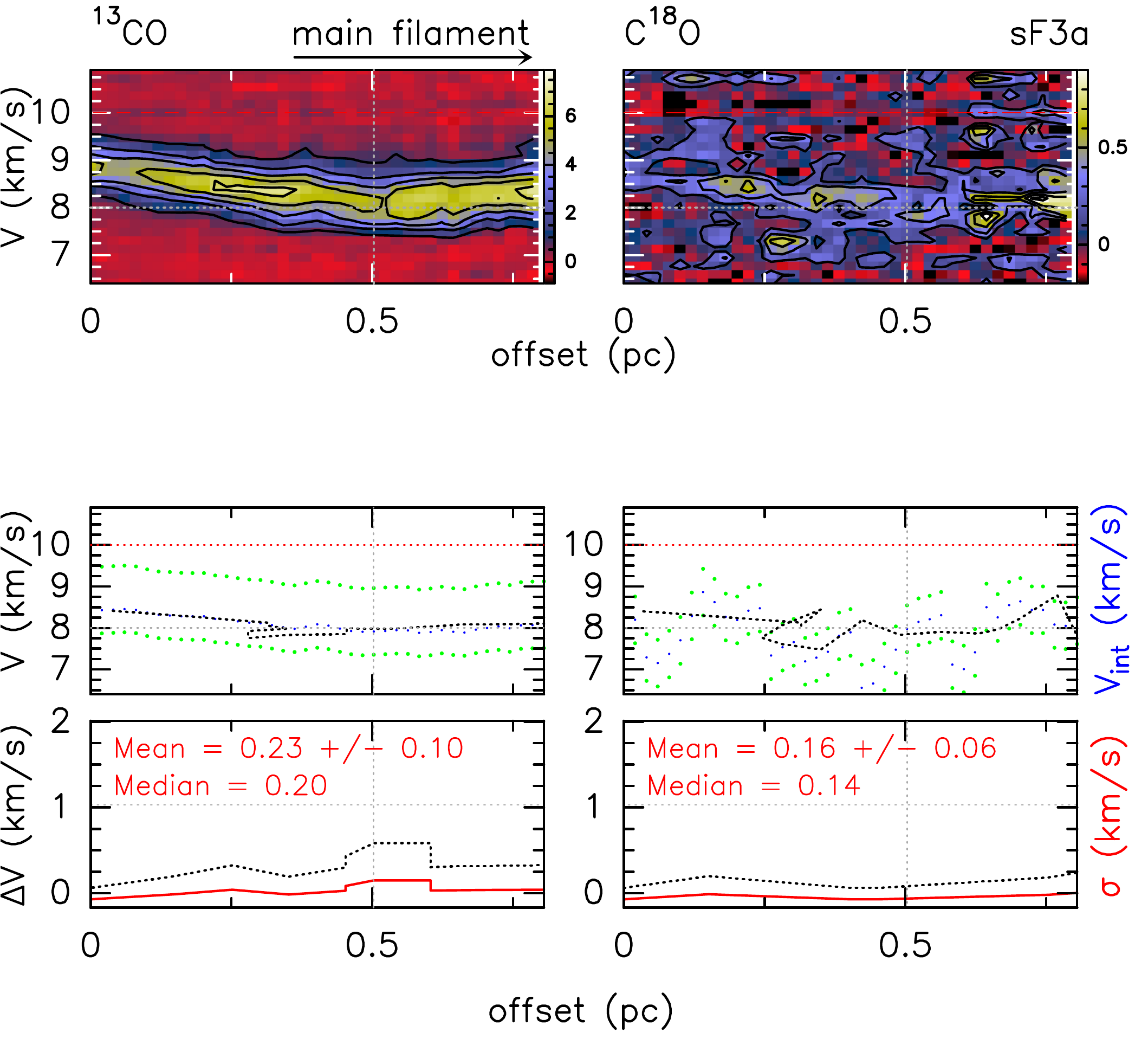} &
\includegraphics[width=0.9\columnwidth]{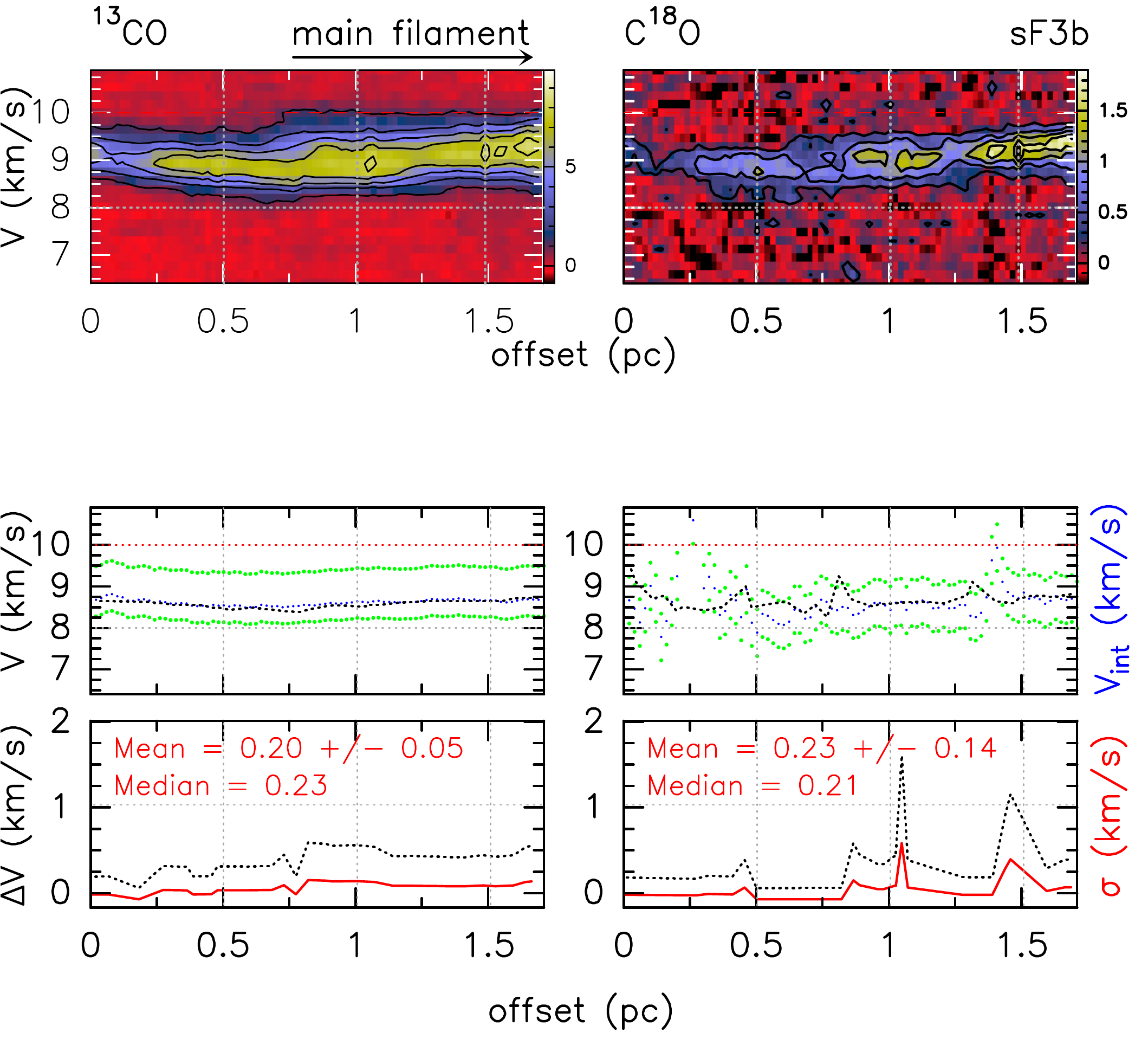} \\
\end{tabular}
}
\caption[]{continued.}
\label{A:pv1}
\end{figure*} 
\begin{figure*}[t!]
\ContinuedFloat
\centering 
\subfloat{
\begin{tabular}{c c}
\includegraphics[width=0.9\columnwidth]{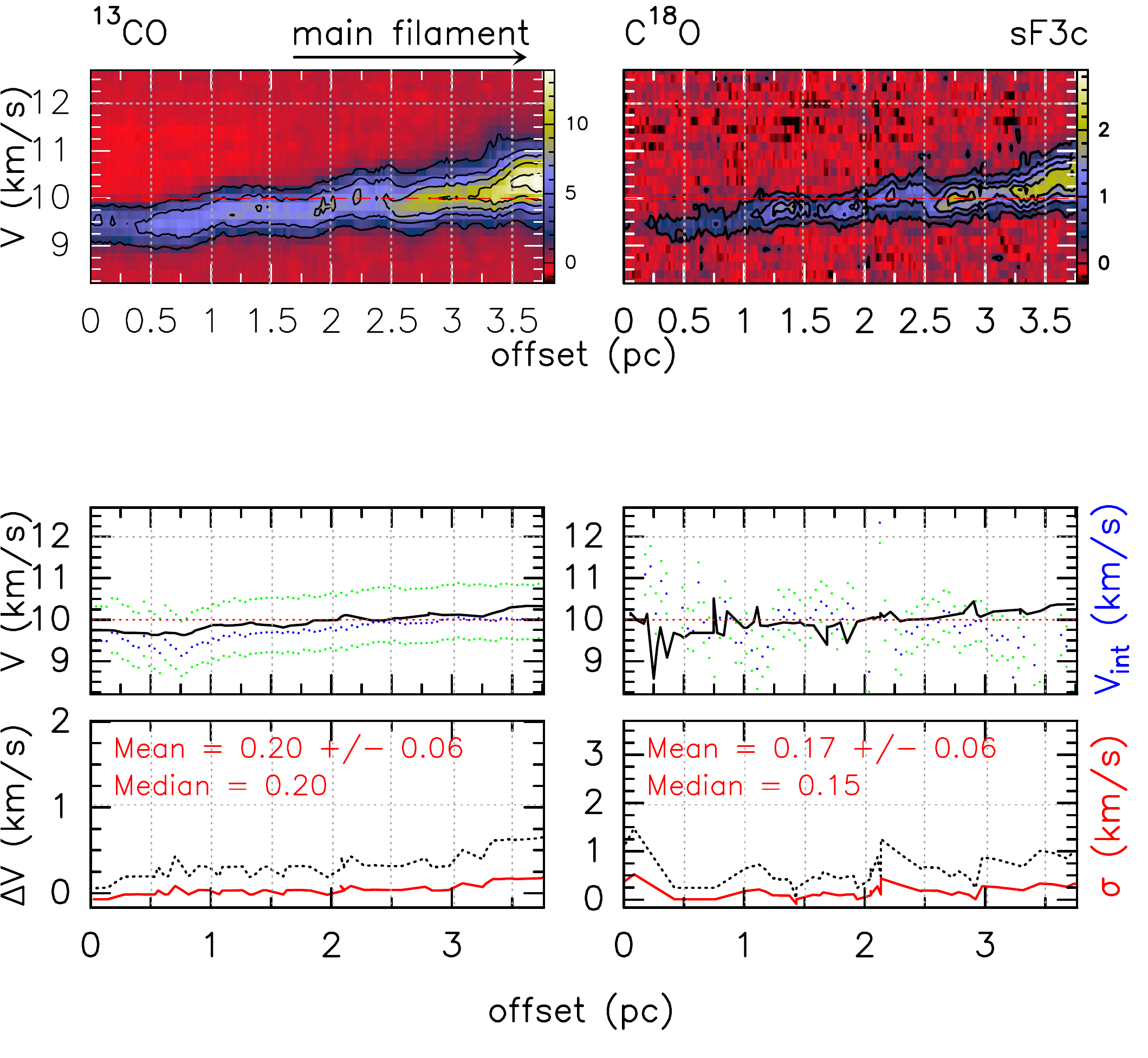} &
\includegraphics[width=0.9\columnwidth]{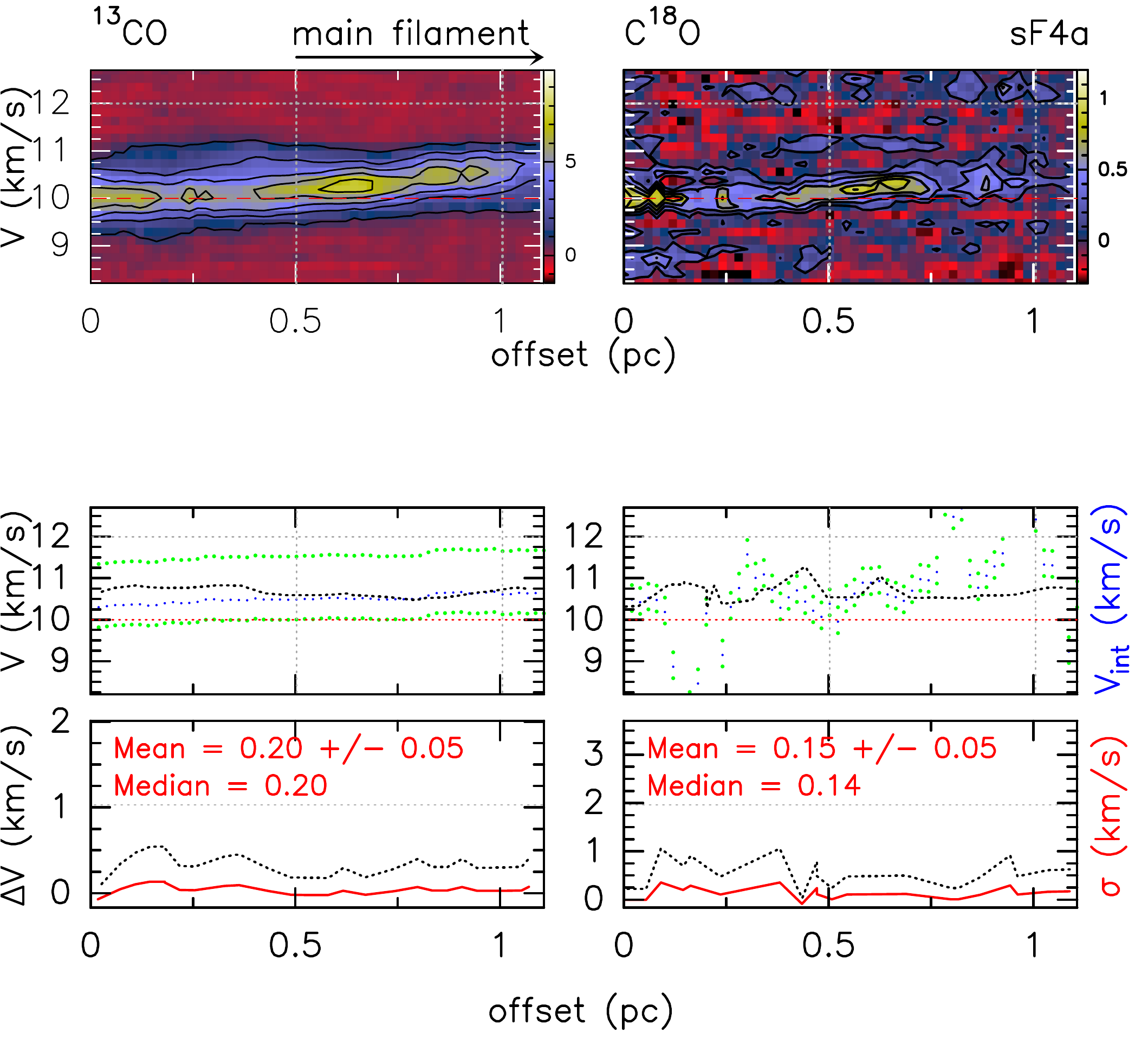} \\
\includegraphics[width=0.9\columnwidth]{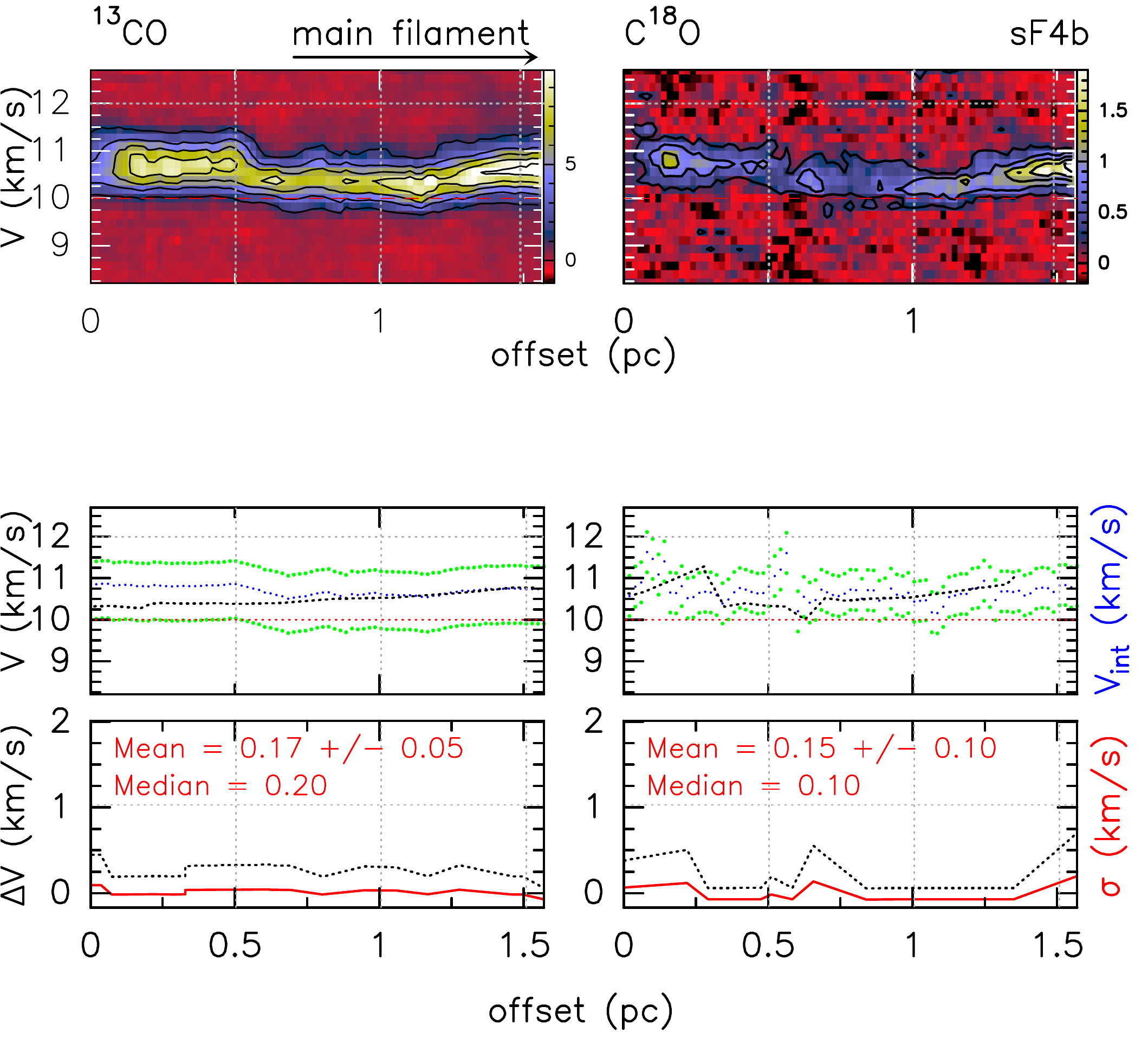} &
\includegraphics[width=0.9\columnwidth]{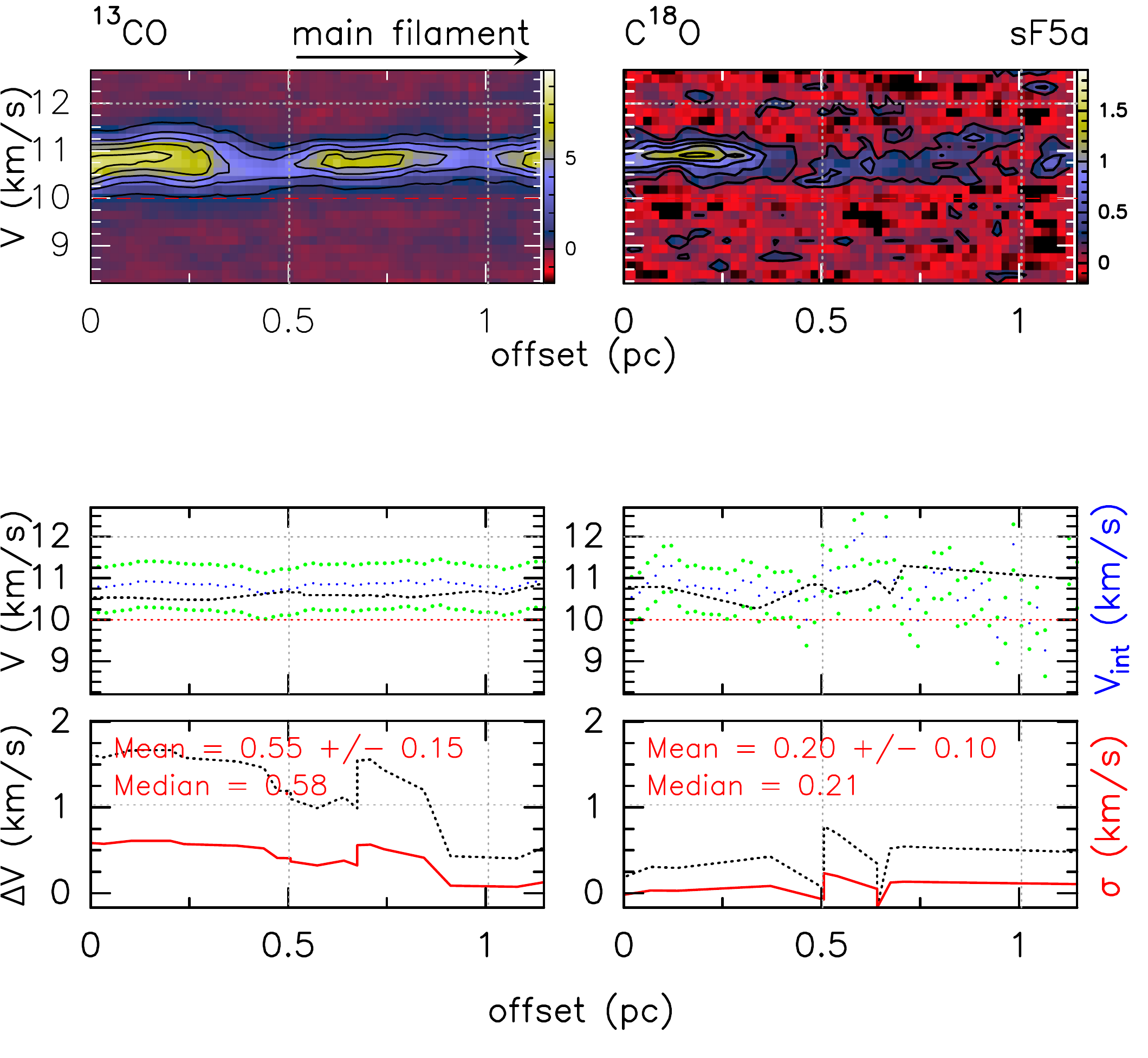} \\
\includegraphics[width=0.9\columnwidth]{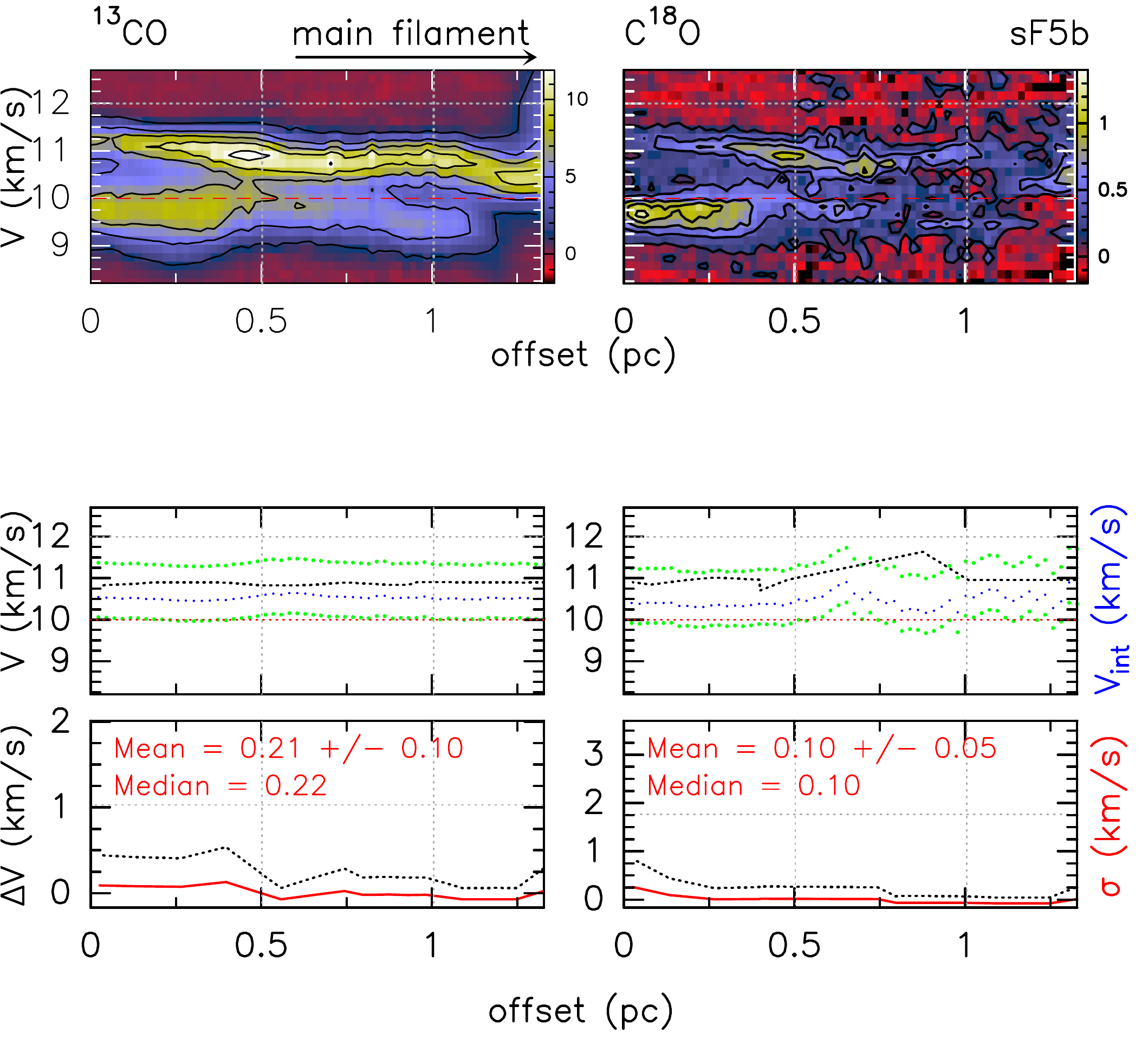} &
\includegraphics[width=0.9\columnwidth]{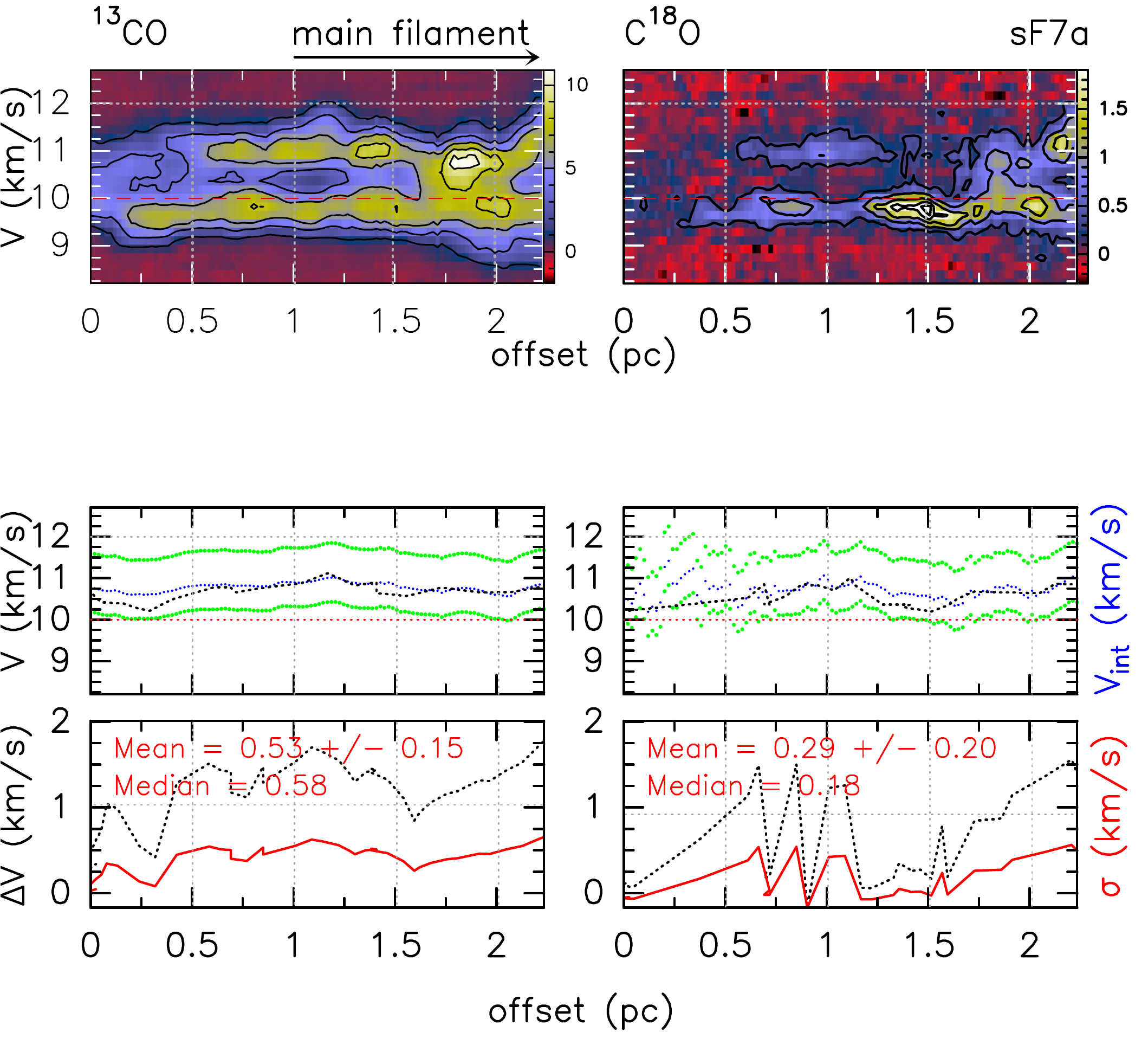} \\
\end{tabular}
}
\caption[]{continued.}
\label{A:pv1}
\end{figure*} 


\begin{figure*}[ht!]
\vspace{1cm}
\centering  
\includegraphics[angle=0, width=0.9\textwidth]{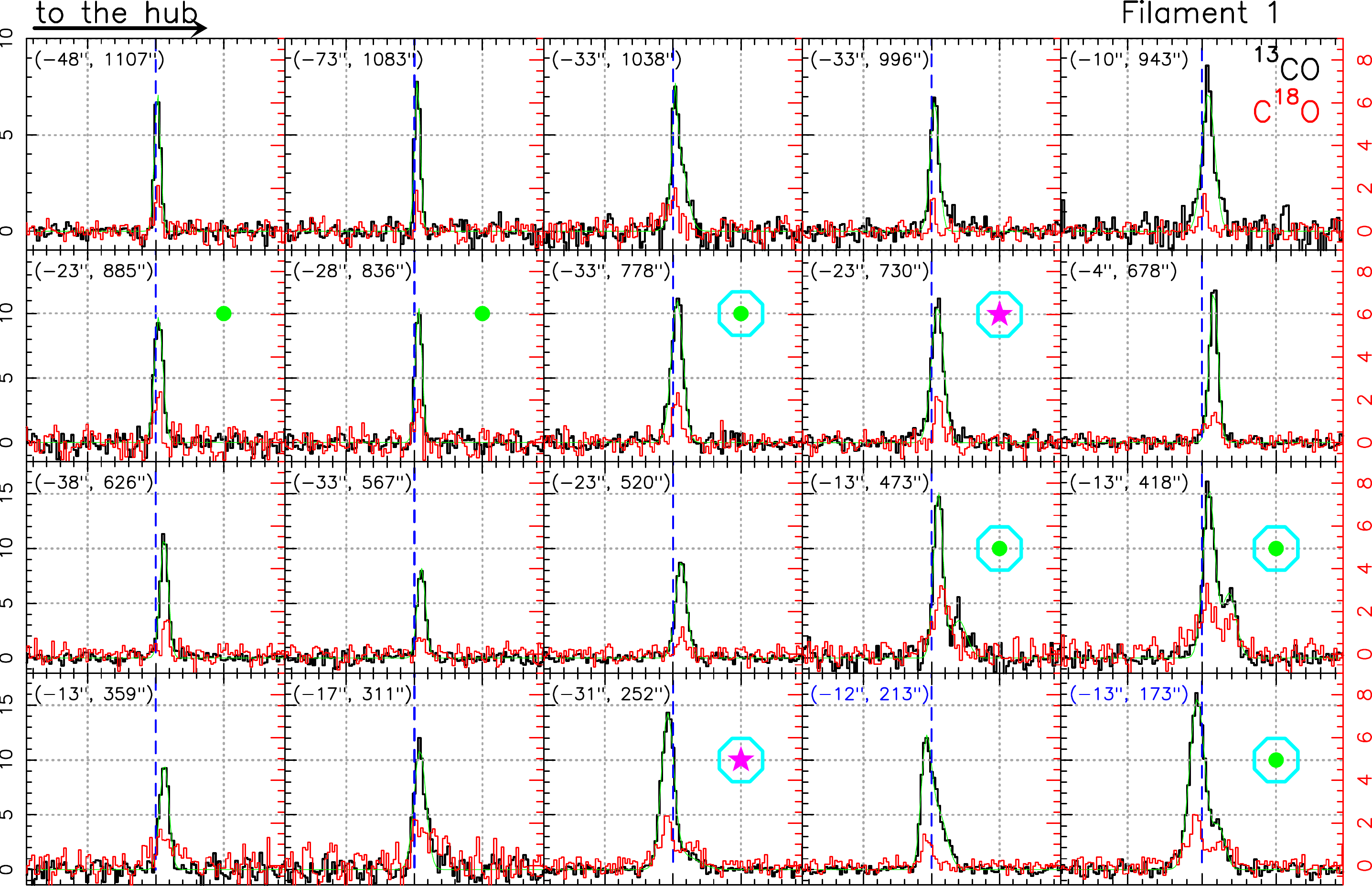} 
\vspace{-3cm}
\hspace{-0.23cm}
\includegraphics[angle=0, width=0.914\textwidth]{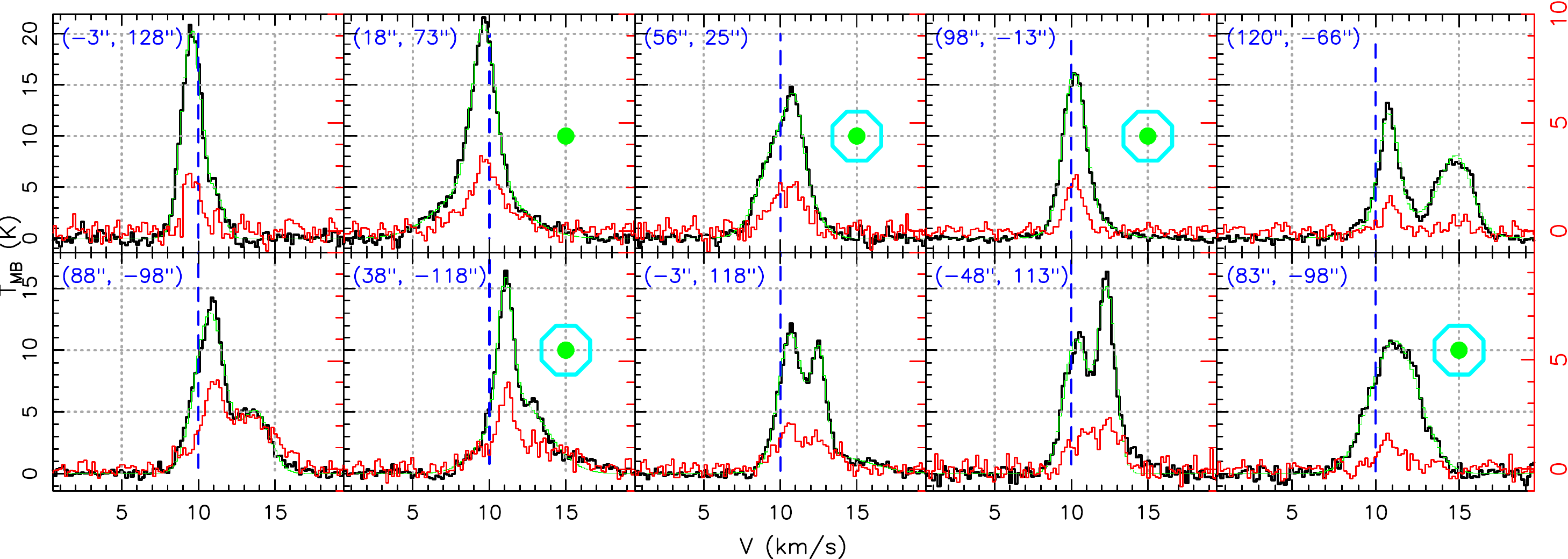}
\vspace{3cm}
\caption{$^{13}$CO (black) and C$^{18}$O (red) spectra along the Filament~1 skeleton. The green solid lines correspond to the Gaussian fits listed in Table~\ref{tab:gauss_parameters_1}. The blue dashed line shows the velocity of 10\kms. The corresponding positions for each spectra are indicated in the left-top corner of the panels. The positions corresponding to the filaments outside the hub are labeled in black. Those positions labeled in blue, corresponds to spectra inside the central hub ($R_{\mathrm{hub}} = 250"$). The colored symbols in the panels indicate the positions corresponding with sources identified be \cite{Rayner+2017}. The pink stars correspond to protostars, the green circles to bound cores, and the red triangles to unbound clumps. The large aqua circles corresponds to the sources identified by \cite{Sokol+2019}. Figs.~\ref{fig:spectra_2} to~\ref{fig:spectra_9} show the spectra along the Filaments~2 to 9 skeleton. Figs.~\ref{fig:spectra_10} and~\ref{fig:spectra_11} show the spectra along the secondary filament skeletons.}
\label{fig:spectra_1}
\end{figure*} 

\begin{figure*}[ht!]
\vspace{0.2cm}
\centering  
\includegraphics[angle=0, width=0.9\textwidth]{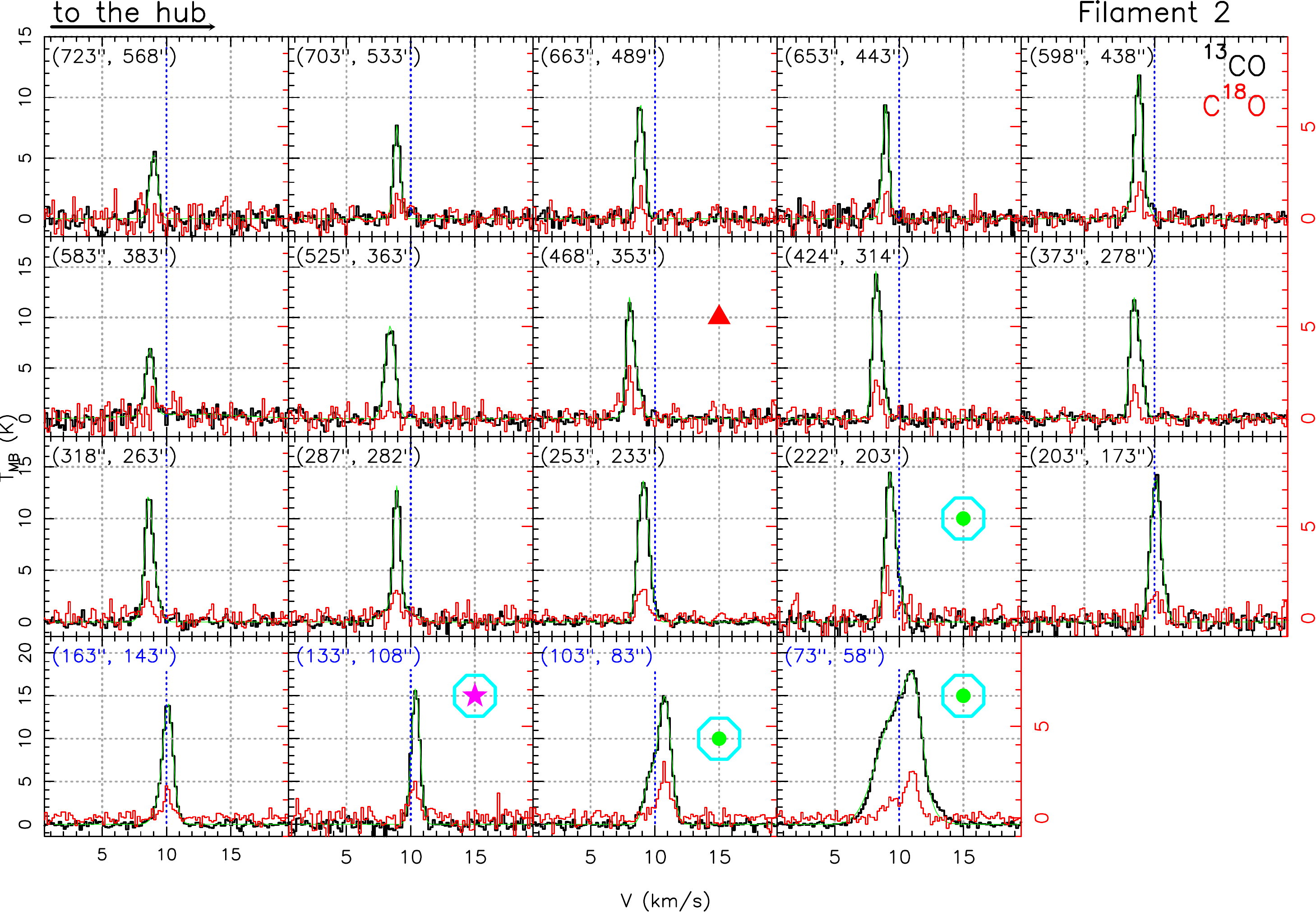} 
\caption{Same as Fig.~\ref{fig:spectra_1} for main filament F2}
\label{fig:spectra_2}
\end{figure*} 

\begin{figure*}[ht!]
\centering  
\includegraphics[angle=0, width=0.9\textwidth]{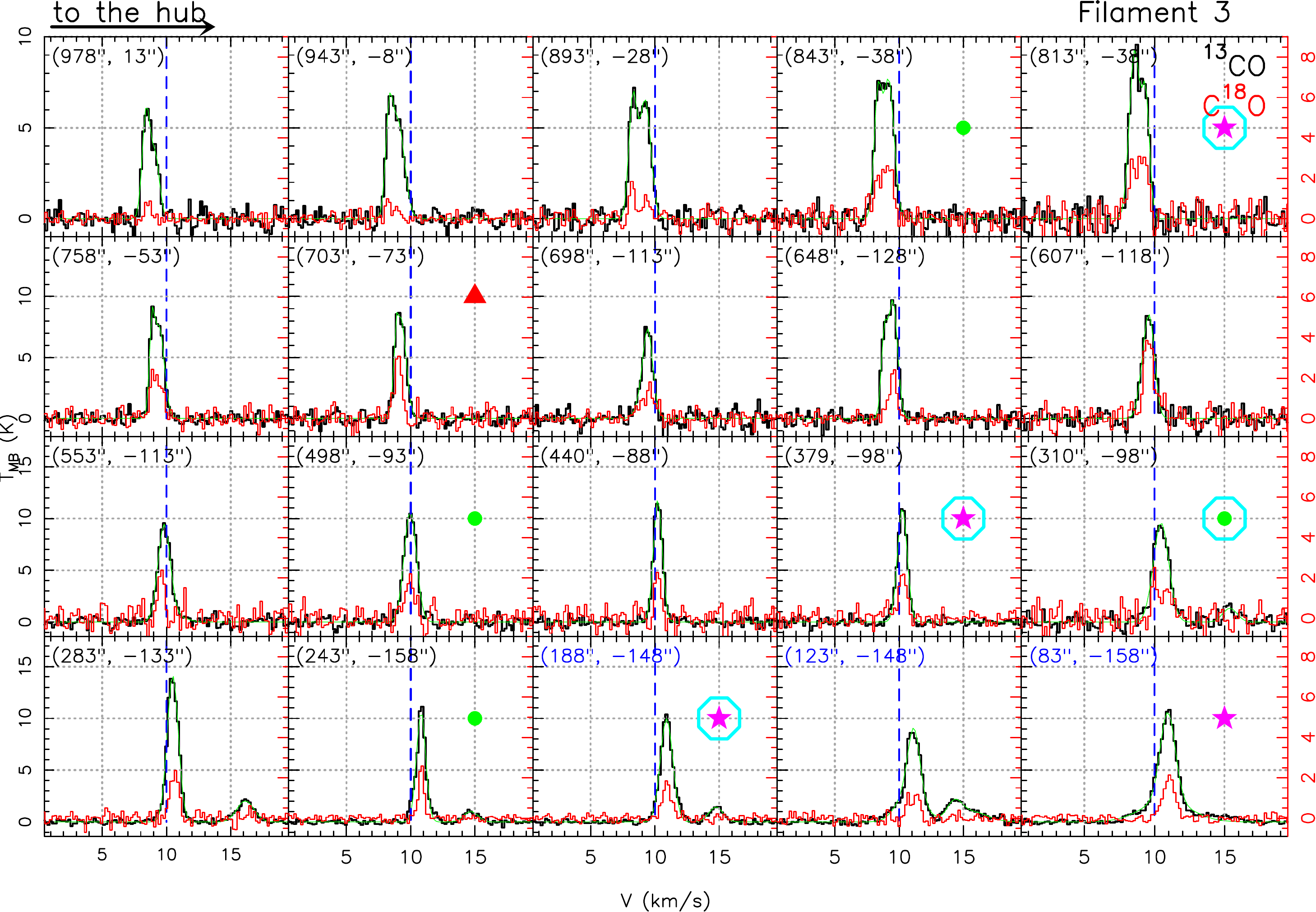}  
\caption{Same as Fig.~\ref{fig:spectra_1} for main filament F3}
\label{fig:spectra_3}
\end{figure*} 

\begin{figure*}[ht!]
\centering  
\vspace{0.5cm}
\includegraphics[angle=0, width=0.9\textwidth]{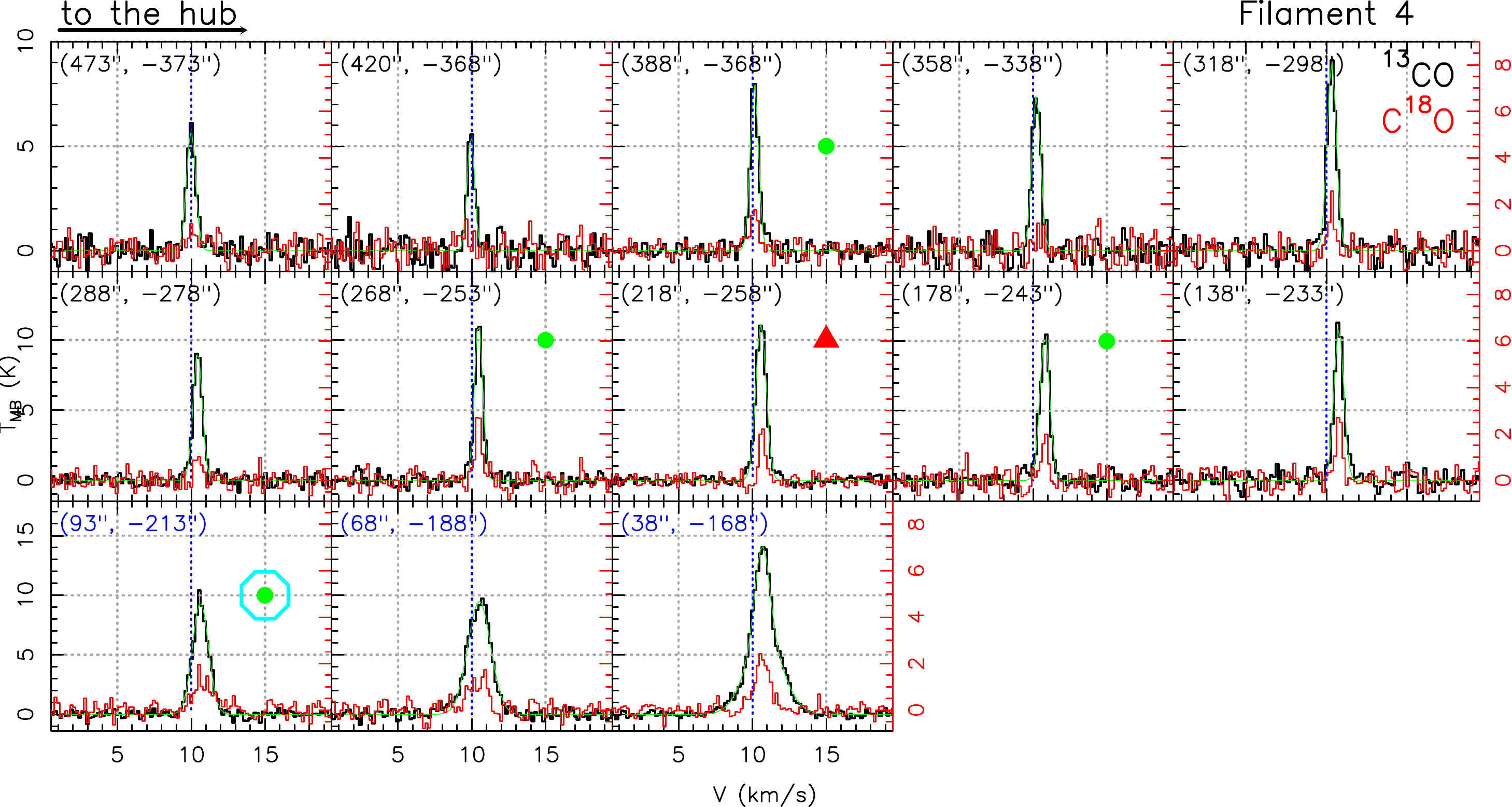} 
\vspace{0.5cm}
\caption{Same as Fig.~\ref{fig:spectra_1} for main filament F4}
\label{fig:spectra_4}
\end{figure*} 

\begin{figure*}[ht!]
\centering  
\vspace{0.5cm}
\includegraphics[angle=0, width=0.9\textwidth]{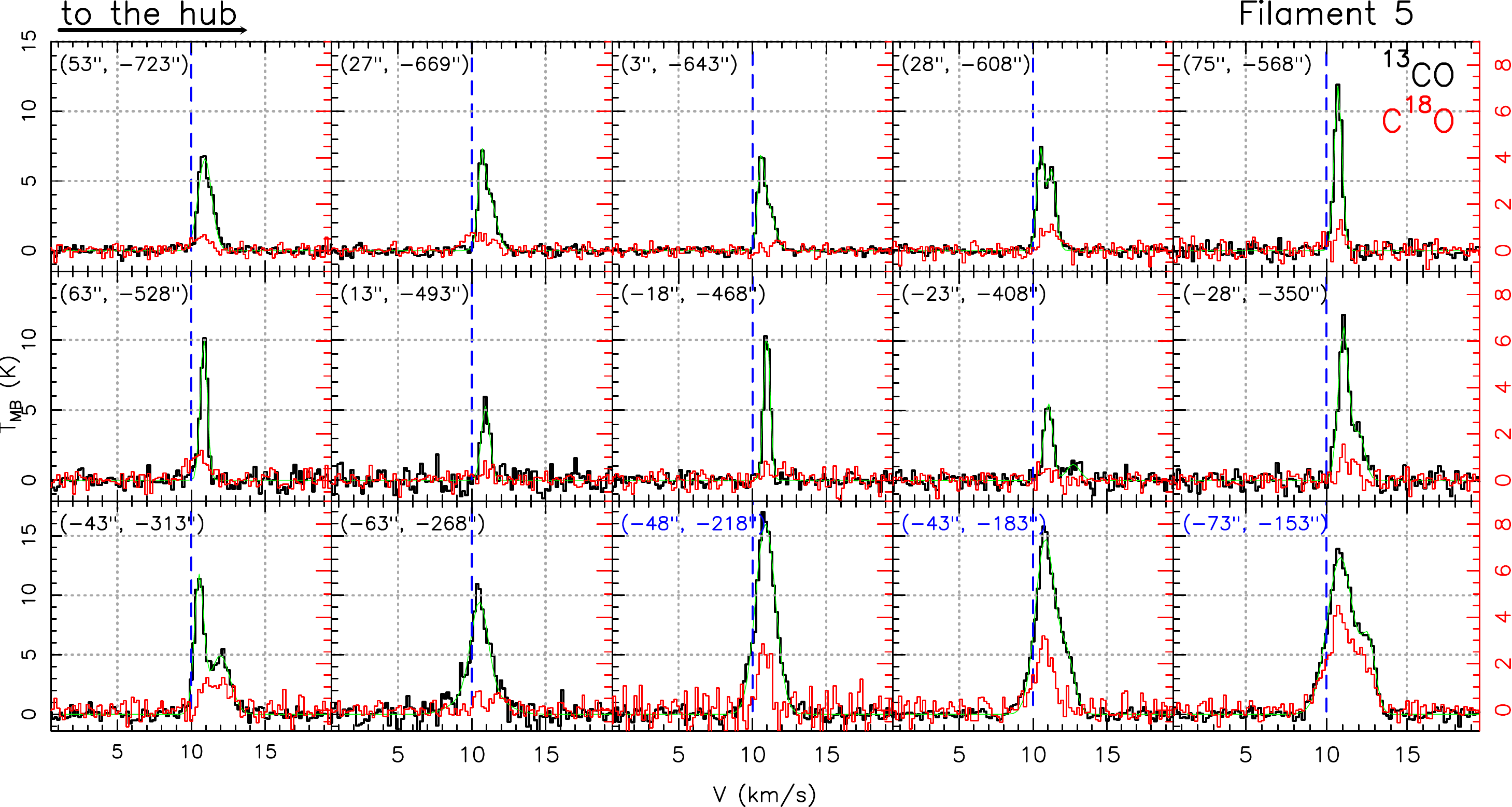} 
\vspace{0.5cm}
\caption{Same as Fig.~\ref{fig:spectra_1} for main filament F5}
\label{fig:spectra_5}
\end{figure*} 

\begin{figure*}[ht!]
\centering   
\vspace{0.5cm}
\includegraphics[angle=0, width=0.9\textwidth]{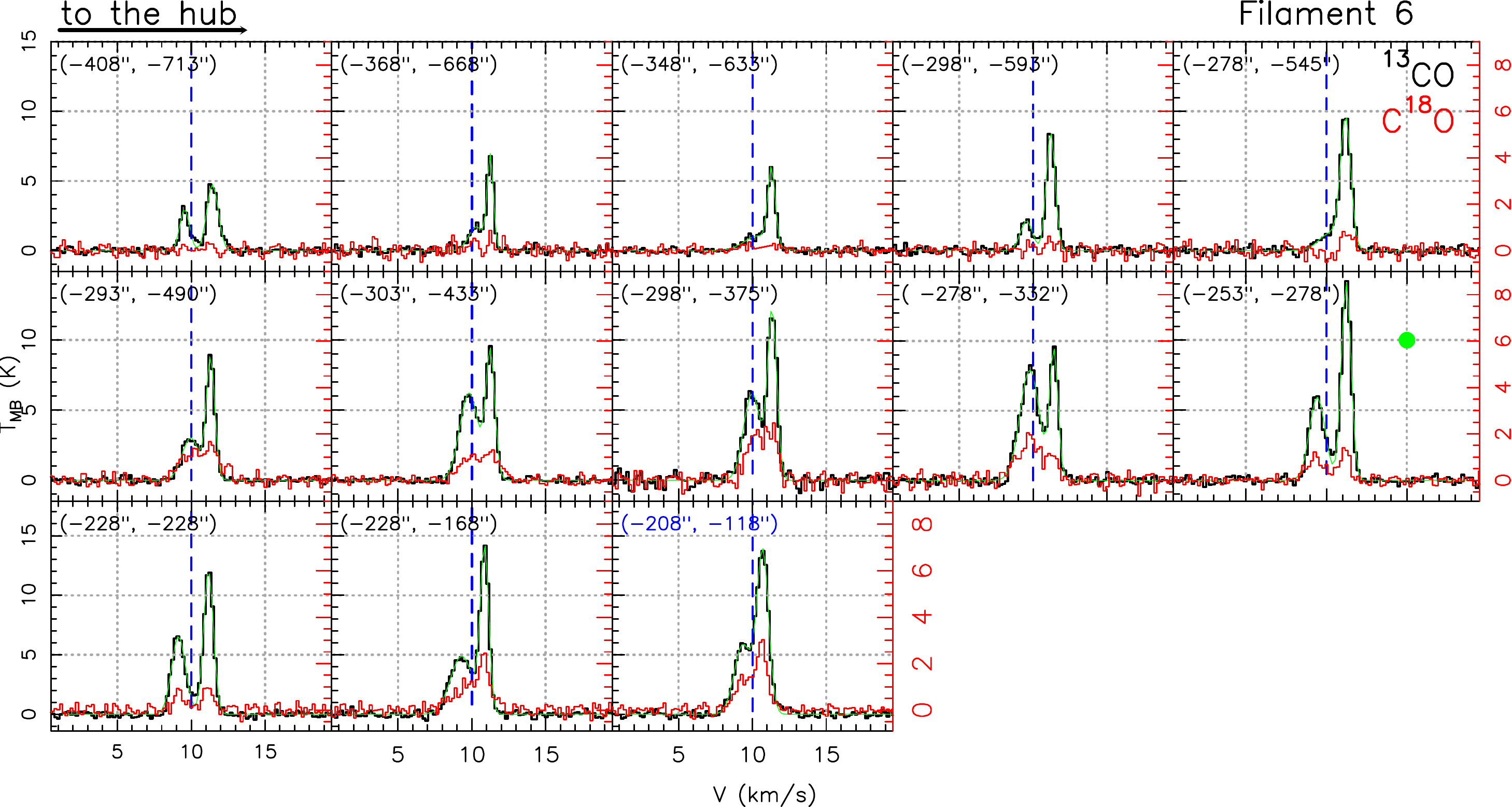}  
\vspace{0.5cm}
\caption{Same as Fig.~\ref{fig:spectra_1} for main filament F6}
\label{fig:spectra_6}
\end{figure*} 

\begin{figure*}[ht!]
\centering   
\vspace{0.5cm}
\includegraphics[angle=0, width=0.9\textwidth]{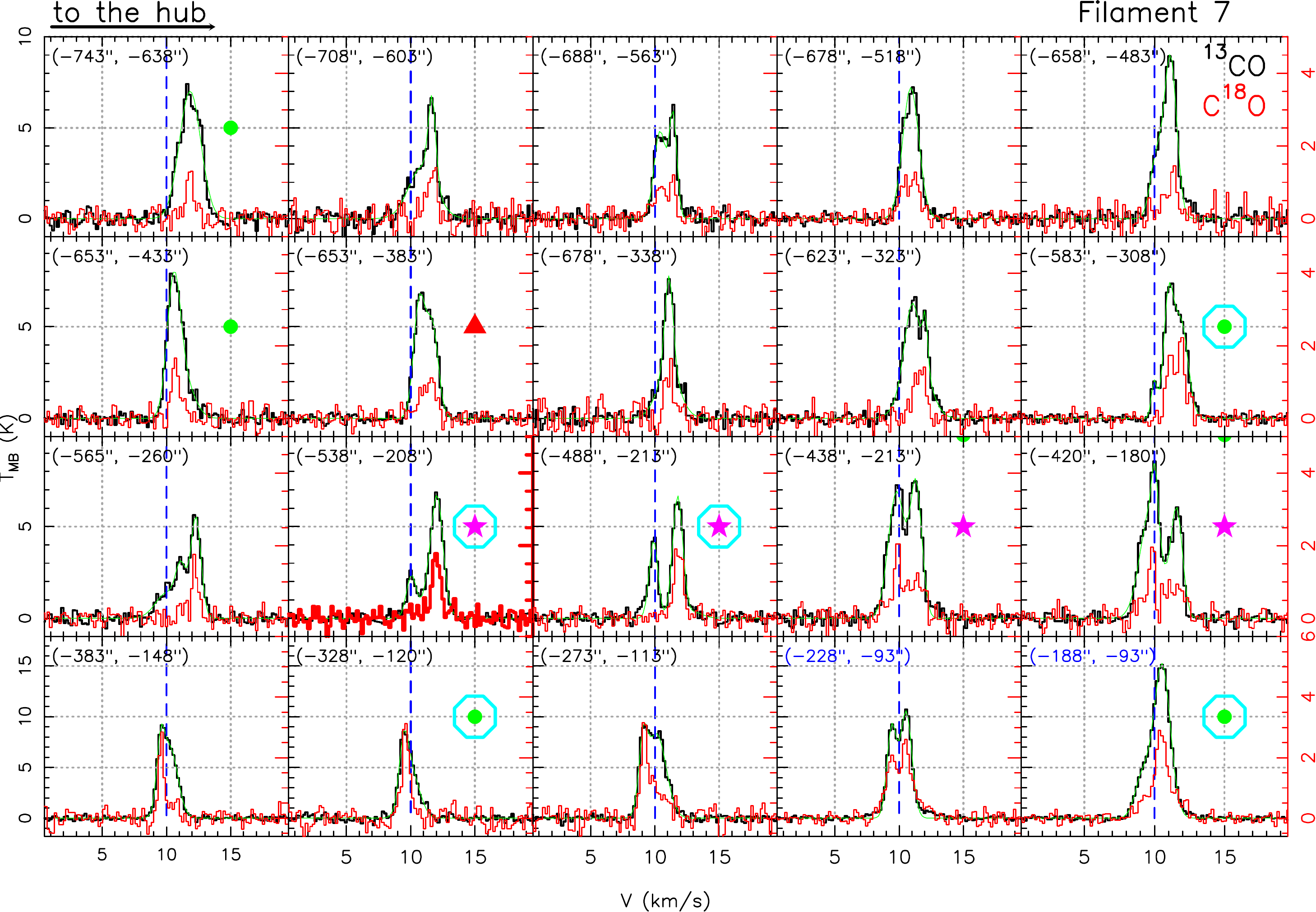}  
\vspace{0.5cm}
\caption{Same as Fig.~\ref{fig:spectra_1} for main filament F7}
\label{fig:spectra_7}
\end{figure*} 

\begin{figure*}[ht!]
\centering 
\vspace{0.5cm}  
\includegraphics[angle=0, width=0.9\textwidth]{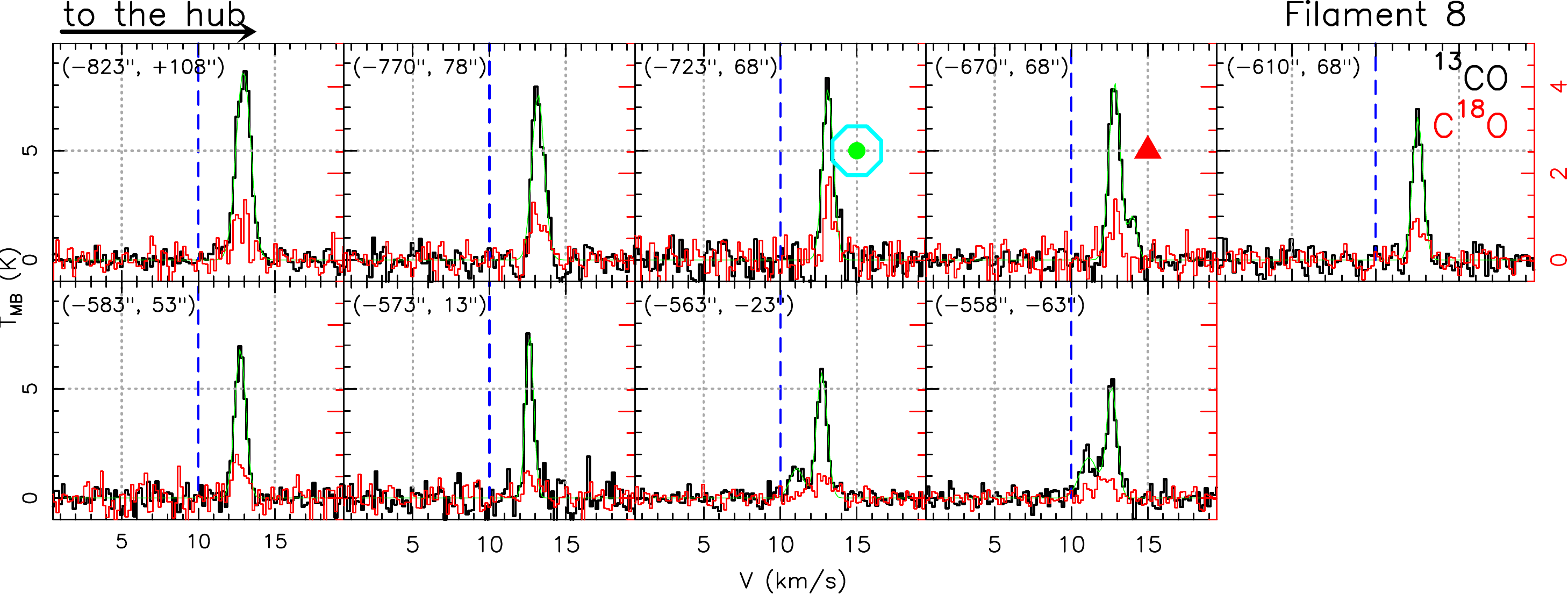} 
\vspace{0.5cm} 
\caption{Same as Fig.~\ref{fig:spectra_1} for main filament F8}
\label{fig:spectra_8}
\end{figure*} 

\begin{figure*}[ht!]
\centering   
\vspace{0.5cm}
\includegraphics[angle=0, width=0.9\textwidth]{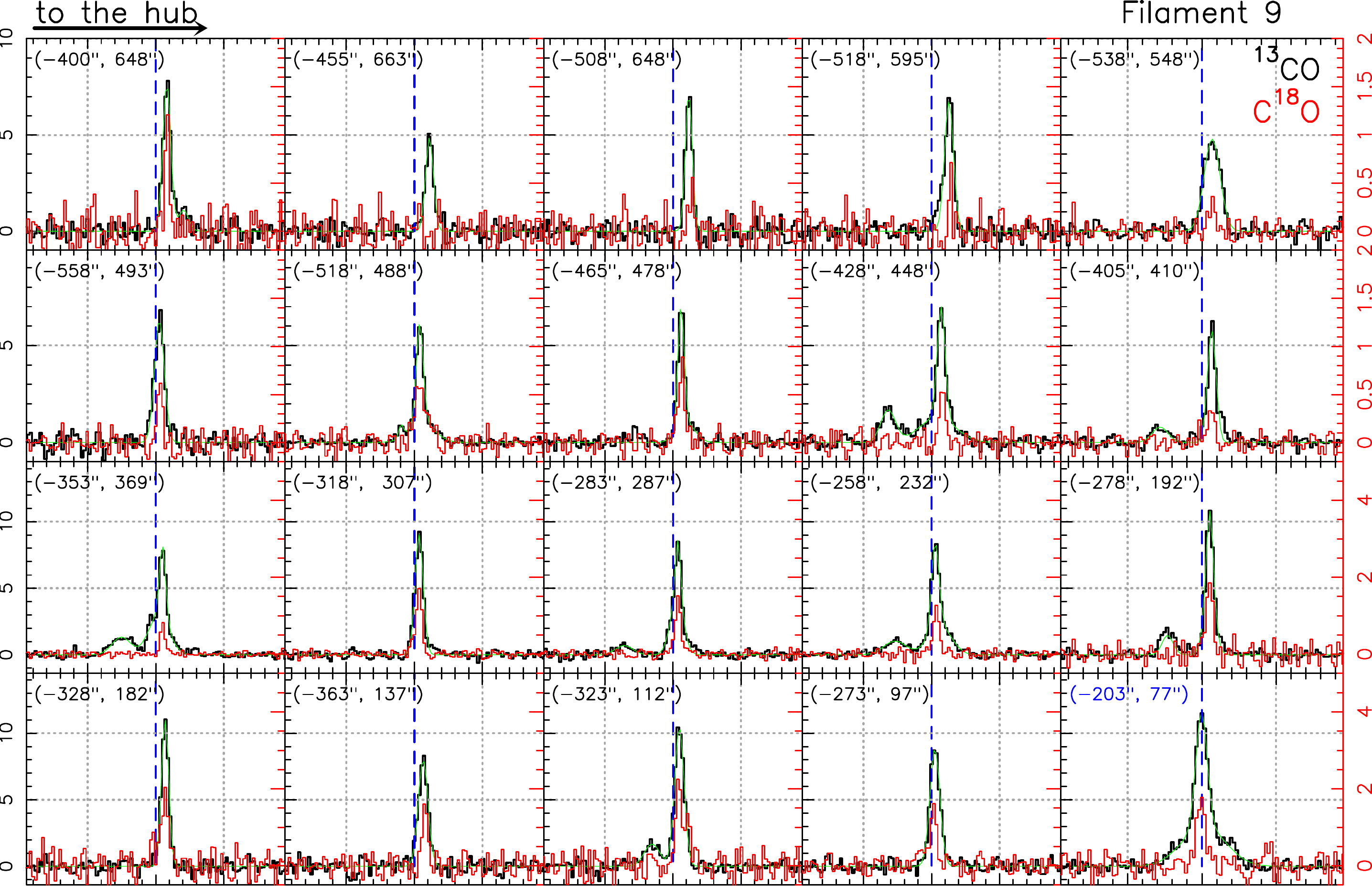} 
\vspace{-1cm}
\hspace{-0.32cm}
\includegraphics[angle=0, width=0.914\textwidth]{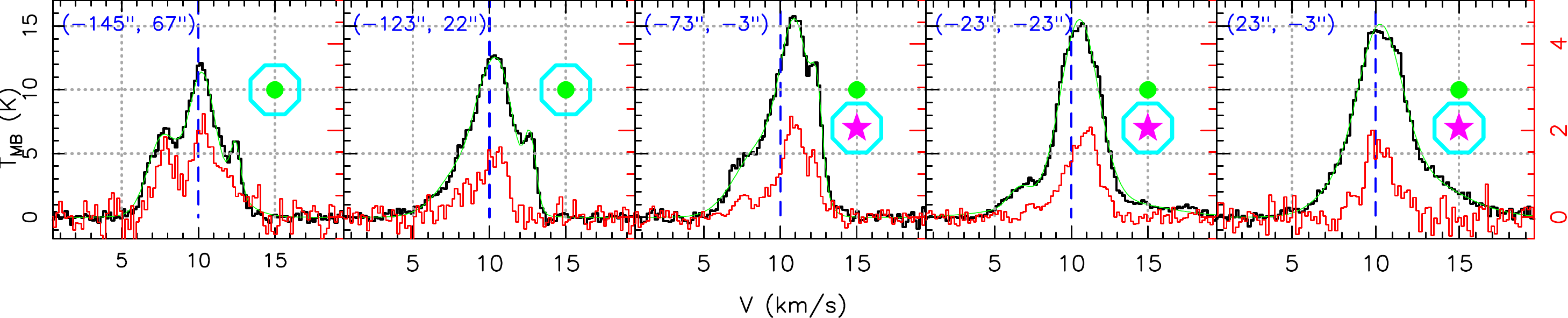}   
\vspace{1.5cm}
\caption{Same as Fig.~\ref{fig:spectra_1} for main filament F9}
\label{fig:spectra_9}
\end{figure*} 

\begin{figure*}[ht!]
\centering   
\vspace{0.5cm}
\includegraphics[angle=0, width=0.6\textwidth]{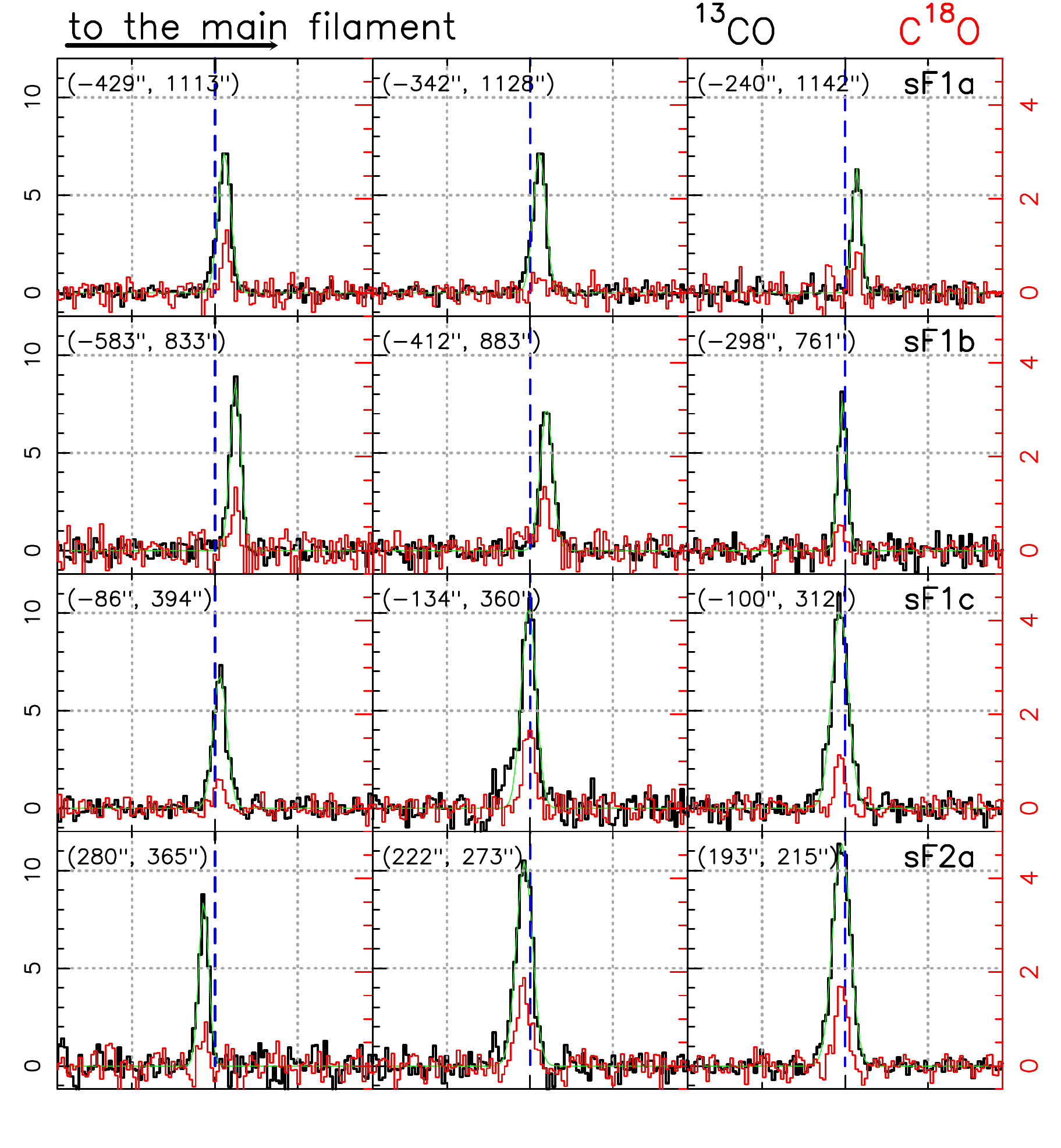}\\ 
\vspace{-1.25cm}
\includegraphics[angle=0, width=0.6\textwidth]{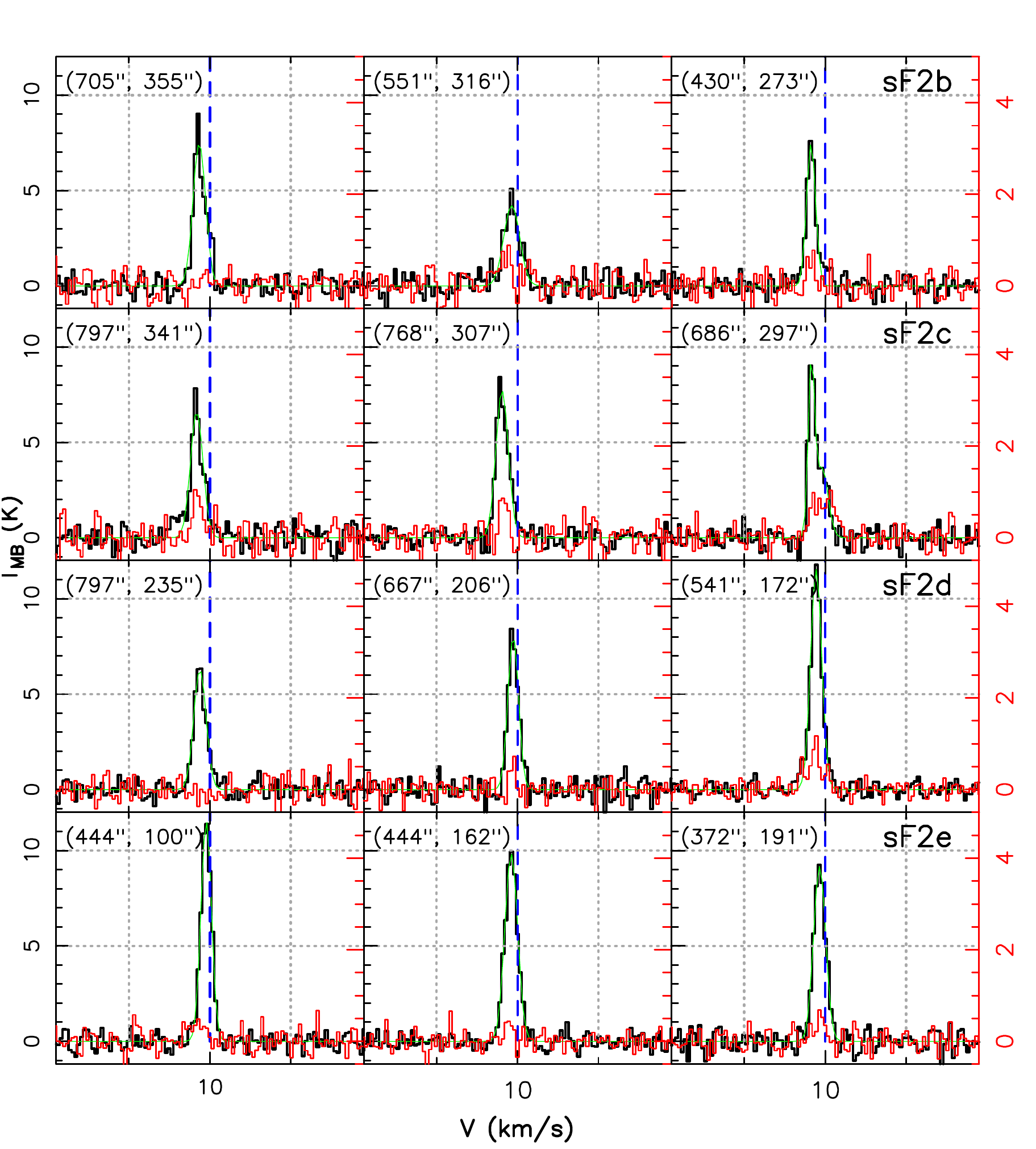}   
\caption{Same as Fig.~\ref{fig:spectra_1} for secondary filaments converging to F2}
\label{fig:spectra_10}
\end{figure*} 

\begin{figure*}[ht!]
\centering   
\vspace{0.5cm}
\includegraphics[angle=0, width=0.6\textwidth]{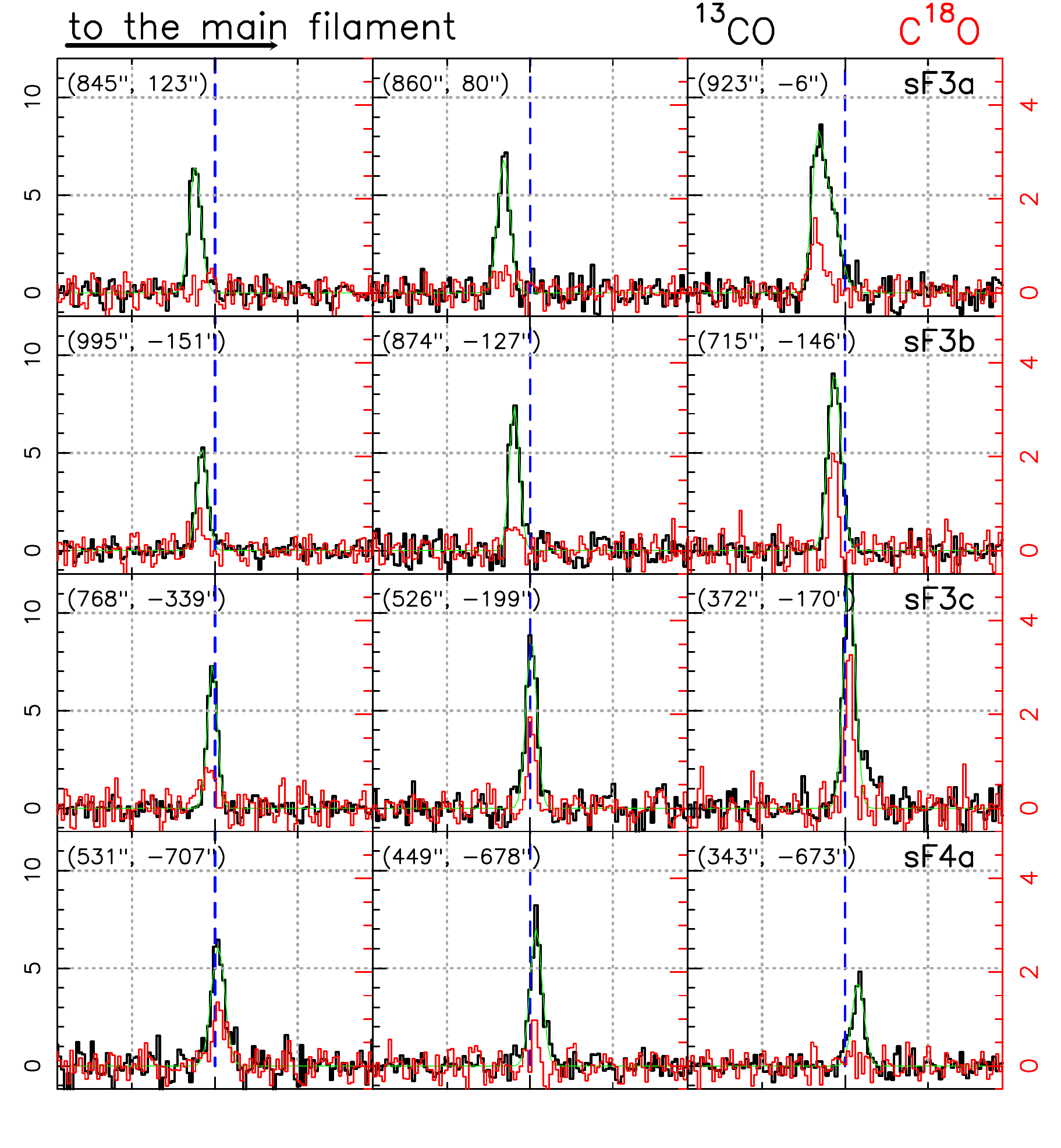}\\ 
\vspace{-1.25cm}
\includegraphics[angle=0, width=0.6\textwidth]{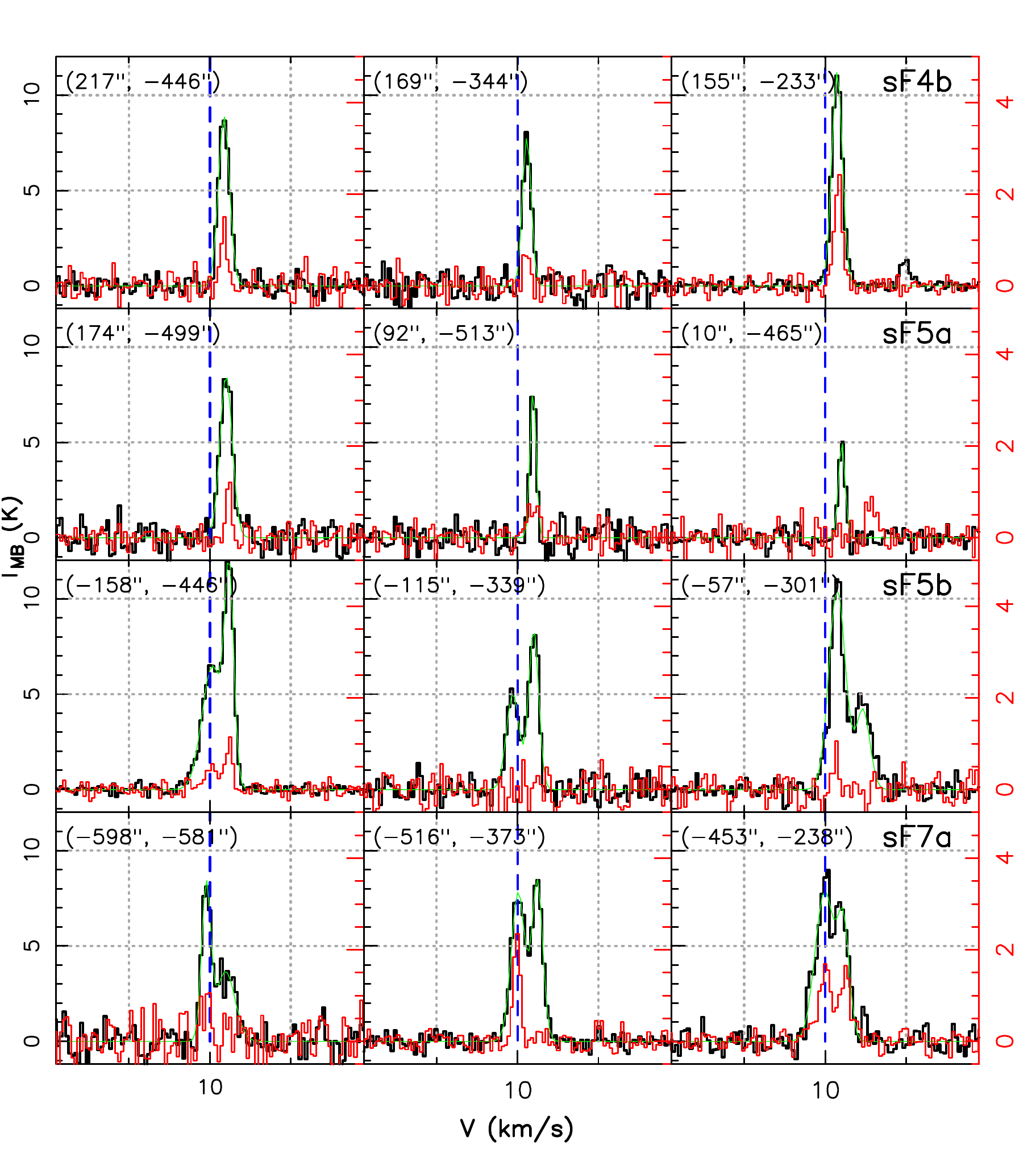}   
\caption{Same as Fig.~\ref{fig:spectra_1} for secondary filaments converging to F3, F4, F5 and F7.}
\label{fig:spectra_11}
\end{figure*} 


\clearpage
\onecolumn
\begin{longtable}{c | c c c c | c c c c}
\caption{\label{tab:gauss_parameters_1} Gaussian fit parameters for offsets along the filaments} \\
\hline\noalign{\smallskip}
\multicolumn{1}{c}{}
&\multicolumn{4}{c}{$^{13}$CO}
&\multicolumn{4}{c}{C$^{18}$O}
\\
\hline\noalign{\smallskip}
offset
&Area
&Vel
&$W$
&$T_\mathrm{peak}$
&Area
&Vel
&$W$
&$T_\mathrm{peak}$
\\
($''$,$''$)
&(K~km~s$^{-1}$)
&(km~s$^{-1}$)
&(km~s$^{-1}$)
&(K)
&(K~km~s$^{-1}$)
&(km~s$^{-1}$)
&(km~s$^{-1}$)
&(K)
\\
\endfirsthead
\caption{continued.} \\
\hline\noalign{\smallskip}
offset
&Area
&Vel
&$W$
&$T_\mathrm{peak}$
&Area
&Vel
&$W$
&$T_\mathrm{peak}$
\\
($''$,$''$)
&(K~km~s$^{-1}$)
&(km~s$^{-1}$)
&(km~s$^{-1}$)
&(K)
&(K~km~s$^{-1}$)
&(km~s$^{-1}$)
&(km~s$^{-1}$)
&(K)
\\
\hline\noalign{\smallskip}
\endhead
\hline\noalign{\smallskip}
\endfoot
\hline\noalign{\smallskip}
\multicolumn{9}{l}{Filament F1 --- outside the hub}\\
\hline\noalign{\smallskip}
(\phn$-$48, $+$1108)     &\phn4.1$\pm$0.1  &10.1$\pm$0.1    &0.5$\pm$0.1   &\phn7.3   &\phn0.9$\pm$0.1   &10.2$\pm$0.1      & 0.4$\pm$0.1   &  2.1\\
(\phn$-$73, $+$1083)     &\phn4.2$\pm$0.2 &10.2$\pm$0.1     &0.5$\pm$0.1   &\phn8.0   &\phn0.5$\pm$0.1   &10.2$\pm$0.1      & 0.2$\pm$0.1   &  2.2 \\
(\phn$-$33, $+$1038)     &\phn2.3$\pm$0.7 &10.1$\pm$0.1     &0.5$\pm$0.1   &\phn4.7   &\phn2.2$\pm$0.3   &\phn9.9$\pm$0.1   & 1.6$\pm$0.3   &  1.3 \\
---                      &\phn5.0$\pm$0.8 &10.5$\pm$0.1     &1.3$\pm$0.1   &\phn3.7   &\phn0.6$\pm$0.1   &10.1$\pm$0.1      & 0.3$\pm$0.1   &  1.9\\
(\phn$-$33, \phn$+$995)  &\phn5.6$\pm$0.2 &10.2$\pm$0.1     &0.8$\pm$0.1   &\phn6.7   &\phn0.7$\pm$0.1   &10.2$\pm$0.1      & 0.4$\pm$0.1   &  1.6\\
(\phn$-$10, \phn$+$943)  &\phn7.8$\pm$0.4 &10.4$\pm$0.1     &1.1$\pm$0.1   &\phn7.2   &\phn1.4$\pm$0.2   &10.2$\pm$0.1      & 0.6$\pm$0.1   &  2.3\\
(\phn$-$23, \phn$+$885)  &\phn7.8$\pm$0.3 &10.2$\pm$0.1     &0.8$\pm$0.1   &\phn9.8   &\phn0.9$\pm$0.2   &10.3$\pm$0.1      & 0.4$\pm$0.1   &  2.1\\
\hline
\end{longtable}

\end{appendix}
\end{document}